\documentclass[10pt]{article}
\usepackage{amsfonts}
\usepackage{amssymb}
\usepackage{amsthm}

\usepackage[pdftex]{graphicx}

\newcommand{\sss}{\setcounter{equation}{0}}
\newtheorem{theorem}{THEOREM}[section]

\newtheorem{lemma}[theorem]{LEMMA}

\newtheorem{remark}[theorem]{REMARK}

\newtheorem{definition}[theorem]{DEFINITION}

\newtheorem{assumption}[theorem]{ASSUMPTION}

\newtheorem{aharonov-bohm-ansatz}[theorem]{Aharonov-Bohm Ansatz}

\newcommand{\ag }{\widetilde{\mathbf A}}
\newcommand{\av }{\overline{\mathbf A}}
\newcommand{\ap }{\mathbf A}
\newcommand{\ere}{ {\mathbb R}}
\newcommand{\ZETA}{{\mathbb Z}}

\newcommand{\CE}{{\mathbb C}}

\def\p2{\mathcal A_{\Phi,2\pi}(B)}
\def\0p2{\mathcal A_{\Phi,2\pi}(0)}
\def\sp2{\mathcal A_{\Phi,2\pi,\hbox{\rm SR}}(B)}

\def\beq{\begin{equation}}
\def\ene{\end{equation}}
\def \ds {\displaystyle}
\newcommand{\bull}{\hfill $\Box$}
\def\qed{\ifhmode\unskip\nobreak\fi\ifmmode\ifinner
\else\hskip5pt\fi\fi\hbox{\hskip5pt\vrule width4pt height6pt
depth1.5pt\hskip1pt}}
\def\v{\mathbf v}

\def\hv{\hat{\mathbf v}}

\def\mo{\mathbf p}

\def\et{e^{-iz H_1}}
\def \12{\tau_{1/ \sqrt{2},\sigma_1,\sigma_2}}
\def \32{\tau_{\sqrt{3}/ 2,\sigma_1,\sigma_2}}
\def\te{\tilde{\varepsilon}}
\def\td{\tilde{\delta}}
\begin{document}
\baselineskip=20 pt
\parskip 6 pt

\title{The Aharonov-Bohm Effect and Tonomura et al. Experiments. Rigorous Results
\thanks{ PACS Classification (2008): 03.65Nk, 03.65.Ca, 03.65.Db, 03.65.Ta.  Mathematics Subject Classification(2000): 81U40, 35P25,
35Q40, 35R30.}
\thanks{ Research partially supported by
 CONACYT under Project P42553­F.}}
\author{ Miguel Ballesteros\thanks{ On Leave of absence from Departamento de M\'etodos
 Matem\'aticos  y Num\'ericos. Instituto de Investigaciones en Matem\'aticas Aplicadas y en
 Sistemas. Universidad Nacional Aut\'onoma de M\'exico. Apartado Postal 20-726,
M\'exico DF 01000.} \thanks{ Electronic Mail: ballesteros.miguel.math@gmail.com}\, and Ricardo Weder$^{\ddag}$\thanks {Fellow,
Sistema Nacional de Investigadores. Electronic mail: weder@servidor.unam.mx  }
\\
 Institut National de Recherche en Informatique et en Automatique
Paris-Rocquencourt.\\
 Projet POems. Domaine de Voluceau-Rocquencourt, BP 105,
78153, Le Chesnay Cedex  France.}

\date{}
\maketitle
\begin{center}
\begin{minipage}{5.75in}
\centerline{{\bf Abstract}}
\bigskip
The  Aharonov-Bohm effect is a fundamental issue in physics. It
describes the physically important electromagnetic quantities in
quantum mechanics. Its experimental verification constitutes a test
of the theory of quantum mechanics itself. The remarkable
experiments of Tonomura et al. [``Observation of
Aharonov-Bohm effect by electron holography," Phys. Rev. Lett. {\bf
48}, 1443 (1982),  ``Evidence for Aharonov-Bohm effect with magnetic field
completely shielded from electron wave", Phys. Rev. Lett. {\bf 56}, 792 (1986)] are widely considered as the only
experimental evidence of the physical existence of the Aharonov-Bohm
effect. Here we give the first rigorous proof that the classical
Ansatz of Aharonov  and Bohm of 1959 [``Significance of electromagnetic potentials in the quantum theory,"
Phys. Rev. {\bf 115}, 485 (1959)], that  was  tested by Tonomura
et al., is a good approximation to the exact solution to the
Schr\"odinger equation. This also proves that the electron, that is
represented by the exact solution, is not accelerated, in agreement
with the recent experiment of Caprez et al. in 2007 [``Macroscopic test of the Aharonov-Bohm effect,"
Phys. Rev. Lett. {\bf 99}, 210401 (2007)], that shows that
the results of the Tonomura et  al. experiments can not be explained
by the action of a force. Under the assumption that the incoming
free electron is a gaussian wave packet, we estimate the exact
solution to the Schr\"odinger equation for all times. We provide a
rigorous, quantitative error bound for the difference in norm
between the exact solution and  the Aharonov-Bohm Ansatz. Our bound
is uniform in time. We also  prove that on the gaussian asymptotic
state  the scattering operator is given by  a constant phase shift,
up to a quantitative  error bound  that we provide. Our results show
that for  intermediate  size electron wave packets, smaller than the
ones used in the Tonomura et al. experiments, quantum mechanics
predicts the results observed by Tonomura et al. with an error bound
smaller than $10^{-99}$. It would be quite interesting to perform
experiments with electron wave packets of intermediate size.
Furthermore, we provide a physical interpretation of our error
bound.

\end{minipage}
\end{center}
\newpage

\section{Introduction}
\sss
In classical electrodynamics  the force produced by a magnetic field on a charged particle is given by the Lorentz force, $F= q \v \times B$, where
$q$ and $\v$ are, respectively, the charge and the velocity of the particle, and $B$ is the magnetic field. In regions where the magnetic
field is zero the Lorentz force is zero and the particle travels in a straight line. In particular, the dynamics of a classical particle is unaffected
by magnetic fields enclosed in regions that are not accessible   to the particle. This also means that in classical electrodynamics the relevant physical
quantity is the magnetic field and that the magnetic potentials are only a
convenient mathematical tool.

The situation is different in quantum mechanics, where the dynamics is described by the Schr\"odinger equation that can not be formulated
directly in terms of the magnetic field. It is required to introduce the magnetic potential. It was  pointed out by Aharonov and Bohm \cite{ab}
that this implies that in quantum mechanics the magnetic potentials have a  real physical significance.
Aharonov and Bohm \cite{ab} proposed an experiment to confirm the theoretical prediction.
They suggested to use a thin, straight solenoid, centered at the origin and with axis in the vertical direction. They supposed that the magnetic field
was essentially confined to
the solenoid. They advised to employ  a coherent electron wave packet that splits in two parts, each one going trough
one side of the solenoid. Both wave packets should be brought together behind the solenoid, to create an
interference pattern due to the difference in phase in the wave function of each part of the wave packet, produced by the magnetic field
 enclosed inside the
solenoid. Actually, the existence of this interference pattern was
first predicted by Franz \cite{f}. The Aharonov-Bohm effect  plays a
prominent role in fundamental physics, among other reasons, because
it describes the physically important electromagnetic quantities in
quantum mechanics, and since it is a quantum mechanical effect, the
verification of its existence constitutes a test of the validity of
the theory of quantum mechanics itself.

The case of a solenoid has been extensively studied  from the theoretical
and experimental points of view. The theoretical analysis is reduced to a two
dimensional problem after making the assumption that the solenoid is
infinite. Nevertheless, experimentally it is impossible to have an infinite
solenoid and, therefore, the magnetic field can not be completely confined into the
solenoid. The leakage of the magnetic field was a highly controversial point. To avoid this problem it  was
suggested to use a  toroidal magnet, that can contain a magnetic field inside without a leak.
The experiments with toroidal magnets where carried over by Tonomura et al. \cite{to3,to1,to2}.
In remarkable experiments they were able to superimpose
 behind the magnet an electron wave packet that traveled inside the hole of the magnet with another electron wave packet that
traveled outside the magnet, and they measured the phase shift
produced by the magnetic flux enclosed in the magnet, giving a
strong evidence of the existence of the Aharonov-Bohm effect. In
fact, the Tonomura et al. experiments \cite{to3,to1,to2}  are widely
considered  as the only experimental evidence of the existence of
the Aharonov-Bohm effect.

In the case of  toroidal magnets, several Ans\"atze have been provided for the solution to the
Schr\"odinger equation and for the scattering matrix without giving error bound
estimates for the difference, respectively, between the exact solution and the exact scattering matrix, and the Ans\"atze. Most
of these works are qualitative, although some of them give numerical values for
their Ans\"atze. Methods like,  Fraunh\"ofer diffraction, first-order Born and high-energy
approximations, Feynman path integrals and the Kirchhoff method in optics were used to propose the
Ans\"atze. The amount of work  related to the Aharonov-Bohm effect is very large. For a
review of the literature up to 1989 see \cite{op} and \cite{pt}. In particular,
in \cite{pt} there is a detailed discussion of the large
controversy  -involving over three hundred papers-
 concerning the existence of the Aharonov-Bohm effect. For a recent update of this controversy see \cite{to,tn}.

 The paper \cite{be} presents a discussion of a version of the Aharonov-Bohm Ansatz for an infinite solenoid. For recent rigorous work in the case
 of an infinite solenoid see \cite{n,w1} where, among other results, it is proven that in the high-velocity limit  the scattering operator is  given by
  a constant phase shift, as predicted by Franz \cite{f} and Aharonov and Bohm
  \cite{ab}. In \cite{ope} rigorous mathematical ground is given
  for the presence of the magnetic potential in the Schr\"odinger
  operator describing the Aharonov-Bohm effect in the case of a
  solenoid.
 In \cite{h}, a semi-classical analysis of the Aharonov-Bohm effect in bound-states in two dimensions is given.
 For a rigorous mathematical analysis of the Aharonov-Bohm effect in three dimensions for toroidal
 magnets -actually in the general case of handle bodies-
 see \cite{bw}, where the high-velocity limit of the scattering operator was evaluated in the case where the direction of the velocity is kept
 fixed as its absolute value goes to infinity. A rigorous error bound was given for the difference between the scattering operator and its high-velocity
 limit for incoming asymptotic states that  have small interaction with the magnet in the high-velocity limit.
The error bound goes to zero as the inverse of the velocity. A
detailed analysis of the Aharonov-Bohm effect in the case of the
Tonomura et al. experiments \cite{to3,to1,to2} was given in
\cite{bw}, as well as other results. The results of \cite{bw} give a
rigorous qualitative proof that quantum mechanics predicts the
interference patterns observed in the Tonomura et al. experiments
\cite{to1,to2,to3} with toroidal magnets. The papers \cite{bw,n,w1},
as well as this paper, use the method introduced in \cite{ew} to
estimate the high-velocity limit of solutions to Schr\"odinger
equations and of the scattering operator. The papers \cite{rou},
\cite{ry1}, \cite{ya1}, and \cite{ya2} study the scattering matrix
for potentials of Aharonov-Bohm type in the whole space.

In this paper we give the first rigorous proof that the classical Ansatz of Aharonov and
Bohm is a good approximation to the exact solution of the Schr\"odinger
equation. We provide, for the first time, a rigorous quantitative mathematical analysis of the Aharonov-Bohm effect with toroidal magnets under the
conditions of the experiments of Tonomura et al. \cite{to3,to1,to2}. We assume that the incoming free electron  is a gaussian wave packet, what from
the physical
point of view is a reasonable assumption. The technical advantage of using a gaussian wave packet for the incoming free electrons is that in this case
 we know very well the dynamics of the free asymptotic gaussian state, and we can carry over the estimates of \cite{bw} in a precise manner.
 We provide a rigorous, simple, quantitative, error bound for the difference in
 norm between the exact solution and the approximate solution given by the
Aharonov-Bohm Ansatz. Our error bound is uniform in time. We also  prove that on the gaussian asymptotic state, the scattering operator is given by
multiplication by $e^{i \frac{q}{\hbar c} \tilde{\Phi}}$ -where $q$ is the
charge of the electron, $c$ is the speed of light, $\hbar$ is Planck's constant, and $\tilde{\Phi}$ is the
magnetic flux in a transversal section of the magnet- up to a quantitative
error bound, that we provide. Actually, the error bound is the same in the cases of the exact solution and
the scattering operator.

Aharonov and Bohm \cite{ab} and Tonomura et al. \cite{to3,to1,to2} suggested to split the electron
wave packet into the part that goes through the hole of the magnet and the part
that goes outside. Tonomura et al. observed that an image was produced behind
the magnet that clearly showed that shadow of the magnet and also the hole and
the exterior of the magnet. They concluded \cite{to1} that this indicates
that there was not interference between the part of the electron wave packet
that went trough the hole and the one that either hit the magnet or traveled
outside. The part of the wave packet that goes outside the magnet can be taken
as the reference wave packet. Therefore, we only model the part of the electron
wave packet that goes through the hole of the magnet. Using the
experimental data of Tonomura et al. \cite{to3, to1, to2} we provide lower and upper bounds on
the variance of the gaussian state in order that the electron wave packet
actually goes through the hole. We also rigorously prove that the results of the Tonomura et
al. experiments \cite{to3,to1,to2}, that were predicted by Aharonov and Bohm,  actually follow from
quantum mechanics.  Furthermore,  our results show that it would be quite interesting
to perform experiments for  intermediate  size electron wave packets (smaller than the ones used in the Tonomura et al. experiments, that where much larger
than the magnet) that satisfies appropriate lower and upper bounds that we provide. One could as well take a larger magnet. In this case, the interaction of the electron wave
packet with the magnet is negligible -the probability that  the electron wave
packet interacts with the magnet is smaller than $10^{-199}$ (See Remark \ref{probabilidad_interaccion} and Section \ref{intermediate_sigma})- and, moreover,
quantum mechanics predicts the results observed by Tonomura et al. with an
error bound  smaller than $10^{-99}$, in norm.

 Our error bound has a physical interpretation. For small variances, it is due to Heisenberg's uncertainty principle. If the variance in
 configuration space is  small, the variance in momentum space is big, and
 then, the  component of the momentum transversal to the axis of the magnet is
 large.  In consequence, the opening angle of the
electron wave packet is large, and there is a large interaction with the magnet. If the variance is large, the opening angle
 is small, but as the electron wave packet is big we have again a large interaction with the magnet.

It has been claimed that the outcome of the Tonomura et al.
experiments \cite{to3,to1,to2} can be explained by the action of a
force acting on the electron that travels through the hole of the
magnet. See, for example, \cite{bo,he} and the references quoted
there. Such a force would accelerate the electron and it would
produce a time delay. In a recent crucial experiment Caprez et al.
\cite{cap} found that the time delay is zero, thus experimentally
excluding the explanation of the results of the Tonomura et al. experiments by the
action of a force. In the Aharonov-Bohm Ansatz the electron  is not
accelerated, it propagates following the free evolution, with the
wave function multiplied by a phase. Since, as mentioned above, we
prove that the Aharonov-Bohm Ansatz approximates the exact solution
with an error bound uniform in time that can be smaller that
$10^{-99}$ in norm, we rigorously prove that quantum mechanics
predicts that no force acts on the electron, in agreement with the
experimental results of Caprez et al. \cite{cap}.

\subsection{Tonomura et al. Experiments} \label{introduction-t-e}
The remarkable experiments of Tonomura at al. \cite{to3,to1,to2} are
widely considered as the only experimental evidence of the physical
existence of the Aharonov-Bohm effect. Tonomura et. al.
 constructed small toroidal magnets such that the
magnetic field is practically zero outside them. In \cite{to2}, the
magnets are impenetrable and, furthermore, they are covered by super
conductive layers that forbid the leakage of magnetic field outside
the magnets. We denote by $\tilde{K}:= \{(x_1,x_2,x_3)\in \ere^3: 0
<\tilde{ r}_1 \leq (x_1^2+x_2^2)^{1/2} \leq \tilde{r}_2, |x_3| \leq
\tilde{ h} \}  $ the magnet ($ \tilde{r}_1 $ is
 the inner radius, $ \tilde{r}_2 $ the outer radius and $ 2 \tilde{h}  $ is the height), and
by  $ \tilde{B}(x) $ the magnetic
field. We suppose that $\tilde{B}(x) $ is zero for $ x  $ outside the magnet.

An  electron wave packet was  sent  towards the magnet. It was superimposed behind
it with a reference electron wave packet  to produce the interference pattern. The
experiments were set up in such a way that the reference electron wave packet was not influenced
by the magnet,  and that the  electron wave packet  and
the reference electron wave packet only interfered behind the magnet, were the interference
patterns were formed. The observed interference patterns provided a strong evidence
of the physical existence of the Aharonov-Bohm effect.

The electron wave packet   was much larger than the magnet. It was 3
micrometers in  size in the direction of the electron propagation
and 20 micrometers in size in a plane perpendicular to the
propagation direction \cite{to4}. It covered the magnet
completely. Recall that it was observed that an image was produced  behind the
magnet that clearly showed the shadow of the magnet and also the
hole and the exterior of the magnet (see \cite{to1,to2}) and that it was
pointed out by Tonomura et al. \cite{to1,to2}, that this indicates
that there was no interference between the part of the electron wave packet
that  went through the hole, and the one that either  hit   the
magnet or traveled outside, because of the clear image of the
shadow of the magnet \cite{to1, to2}.
As mentioned before, we will concentrate our
 analysis on the part of the  wave packet that goes through the hole, and we will take it as the electron wave packet
 itself. It is either the part of the  electron wave packet that goes trough the
 hole, or a smaller electron wave packet that really goes trough the hole.

\subsection{Aharonov-Bohm Ansatz for the Exact Solution}\label{introduction-a-b-a}

At the time of emission, i.e., as  $t \rightarrow -\infty$, the electron wave packet is
far away from the magnet  and  it does not  interact with it, therefore,
it can be assumed that it follows the free evolution,

\beq \label{I.2.1}
i\hbar \frac{\partial}{\partial t} \phi (x,t) =
H_0 \phi(x,t), x \in  \ere^3, t \in \ere. \ene where $H_0$ is the
free Hamiltonian.
\beq
\label{I.2.2}
H_0 : =  \frac{1}{2 M} {\mathbf
P}^2.
\ene
$M$ is the mass of the electron and ${\mathbf P}:=-i\hbar
\nabla $ is the momentum operator. We represent the emitted electron wave packet  by the free evolution of a
gaussian wave function, $\varphi_\v$, with velocity $\v$,

\beq
\label{I.2.3}
\varphi_{\v} : =  e^{ i \frac{M}{\hbar} \v \cdot x  }\, \varphi, \, \hbox{\rm where} \quad
  \varphi := \frac{1}{(\sigma^2 \pi )^{3/4}}
 e^{- \frac{x^2}{2 \sigma^2}},
\ene
with variance $\sigma$ smaller than the inner radius of the magnet. We have chosen
the variance transverse to the velocity  of propagation, $\v$, equal to the
longitudinal variance in the direction of
 propagation. In fact, the size of the longitudinal variance is not essential
 for our arguments and we have chosen it equal to the transversal variance
 only for simplicity. Notice that in the momentum representation, $ e^{i
   \frac{M}{\hbar} \v \cdot x}$ is a translation operator by the vector $ M \v
 $, what implies that
the wave function (\ref{I.2.3}) is centered at the classical momentum $M \v $ in the momentum
representation,

$$
\hat{\varphi}_{\v}(p)= \hat{\varphi}(p-M\v),
$$
where for any state represented by the wave function $\phi(x)$ in the
configuration representation, the momentum representation is given by the
Fourier transform,

$$
\hat{\phi}(p):= \frac{1}{(2\pi \hbar)^{3/2}}\, \int_{\ere^3}\, e^{\ds
  -i\frac{p}{\hbar}\cdot x }\, \phi(x)\, dx.
$$

By the previous analysis, the electron wave packet  is represented at the time of emission by the following gaussian
wave packet that is a solution to the free Schr\"odinger equation (\ref{I.2.1})

\beq
\psi_{\v,0}(x,t):= e^{-i \frac{t}{\hbar} H_0}  \, \varphi_\v(x).
\label{I.2.4}
\ene

The (exact) electron wave packet, $\psi_\v(x,t)$, satisfies the interacting Schr\"odinger equation for all times,

\beq
i\hbar
\frac{\partial}{\partial t} \psi_\v (x,t) = H \psi_\v(x,t), x \in  \Lambda :=\ere^3 \setminus
\tilde{K}, t \in \ere,
\label{I.2.5}
\ene
where

\beq
H := H(A) :=   \frac{1}{2 M} ({\mathbf P}- \hbar  A
)^2
\label{I.2.6}
\ene
is the Hamiltonian  and
 $ A = \frac{q}{\hbar c} \tilde A $, where
 $c$ is the speed of light,  $q$ is  the charge of
the electron, $\hbar$ is Plank's constant, and $  \tilde{A}$ is a magnetic potential with $\hbox{\rm
curl}  \tilde{A}= \tilde{B}$ where $\tilde{B}$ is the magnetic field. We
define the Hamiltonian (\ref{I.2.6}) in $L^2(\Lambda)$  with
Dirichlet boundary condition at $\partial \Lambda$, i.e. $\psi=0$ for $x
\in \partial \Lambda$. This is the standard boundary condition that corresponds to an impenetrable magnet.
It implies that the probability that the electron is at the boundary of the magnet is zero.  Note that the Dirichlet boundary condition is invariant
under gauge transformations. In the case of the impenetrable magnet  the
existence of the Aharonov-Bohm effect is more striking, because in this
situation there is zero interaction of the electron with the magnetic field
inside the magnet. Note, however, that once a magnetic potential is chosen the particular self-adjoint boundary condition  taken at
$\partial \Lambda$ does not play an essential role in our calculations.  Furthermore, our results hold also for a penetrable magnet
where the interacting Schr\"odinger equation (\ref{I.2.5}) is defined in all
space. Actually, this later case is slightly simpler because we do not need to
work with two Hilbert spaces, $L^2(\ere^3)$ for the free evolution, and
$L^2(\Lambda)$ for the interacting evolution, what simplifies the proofs.
In consequence, the electron wave packet  is the unique solution, $\psi_{\v}$, to the
interacting  Schr\"odinger equation (\ref{I.2.5}) that is asymptotic to the
free gaussian wave packet, $\psi_{\v,0}$, as $t \rightarrow -\infty$,

\beq
\psi_{\v}(x,t) \approx \psi_{\v,0}(x,t),\quad t \rightarrow -\infty.
\label{I.2.7}
\ene

Aharonov and Bohm  \cite{ab} proposed an approximate solution to the
Schr\"odinger equation over simply connected regions (regions with no
holes) where the magnetic field is zero, by a change of gauge
formula from the zero vector potential. Of course, it is not
possible to have a gauge transformation from the zero potential
everywhere because that would imply that the magnetic flux on a
transversal section of the magnet would be zero. Hence, the gauge
transformation has to be discontinuous somewhere. As mentioned in Section  \ref{introduction-t-e}, in the case of
Tonomura et al. \cite{to3,to1,to2} experiments the magnet is a
cylindrical torus, $\tilde{K}$.

We take as the surface of discontinuity of the gauge transformation
$$
\mathcal S:= \{ (x_1,x_2,x_3)\in \ere^3: (x_1^2+x_2^2)^{1/2} > \tilde{r}_2,
x_3=0\}
$$
and we define the gauge transformation in the domain, $ \mathcal{D} $, given by

$$
\mathcal{D}:= \Lambda\setminus \mathcal S.
$$
 Without loss of
generality we can suppose that the support of $A $ is contained on the convex
hull of $\tilde{K}  $ (see Section \ref{aharonov-bohm-anzats}).  For every
$ x \in \mathcal{D} $, and a fixed point $ x_0 $ in $ \mathcal{D} $ with
vertical component less than $-\tilde{h}  $, we define the gauge transformation
as follows,

 $$
\lambda_{A,0}(x):= \int_{x_{0}}^x\ A,
$$
where the integral is over a path in
$ \mathcal{D} $. Note that for any $ x\in \mathcal{D}$ with $x_3 > 0$ the
integration contour has to go necessarily through the hole of the magnet.

For any solution to the Schr\"odinger equation (\ref{I.2.5}), $ \phi(x,t)$, that
stays in  $\mathcal{D}$, Aharonov and Bohm \cite{ab} propose that the solution
is given by the following Ansatz, motivated  by the
change of gauge formula from the zero vector potential,
\beq
\label{I.2.8}
\phi_{AB}(x,t):= e^{i \lambda_{A,0}(x) } e^{-i \frac{t}{\hbar}H_0} e^{-i
  \lambda_{A, 0}(x) } \,\phi(x,0).
\ene
Note that if the initial state at $t=0$ is taken as
$ e^{-i\lambda_{A, 0}(x) } \,\phi(x,0)$ the Aharonov-Bohm Ansatz is the
multiplication of the free solution by the Dirac magnetic factor
$e^{i \lambda_{A,0}(x) }$ \cite{di}.

The Aharonov-Bohm Ansatz is expected to be a good approximation to
the exact solution if the electron wave packet stays in a connected
domain, away from the surface $\mathcal{S}$ where the gauge
transformation is discontinuous. This Aharonov-Bohm Ansatz is valid
for solutions whose initial data is given at time equal to zero.

For the incoming electron wave packet that satisfies (\ref{I.2.7}) the initial data is
given as time tends to $-\infty$ and then, the Aharonov-Bohm Ansatz has to be
modified. To formulate the appropriate Ansatz we define the wave operators,

$$
 W_\pm(A) :=  W_\pm := \hbox{s-}\lim_{t\rightarrow \pm \infty}\, e^{i\frac{t}{\hbar} H(A)}\, J\, e^{-i\frac{t}{\hbar}H_0}.
$$
where $J$ is the identification operator from $L^2(\ere^3)$ into $L^2(\Lambda)$
given by multiplication by the characteristic function of $\Lambda$, i.e., $J
\phi(x):= \chi_\Lambda(x)\, \phi(x)$ where, $\chi_\Lambda(x)=1, x \in \Lambda,
\chi_\Lambda(x)= 0, x \in \ere^3\setminus \Lambda$. It is proved in \cite{bw} that the strong limits  exist and that we can
replace the operator $J$ by the operator of multiplication by any smooth
characteristic cutoff function  $\chi(x) \in C^\infty$ such that $\chi(x)=0, x \in
\tilde{K}$ and $\chi(x)=1$ for $x$ in the complement of a bounded  set that
contains $\tilde{K}$  on its interior.

The solution to the Schr\"odinger
equation that is asymptotic to the free solution $e^{-i\frac{t}{\hbar}H_0 }\varphi_\v$ as $t \rightarrow -\infty$
is given by

\beq
\psi_\v:= e^{-i\frac{t}{\hbar}H(A)} \,W_{-}\varphi_\v.
\label{I.2.9}
\ene
It satisfies,

\beq
\lim_{t \rightarrow - \infty}\left\|\psi_\v - J \, \psi_{\v,0}\right\|=0.
\label{I.2.10}
\ene

Using this  fact we prove in Section \ref{aharonov-bohm-ansatz} that the
Aharonov-Bohm Ansatz for the exact solution to the Schr\"odinger equation  (\ref{I.2.5}) with
initial data  as time tends to $-\infty$  is given by,

\beq
\label{I.2.11}
\psi_{AB,\v}(x,t)= e^{i \lambda_{A,0}(x)} e^{-i
\frac{t}{\hbar}H_0} \varphi_\v,
\ene
what, again, is the multiplication of the free incoming solution by the Dirac magnetic factor
$e^{i \lambda_{A,0}(x) }$ \cite{di}.

It is expected that if the electron wave packet stays in a connected region of space, away from the surface of discontinuity $\mathcal S$,
the Aharonov-Bohm Ansazt should be a good approximation to the exact solution, i.e., that,
\beq
\psi_\v \approx \psi_{AB,\v}.
\label{I.2.12}
\ene

The Aharonov-Bohm  Ansatz, $ \psi_{AB,\v},$ is what is observed in the Tonomura et.
al. experiments \cite{to3,to1, to2}: as the support of the  vector
potential  $A  $ is contained in the convex hull of $
\tilde{K} $, for every $ x   $  whose vertical component is bigger
than $ \tilde{h} $, $\lambda_{A,0}(x)$ is equal to the constant
$\frac{q}{\hbar c}\tilde{\Phi}  $, where $\tilde{\Phi}  $ is the flux of the
magnetic field over a transverse section of the magnet.  Then, for
$x_3  > \tilde{h}$, the  Aharonov-Bohm Ansatz is given by

\beq
\psi_{AB,\v}(x)= e^{i\frac{q}{\hbar c}\tilde{\Phi} } e^{-i \frac{t}{\hbar} H_0}
\varphi_\v, \quad x_3  > \tilde{h}.
\label{I.2.13}
\ene

This is exactly what it was observed  in the Tonomura
 et al. experiments \cite{to3,to1,to2}.

The scattering operator is defined as

$$
S:= W_+^\ast\, W_-.
$$
For  large positive times, when the exact electron wave packet is far away from the magnet, and it is localized in the region with large positive $x_3$,
it can be again approximated with an outgoing solution  to the free Schr\"odinger equation,
\beq
 \psi_{+,\v,0}:= \,e^{-i\frac{t}{\hbar}
H_0 } \varphi_{+,\v},
\label{I.2.14}
\ene
such that,
\beq
\lim_{t \rightarrow \infty}\left\|\psi_\v - J\, \psi_{+,\v,0}\right\|=0.
\label{I.2.15}
\ene
The initial data of the incoming and the outgoing solutions to the free
Schr\"odinger equation are related by the scattering operator (see Section \ref{section3.2}),

\beq \varphi_{+,\v} = S \varphi_{\v}. \label{I.2.16} \ene By
equations (\ref{I.2.10}) and (\ref{I.2.12}-\ref{I.2.16}) the
Aharonov-Bohm Ansatz suggests  that
\beq \varphi_{+,\v}= S
\varphi_\v  \approx e^{i\frac{q}{\hbar c}\tilde{\Phi} }\,
\varphi_{\v}, \label{I.2.17} \ene
i.e., that on the gaussian
asymptotic state, $\varphi_\v$, the scattering operator is given by
multiplication by $e^{i \frac{q}{\hbar c} \tilde{\Phi}}$, to a good
approximation. This also is precisely what was observed in the
Tonomura et al. experiments \cite{to3,to1,to2}. Furthermore, in the
Aharonov-Bohm Ansatz (\ref{I.2.11})    the electron  is not
accelerated, it propagates along the free evolution, with the wave
function multiplied by a phase. This implies  that in the
Aharonov-Bohm Ansatz no force acts on the electron, and hence, it is
not accelerated. This is precisely what was observed in the Caprez
et al. \cite{cap} experiments.

\subsection{The  Main Results}\label{mr}
As under the free evolution the electron wave packet is concentrated along the classical trajectory, we can expect that if the velocity $\v$
-that is directed along the positive vertical axis- is large enough, the exact electron wave packet will keep away, for all times, from the surface,
$\mathcal S$, where the gauge transformation is discontinuous. In consequence, the Aharonov-Bohm Ansatz should be a good approximation, and equations
(\ref{I.2.12}) and (\ref{I.2.17}) should hold.
In the following theorem (see also Theorem \ref{th6.13.viejo}) we prove that this is true under the conditions of the Tonomura et al. experiments
\cite{to3,to1,to2}, provided that
appropriate, quantitative, lower and upper bounds on the variance, $\sigma$, of
the gaussian wave function are satisfied. The requirement for the variance $\sigma$ to lie within the interval
 below assures that interaction of the electron with the magnet and the surface  $ \mathcal S$ is small.

\begin{theorem}{Aharonov-Bohm Ansatz, Scattering Operator and Tonomura et al. Experiments}
\label{th1.1}\\
Suppose that the magnets and energies are the ones of the experiments of
Tonomura et al.. Then, for every gaussian wave function, $\varphi$, with variance $ \sigma
\in [\frac{4.5}{mv}, \frac{\tilde{r}_1}{2}]$
and every $ t \in \ere $, the solution
to the Schr\"odinger equation,  $e^{-i\frac{t}{\hbar}H(A) } \,W_{-}\,\varphi_\v$, that behaves as
$e^{-i \frac{t}{\hbar} H_0}  \, \varphi_\v$  as
$t \rightarrow -\infty$   is given at the time $ t  $ by
\beq\label{I.3.1}
 \psi_{AB,\v}:=e^{i\lambda_{A,0}(x) } e^{-i\frac{t}{\hbar} H_0} \varphi_\v,
\ene
up to the following error,
\beq \label{I.3.2}
\begin{array}{l}
\| e^{- i \frac{t}{\hbar}H}W_{-}(A)\varphi_{\v}- e^{i \lambda_{A,0} }
e^{-i \frac{t}{ \hbar} H_0}  \varphi_\v
\| \leq \\\\ 7 e^{- \frac{r_{1}^2}{2 \sigma^2}}   +   177 \times 10^{3} e^{- \frac{33}{34}
  \frac{(\sigma m v)^2}{2} }  + 10^{-100},
\end{array}
\ene
where, $m:=  M/ \hbar$. Furthermore,  the scattering operator satisfies
\beq \label{I.3.3}
\begin{array}{l}
 \|  S\,\varphi_\v - e^{i\frac{q}{\hbar c}\tilde{\Phi}}\varphi_\v \|\leq \\\\
7 e^{- \frac{r_{1}^2}{2 \sigma^2}}   +   177 \times 10^{3} e^{- \frac{33}{34}
  \frac{(\sigma m v)^2}{2} }  + 10^{-100}.
\end{array}
\ene
\end{theorem}

 The main factors that produce the error bound in equation
(\ref{I.3.2}, \ref{I.3.3}) are the terms,

\begin{itemize}
\item Size of the electron wave packet factor,
\beq \label{I.3.5}
 e^{- \frac{r_{1}^2}{2 \sigma^2}}.
\ene

\item Opening  angle of the electron wave packet  factor,
\beq \label{I.3.6}
 e^{- \frac{33}{34}
\frac{(\sigma m v)^2}{2} }.
\ene
\end{itemize}
When the variance $\sigma $  is close to the inner radius of the magnet (the electron wave packet is big),
(\ref{I.3.5}) is close to $ 1 $  and (\ref{I.3.6}) is extremely small
(because in this case $ \sigma m v  $ is big ). Then, when the electron wave packet is big compared to the
inner radius, (\ref{I.3.5}) is the important term, what justifies our name.
When the  variance is  small (such that $ \sigma m v  $ is close to $ 1 $)
the factor (\ref{I.3.6}) is close to one and (\ref{I.3.5}) is
extremely small ( $\frac{r_1}{\sigma}  $ is big) and so, the important factor is
(\ref{I.3.6}). Note that when the variance in position, $ \sigma $, is small,
by Heisenberg uncertainly principle the variance in momentum is big. In particular, the transversal component of momentum is large and
the electron wave packet spreads  a lot as it propagates, what makes  the opening angle of the electron wave packet  large . This justifies the name
that we give to  (\ref{I.3.6}).
Note that in both cases the part of the electron wave packet that hits the obstacle is big. When $\sigma$ is big, because the wave packet is big, and
when $\sigma$ is small, because the opening angle is big,  and even if the wave packet was initially  small,
it  spreads rapidly  as it propagates inside the magnet and, in  consequence, a large part of the wave packet
hits the obstacle.
For variances, $\sigma$, that are neither to small nor too  big the part of the electron wave packet that hits the obstacle is small and
the error is very small.
In Section \ref{physical-interpretation-error-bounds} we discuss in detail the physical interpretation of our error bound and
we present a detailed quantitative analysis  for a large range of $\sigma$.

In particular, we give a  rigorous proof that if $ 1.1592 \times 10^{-9}  \leq \sigma \leq 7.7955 \times 10^{-6}  $  the error bound
is smaller than $ 10^{-99}$.
As mentioned above, it would be quite interesting to perform an experiment with electron wave packets that satisfy our bounds. One could as well take
a larger magnet. In this case the probability that  the electron wave
packet interacts with the magnet is smaller than $10^{-199}$ (See Remark \ref{probabilidad_interaccion} and Section \ref{intermediate_sigma}),
 and quantum mechanics predicts with a very small error bound the interference fringes
observed in the experiments of  Tonomura et al. \cite{to3,to1, to2}, and the absence of a force on the electron, as observed in the Caprez et al.
experiment \cite{cap}.

The paper is organized as follows. In Section \ref{notation} we introduce notations and definitions that we use along the paper.
In Section 3 we study the time
evolution of the electron wave packet. We define the wave and the scattering operators, and we introduce the  solutions to the Schr\"odinger
equation with initial condition as time goes to $-\infty$. We estimate the  solution to the Schr\"odinger equation when it is   incoming,
interacting, and outgoing. In Section 4 we use the freedom that we have in the selection of  the magnetic field, the magnetic potential and the
smooth characteristic cutoff
function to make a choice that is convenient for the computation of the error bounds. In Section 5 we make a choice of the free parameters
 under the experimental
conditions of Tonomura et al. \cite{to3}. In Section \ref{time-continued} we continue our study of the time evolution of the electron wave packet
when it is incoming, interacting,
and outgoing.  In Section \ref{aharonov-bohm-ansatz} we consider the Aharonov-Bohm Ansatz for initial data at time zero and for initial data at
time $-\infty$. In Section \ref{final}  we estimate the difference between the exact solution to the Schr\"odinger equation and the Aharonov-Bohm
 Ansatz as the electron is incoming, interacting, and outgoing. In particular,
in Theorem \ref{th6.13} we prove our main result that is quoted as Theorem \ref{th1.1}
in the Introduction. In Section \ref{physical-interpretation-error-bounds}
  give a detailed analysis of the physical interpretation of our error bound with quantitative results. In Section \ref{conclusions} we give the conclusions of our
paper. In appendix A we prove estimates for the free evolution of gaussian states that we use in our work. In Appendix B we prove upper bounds for
integrals that we need to compute our error bound.

\section{Notations and Definitions}\label{notation}
\sss
In this section we collect notations and definitions that are used along the paper.

The magnet $ \tilde{ K}$ - see Section \ref{introduction-t-e} - is defined by the following formula,
\begin{equation}
\tilde{K}:= \left\{(x_1,x_2,x_3)\in \ere^3: 0 <\tilde{ r}_1 \leq
(x_1^2+x_2^2)^{1/2} \leq \tilde{r}_2, |x_3| \leq \tilde{ h}
\right\}. \label{n.1}
\end{equation}
We call $ \tilde{D}  $ the convex hull of   $ \tilde{ K}$. We use the
notation,

\beq\label{n.2}
\Lambda := \ere^3 \setminus \tilde{K}.
\ene

We employ the symbol  $\chi = \chi(x)=\chi (x, \sigma)$ for a twice continuously
differentiable cut-off function that
depends on  the variance of the wave
packet, $ \sigma $, - see (\ref{I.2.3}). The
support of $ 1 - \chi  $ is contained in the set

\begin{equation}
K:=K(\sigma):= \left\{(x_1,x_2,x_3)\in \ere^3: 0 < r_1 \leq
(x_1^2+x_2^2)^{1/2} \leq r_2, |x_3| \leq h(\sigma ) \right\},
\label{n.3}
\end{equation}
where $  r_{1}$ and $ r_{2}$ are some positive numbers  such that $
r_{1} <\tilde{r}_{1}   $, $ r_{2} > \tilde{r}_{2}   $, $
\tilde{r}_{1} - r_{1}  = r_{2} - \tilde{r}_{2} $ and $ h=h(\sigma ) : \ere_{+}
\to \ere_{+}   $ is an increasing function such that $h(\sigma) >
\tilde{h}$ for all $\sigma$  in  $ \ere_{+} $. We will write either
$ h $ or $ h(\sigma) $ for the same
object.  \\
We designate by
\begin{equation}\label{n.4}
\epsilon := \tilde{r}_{1}- r_{1}= r_2-\tilde{r}_2, \quad \delta(\sigma):= \delta:=
h(\sigma) - \tilde{h},
\end{equation}
and  by  $D := D(\sigma) $ the convex hull of $ K $.

For every $ \zeta, \ \tilde{\omega}, \ \sigma \in \ere_{+}$ such that $
0  < \tilde{\omega}^{-1} < \sigma m v$, we denote by $ z_{\tilde{\omega}, \sigma}(\zeta)
$ the unique solution of the equation,
\begin{equation}\label{n.5}
(z_{\tilde{\omega}, \sigma}(\zeta)- \zeta)\frac{\sigma
mv}{(\sigma^{4}m^{2}v^{2}+z_{\tilde{\omega}, \sigma}(\zeta)^{2})^{1/2}} =
\tilde{\omega}^{-1},
\end{equation}
and for every $ \sigma_{1}, \sigma_{2} \in (0, r_{1})  $ (see (\ref{n.3})) we define
\begin{equation}
\label{n.6}
z_{\tilde{\omega},\sigma_{1}, \sigma_{2}}(\zeta):= \max(z_{\tilde{\omega},
\sigma_{1}}(\zeta), z_{\tilde{\omega}, \sigma_{2}}(\zeta) ), \quad
r_{\sigma_{1}, \sigma_{2}} := \min_{i \in \{ 1,2 \} } \{  \lambda > 0
: \frac{(r_{1}\sigma_i mv)^{2}}{\sigma_{i}^{4} (mv)^{2} + \lambda^2 } =
1 \}.
\end{equation}

For every $ \sigma \in  \ere_{+} $, we  define
\beq
\label{n.7}
\tilde{\omega}(\sigma)  :=\frac{1} {\min\left( \sqrt{\frac{33}{34}}\, \sigma mv ,
\sqrt{2000}\right)}, \quad z(\sigma)
:=z_{\tilde{\omega}(\sigma),\sigma}(h(\sigma)), \quad \sigma_{0} :=
\sqrt{\frac{34 }{33}}\,\,\frac{\sqrt{2000}}{mv}.
\ene
Note that (see equation (\ref{a.28}) in Appendix A )
\beq
z(\sigma) > h(\sigma).
\label{n.7b}
\ene

For every $ \sigma \in \ere_{+}  $ and every $ z, \zeta, s
\in \ere  $ we use the following notation,
\beq
\rho=\rho(z)=\rho(\sigma, z ):=\frac{ \sigma m v}{(\sigma^4
m^2v^2+z^2)^{1/2}} ,
\label{n.8}
\ene
\beq
 \theta_{inv}(\sigma, z,s, \zeta ) :=(\zeta -s)  \frac{ \sigma m v}{(\sigma^4 m^2v^2+z^2)^{1/2}},
\, \, \, \theta_{inv}(\sigma, z) := \theta_{inv}(\sigma,z,z, h(\sigma)),
 \label{n.9}
\ene
and
\begin{equation}
\Upsilon(\sigma, z, s, \zeta ):= \int_{ \theta_{inv}(\sigma, z,s,
-\zeta ) }^{ \theta_{inv}(\sigma, z,z, \zeta )} e^{-\tau^2}\,
d\tau, \quad \Upsilon(\sigma, z):=\Upsilon(\sigma, z, z, h(\sigma) ),
\label{n.10}
\end{equation}

\begin{equation}
\Theta(\sigma,z,s,\zeta):=  \int_{ \theta_{inv}(\sigma, z,s, -\zeta
) }^{ \theta_{inv}(\sigma, z,z, \zeta )} \, \tau^2\, e^{-\tau^2}\,
d\tau,  \quad \Theta(\sigma, z):=\Theta(\sigma, z, z, h(\sigma) ).
\label{n.11}
\end{equation}

We utilize the symbols $ \hbar $, $ c  $, $ M   $ and  $q $ for the
Planck constant, the speed of light and the mass and charge of the
electron, respectively. We define,

 $$
m:= \frac{M}{\hbar}.
$$

We denote by $ \v \in \ere ^3 $ the velocity - see (\ref{I.2.3}) - and we
designate by  $ v := |\v|
$, and $ \hv := \v / v $, respectively,  the modulus and the direction of the velocity. We suppose that $ \hv = (0,0,1)  $.
We designate by  $ \mo := -i \nabla_x   $. The momentum operator is
${\mathbf P}:=\hbar \mo$.

We use the letters $\tilde{B} $ and $ \tilde{A} $ for the
magnetic field and the magnetic potential, respectively.
The details of the distribution of the magnetic field inside $\tilde{K}$ are not relevant for the dynamics of the electron that propagates outside
$\tilde{K}$, as long as $\tilde{B}$ is contained inside $\tilde{K}$. Actually, what is relevant is the flux of $\tilde{B}$ along a transversal section
of $\tilde{K}$ modulo $2\pi$. See \cite{bw} for this issue. We use this freedom to choose $\tilde{B}$ and $\tilde{A}$ in a technically convenient way
Then, unless we specify something else, we assume that the support of $ \tilde{B} $ is contained in $
\tilde{K}$, that  the support of $ \tilde{A}$ is contained in the convex hull
of $\tilde{K}$ (what is always possible), and that both are continuously differentiable.
In Section \ref{section-the-magnetic-field-and-the-magnetic-potential-and-the-Cutoff-Fuction}, for any given flux in the transversal section of the
magnet we explicitly construct a magnetic field and a magnetic potential that satisfy our assumptions.
We define $ A:
=\frac{q}{ \hbar c
  }\tilde{A}  $,  $ B: =\frac{q}{ \hbar c
  }\tilde{B}$, and
\beq
  \eta(x,\tau):= \int_0^\tau (\hv \times B)(x+\rho \hv)\, d
\rho.
\label{n.12}
\ene
We denote by $\tilde{\Phi}$ the flux of the magnetic field $\tilde{B}$ over a
transversal section ($\mathrm TS  $)  of the magnet,

\beq \label{n.13}
\tilde{\Phi}:= \int_{\mathrm TS}\,\tilde{ B}.
\ene

Then, the flux of $B$ over a transversal section of the magnet is given by,

\beq \label{n.14}
\Phi:= \int_{\mathrm TS }B  =\frac{q}{ \hbar c }\tilde{\Phi}.
\ene

By Stokes theorem, for every $ x =(x_1,
x_2, x_3) \in \ere^3  $ such that $ \sqrt{x_1^2 + x_2^2} \leq \tilde{r}_1  $ we
have that,

\beq \label{n.15}
\tilde{\Phi}= \int_{-\infty}^{\infty} \hv \cdot \tilde{ A} (x + \tau \hv)d
\tau, \quad  \Phi= \int_{-\infty}^{\infty} \hv \cdot  A (x + \tau \hv
)d \tau.
\ene

Given a function $F$ with domain $  \mathcal D \subset \ere^n, n=1,2,\cdots $
 that takes values on a normed space
$ \tilde{C}  $ with norm $ \| \cdot \|  $, we denote by $ \| F
\|_{\infty} := \hbox{ess sup} \{
 \| f(x) \| :  x \in  \mathcal D   \} $.

The vector $ \bar{M} = \bar{M}(\chi,A,\v)= (M_{1}(\chi,A,\v),
\cdots , M_{5}(\chi,A,\v)) := (M_{1}, \cdots , M_{5}) \in \ere ^5 $
is given by ($ \v  $ and $A $ and  $\chi $ are defined above in this
section ),

\beq
\label{n.16}
\begin{array}{l}
 M_{1} := \| \mo^{2} \chi  \|_{\infty} +  \| \chi \mo \cdot A
\|_{\infty}
 +  \|2 (\mo \chi ) \cdot A  \|_{\infty} +  \|\chi  A^{2}  \|_{\infty}, \\\\
M_{2}:= \|2(\mo \chi) \|_{\infty} + \| 2 \chi A  \|_{\infty}, \\\\
M_{3} = \|(\mo \chi) \cdot \hat{v}  \|_{\infty} + \| \chi A \cdot
\hat{v} \|_{\infty},\\\\

 M_{4}: = \| \chi(x) (\mo \cdot A)(x + t \hat{v})   \|_{\infty} +
\| \chi(x)A^{2}(x+ t \hat{v}) \|_{\infty} + 2 \|A(x+t\hat{v}) \cdot
(\mo \chi )(x)   \|_{\infty} +2 \| \chi(x) A(x + t\hat{v})) \cdot
\eta (x,t)
 \|_{\infty}, \\\\
 M_{5} := 2 \| \chi (x) A(x + t \hat{v})  \|_{\infty}.
\end{array}
\ene

The norms in $ M_{4} $ and $ M_{5} $ are taken with respect to $ x
\in \ere^{3} $ and $ t  \in \ere $.

We define the linear function $ {\cal A} :  \ere^{5} \to \ere^{5}
$ by the following: given a vector $ w:= (w_{1}, \cdots w_{5})\in \ere^5 $, we
take $ {\cal A}(w):= ({\cal A}(w)_{1},\cdots , {\cal A}(w)_{5}  )  $ as,

\beq
\begin{array}{l}
 \mathcal{A}(w)_{1}  := \frac{1}{\sqrt{2}m v} w_{1} +
\sqrt{2}w_{3}, \quad
 \mathcal{
 A}(w)_{2}:= \frac{4}{\pi^{1/4}}[\frac{ \sqrt{2}}{2 m v}w_{1}  +
 \frac{2+ \sqrt{2}}{2}w_{2} + \sqrt{2}w_{3} ],\\\\

  \mathcal{A}(w)_{3}:= \frac{\frac{1}{\sqrt{2}}+ \frac{\sqrt{3} \pi^{1/4}  }{2}
 }{\sqrt{m v} \pi^{1/4} } w_{2}, \,\quad    \mathcal{A}(w)_{4} :=
 \frac{1}{\sqrt{2} m v} w_{4}, \quad \mathcal{A}(w)_{5}:= \frac{\frac{1}{\sqrt{2}}+ \frac{\sqrt{3} \pi^{1/4}  }{2}
 }{\sqrt{m v} \pi^{1/4} } w_{5}.
\end{array}
\label{n.17}
\ene
The symbols used  on the  formulae below  where defined in this
section. Given  $ S_{1}, \,  v \in \ere_{+}  $,  $ w \in \ere^{5}$ and
 $ j \in \{ -\infty, 0, \infty   \} $, we define the function
 $\ag^{j}_{w,v}= \ag^{j}_{w} :  \ere \times \ere_{+}
\to \ere $  by

\begin{equation}\label{n.18}
\begin{array}{l}
  \ag^{- \infty}_{w}(z,\sigma) = \ag^{-\infty}(z,\sigma) := \\\\
 \max(z,S_{1})\frac{{\cal
A}(w)_{1}}{2} + \max(z, S_{1})^{-1/2}(h r_{2}^{2} (\sigma m
v)^{3})^{1/2} \frac{{\cal
    A}(w)_{2}}{2} + \frac{\max(z, S_{1})}{\sigma^{1/2}} \frac{{\cal
    A}(w)_{3}}{2} - z \frac{{\cal A}(w)_{1}}{2} - \frac{z}
{\sigma^{1/2}} \frac{{\cal A}(w)_{3}}{2},
\\\\ \ag^{0}_w(z,\sigma)= \ag^{0}(z, \sigma):=
\ag_{w}^{-\infty}(z,\sigma) +
z \mathcal{A}(w)_4 + \frac{z}{\sigma^{1/2}}\mathcal{A}(w)_5,
\\\\
\ag^{\infty}_w(z, \sigma)= \ag^{\infty}(z, \sigma):=
3 \ag_{w}^{-\infty}(z,\sigma) + z \mathcal{A}(w)_4 +
\frac{z}{\sigma^{1/2}}\mathcal{A}(w)_5.
\end{array}
\end{equation}
We will  not make explicit the dependence on  $ S_{1} $ because it will
be fixed in our estimates. Actually, $S_1$ is a free parameter that we introduce
to optimize the error bound for the incoming electron wave packet in Theorem
\ref{incoming-wave-packet}. We fix $S_1$ in Section \ref{parameters-election}.
       Note, furthermore, that   $\ag^{-
  \infty}_{w}(z,\sigma)$ is independent of $w_3$ and of $w_4$. We define it as a
function of $w \in \ere^5$ to  simplify the  statement of our results.

We define the following quantities,
\beq
\begin{array}{l}
C_{pp}(\sigma) =  C_{pp}(\sigma, B, \chi):= \frac{1}{\pi^{1/4} mv
}(\| \triangle \chi \|_{\infty} + 2 \| \eta (x,t)\cdot (\mo \chi
)(x)  \|  ) +
\frac{2}{\pi^{1/4}} \| \mo \chi(x) \cdot \hat{\v} \|_{\infty}, \\\\
C_{ps}(\sigma)=  C_{ps}(\sigma, B, \chi):= \frac{1}{\pi^{1/4} mv
}(\|  \chi   \mo
\cdot \eta(x,t)  \|_{\infty} +  \|  \chi(x) \eta(x,t)   \|_{\infty}^{2}  ), \\\\
C_{sp}(\sigma)=  C_{sp}(\sigma, B, \chi):= \frac{2}{\pi^{1/4}\sigma mv
}(\| \mo
\chi(x) \|_{\infty}),\\\\
C_{ss}(\sigma)=  C_{ss}(\sigma, B, \chi):= \frac{2}{\pi^{1/4} \sigma mv
}(\|\chi
\eta(x,t)  \|_{\infty}),\\\\
 \mathcal{R}(\zeta)= \mathcal{R}(\zeta, Z)= \mathcal{R}(\zeta, Z, A):= \| A \|_{\infty}\frac{\pi^{1/2} (\sigma^4m^2v^2+
  \zeta^2)^{1/2}}{ \sigma m v  } e^{- \frac{1}{2}(h-Z)^2 \frac{(\sigma m v)^2}{ \sigma^4m^2v^2
      + \zeta^2} }.
\end{array}
\label{n.19}
\ene

\section{Time Evolution of the Electron Wave Packet}\label{time_evolution_electron}
\sss

\subsection{  Wave and Scattering Operators}
\label{section3.2}

The Hamiltonian operator (\ref{I.2.6}) is self-adjoint when it is defined on the domain
$ D(H):= \mathcal H_2
(\Lambda) \cap \mathcal H_{1,0}(\Lambda)$, where  by
$\mathcal H_s(\Lambda), s=1,2,\cdots$ we denote the Sobolev spaces
and by $\mathcal H_{1,0}(\Lambda)$ we denote the closure in the norm
of $\mathcal H_1(\Lambda)$ of the set $C^\infty_0(\Lambda)$ of all
infinitely differentiable functions with compact support in $\Lambda$
\cite{ad}. Note that as the functions in $\mathcal H_{1,0}(\Lambda)$
vanish in trace sense at $\partial \Lambda$, $H$ is the positive
self-adjoint realization in $L^2(\Lambda)$ of the formal
differential operator $\frac{1}{2 M} (\mathbf P-  \hbar A )^2$
with Dirichlet boundary condition at the boundary of
$\Lambda$ \cite{ka,rs2}.
The free Hamiltonian (\ref{I.2.2}) is self-adjoint when it is defined on the domain
$ D(H_0):= \mathcal H_2(\ere^3) $.
Let $J$ be the identification operator from $L^2(\ere^3)$ into
$L^2(\Lambda)$  given by multiplication by the characteristic
function of $\Lambda$, i.e.,
\beq
J\phi(x)= \chi_\Lambda(x)\, \phi(x),
\label{3.1}
\ene
where $\chi_\Lambda(x)=1, x \in \Lambda, \chi_\Lambda(x)=0, x \in \ere^3\setminus \Lambda$. As mentioned in the introduction,  the wave operators are
defined as follows \cite{rs3},
\beq
 W_\pm(A) =  W_\pm := \hbox{s-}\lim_{t\rightarrow \pm \infty}\, e^{i\frac{t}{\hbar} H}\, J\, e^{-i\frac{t}{\hbar}H_0}.
\label{3.2}
\ene
It is proved in \cite{bw} that the strong limits (\ref{3.2}) exist, that they are partially isometric, and that we can
replace  $J$ by the operator of multiplication by any  smooth characteristic function, $\chi(x) \in C^2$ such that $\chi(x)=0, x \in
\tilde{K}$ and $\chi(x)=1$ for $x$ in the complement of a bounded  set that
contains $\tilde{K}$  on its interior.

\beq
 W_\pm(A) =  W_\pm = \hbox{s-}\lim_{t\rightarrow \pm \infty}\, e^{i\frac{t}{\hbar} H}\,\chi\, e^{-i\frac{t}{\hbar}H_0}.
\label{3.3}
\ene
It is also known \cite{ku} that the wave operators are asymptotically complete, i.e.,
that the ranges of $W_{\pm}$  are the same, and that they coincide with the subspace of absolute continuity of $H$. Moreover, the $W_{\pm}$ are
 unitary  from
$L^2(\ere^3)$  onto the subspace of absolute continuity of $H$, and  they satisfy the intertwining relations,

\beq
e^{-i\frac{t}{\hbar} H }\, W_{\pm}= W_{\pm}\, e^{-i\frac{t}{\hbar} H_0 }.
\label{3.4}
\ene
Recall that the scattering operator is defined as \cite{rs3},
\beq
S:= W_+^\ast\, W_-.
\label{3.5}
\ene

\subsection{Initial Conditions at Minus Infinity}\label{in-con-minus-inf}
In scattering experiments we know the wave packet of the
electron at the emission time. Thus, if we want to know the  evolution of the emitted
electron for all times, we have to solve the interacting Schr\"odinger equation (\ref{I.2.5}) with initial
conditions at minus infinity. As mentioned in the introduction this is accomplished with  wave operator $ W_{-} $.
The incoming electron wave packet is described at the time of emission ($t \rightarrow -\infty$) by a solution to the free Schr\"odinger equation,
(\ref{I.2.1}),
\beq
 e^{- i \frac{t}{\hbar} H_0}\,\phi_-.
 \label{3.6}
 \ene
 As $ e^{-i\frac{t}{\hbar} H }$ is unitary, for all $\phi_- \in L^2(\ere^3)$

\beq
\lim_{t \rightarrow \pm \infty}\left\|e^{-i\frac{t}{\hbar}H } \,W_{\pm}\phi_- - J \, e^{-i\frac{t}{\hbar}H_0 } \phi_-  \right\|=0.
\label{3.7}
\ene

Then, the  solution to (\ref{I.2.5}) that behaves as  (\ref{3.6}) as $ t \rightarrow -\infty$ is given by,

\beq
e^{-i \frac{t}{\hbar}H }W_{-}\,\phi_-.
\label{3.8}
\ene
And, moreover,

\beq
\lim_{t \rightarrow  \infty}\left\|e^{-i\frac{t}{\hbar}H } \,W_{-}\,\phi_- - J \, e^{-i\frac{t}{\hbar}H_0 } \phi_+\right\|=0,\quad
\mathrm{where} \, \phi_+:= W_+^\ast W_- \, \phi_-.
\label{3.9}
\ene
This means that -as to be expected- for large positive times, when the exact electron wave packet is far away from the magnet, it behaves as the
outgoing solution to the free Schr\"odinger equation (\ref{I.2.1})
\beq
 e^{-i\frac{t}{\hbar}H_0 } \phi_+,
\label{3.10}
\ene
where the data at $t=0$ of the incoming and the outgoing  free wave packets (\ref{3.6}, \ref{3.10}) are related by the scattering operator,

$$
\phi_+ = S\phi_-.
$$

\subsection{The Incoming Electron Wave Packet}
We first introduce  concepts that will be used latter in our estimates.

We define the re-scaled boosted Hamiltonians \cite{bw,w1}  as follows (see (\ref{I.2.2}), (\ref{I.2.6})),
\beq
H_1 = H_1(\v):= \frac{1}{\hbar v } e^{-im\v\cdot x} \, H_0\,
e^{im\v\cdot x}, \,\,
H_2 = H_2(A,\v):= \frac{1}{ \hbar v} e^{-im\v\cdot x} \, H(A)\,
e^{im\v\cdot x}.
\label{f.1}
\ene
Recall that $ m = \frac{M}{\hbar} $ and $ \v$ is the
velocity (see (\ref{I.2.3})).
Let us denote by
\beq
W_{\pm,\v}:= e^{-im\v\cdot x}\, W_\pm\, e^{im\v\cdot x}
\label{3.11}
\ene
the  boosted wave operators.
We have that,
\beq
W_{\pm,\v} = \hbox{s} \, \lim_{\zeta \rightarrow \pm \infty} e^{i \zeta H_2 }\, \chi(x)\, e^{-i \zeta H_1},
\label{3.12}
\ene
where  $ \zeta $ represents the classical $ x_3-$coordinate of the electron at the time $ t = \zeta/v  $.

We notice that,

\beq
\label{3.13}
e^{-i \zeta H_{2}} = e^{-i m\v \cdot x}e^{-i\frac{\zeta}{\hbar v} H(A)}e^{i m \v \cdot
  x},\, e^{-i \zeta H_{1}} = e^{-i m\v \cdot x}e^{-i\frac{\zeta}{\hbar v} H_{0}}e^{i m \v \cdot x}=
  e^{-i \frac{\zeta}{2mv}(\mo+m\v )^2}.
\ene

The following theorem  gives us an estimate of the exact electron wave
packet  $ e^{- i  \frac{Z}{v \hbar}H } W_{-}(A)  \varphi_{  \v}$ for distances
$Z \leq \ -z(\sigma) <- h(\sigma)$, i.e.,  where  it is incoming.

\begin{theorem}
\label{incoming-wave-packet}
 Let $ w= (w_1, \cdots , w_5) \in \ere^5 $ be such that $ w_{i} \geq
M_{i}(\chi,A,\v)  $ for $ i \in \{ 1,2 ,3\} $. Assume that $\sigma mv \geq
 \sqrt{34/ 33}$. Then, for any $Z
\in \ere^+ $ such that $ Z \geqq z(\sigma) > h(\sigma)$,

\beq
\| e^{  \mp i  \frac{Z}{v \hbar}H } W_{\pm}  \varphi_{  \v} - \chi
e^{\mp i\frac{Z}{v \hbar}H_{0}
  }\varphi_{ \v} \| \leq e^{-\frac{1}{2 \tilde{\omega}(\sigma)^2}}
 \ag_{w}^{- \infty}(z(\sigma ),\sigma).
 \label{3.14}
 \ene
\end{theorem}

\noindent{\it Proof:} First we prove (\ref{3.14}) for $ W_{+}(A) $. By
Duhamel's formula and (\ref{3.13}) we have that,
\beq
\left\|\left(W_{+,\v}- e^{iZ H_2}\, \chi e^{-iZ
 H_1}\right) \varphi
\right\| \leq \frac{1}{2mv }\int_Z ^\infty \, \left[\left\| m_1\,
\et \, \varphi \right\| +2 \left\|m_2 \cdot \mo \,\et \varphi
\right\|+ 2mv\left\| m_2\cdot\hv \et \varphi \right\|\right]\, dz ,
\label{3.15}
\ene

where,
\beq m_1:= (\mo^2 \chi)- \chi (\mo \cdot A)- 2 (\mo
\chi)\cdot A+A^2 \chi. \label{f.4} \ene \beq m_2:= (\mo \chi)- \chi
A
\label{3.16}.
\ene

Equation (\ref{3.14}) for $W_+(A)$ follows from (\ref{3.15}),  Lemmata \ref{lem-a.3} and  \ref{lemm-a.5} in Appendix A, the facts that the function
$\theta_{inv}(\sigma, Z)  $ is decreasing as a function of $ Z $,   for $ Z \geq
0$,  that $1/ \tilde{w}(\sigma)= -\theta_{inv}(\sigma,z(\sigma))$     and the following
estimates:
$$
\int_{\max(Z, S_{1})}^{\infty} \left(\frac{1}{\sigma^4 m^2v^2 + \zeta^2 }\right)^{3/4} \leq
2 \max(z, S_{1})^{-1/2},
$$
$$
|\theta_{inv}(\sigma, Z)|  \leq \sigma mv.
$$
The last inequality follows from the definition of $\theta_{inv}(\sigma, Z)$,
since $Z \geqq z(\sigma) > h(\sigma)$ (see equation (\ref{n.7b})).

We now consider the case of $W_-(A)$. Note that by the uniqueness of the solutions to the Schr\"odinger
equation we have that,

\beq
\overline{e^{-i Z H_2(A,\v )}\,\psi }=
e^{i Z H_2(-A,-\v)}\, \overline{\psi}.
\label{3.17}
\ene
This is the invariance under time reversal and charge conjugation.
 Hence,
 \beq
W_{-,-\v}(-A) \psi= \overline{W_{+,\v}(A) \overline{\psi}},
\label{3.18}
\ene
and then,
\beq
\left(W_{-,-\v}(-A)- e^{- iZ
H_2(-A,-\v)}\, \chi e^{ iZ H_1(-\v)}\right) \varphi=
\overline{\left(W_{+,\v}(A)- e^{ iZ H_2(A,\v)}\, \chi e^{ -iZ
H_1(\v)}\right) \varphi}.
\label{3.19}
\ene
It follows that
(\ref{3.14}) for $W_{-}(-A)$  and $ \varphi_{-\v}  $ follows from
(\ref{3.14}) for $W_{+}(A)$ and $ \varphi_{\v}  $, and the fact that
$ \bar{M}(\chi, A,\v) = \bar{M}(\chi, -A, -\v)   $.

Let $ L: \ere^3  \to \ere^3   $ be defined as $ L(x) = -x    $, for
$x \in \ere^3 $. Note that,

\beq
  (e^{-i \zeta
H_2(A,\v)}\psi) \circ L=   e^{-i \zeta H_2(-A\circ L,-\v )}\,(\psi \circ L).
 \label{3.20}
\ene
Equation (\ref{3.20}) implies that,
\beq
\left( W_{-,\v}(A)  \varphi\right)\circ L = W_{-,-\v}(-A\circ L) \left(
  \varphi\, \circ L  \right),
\label{3.20b}
\ene
where we used that as $\chi (x)=1$ for $x$ in the complement of a bounded
set,
$$
\hbox{s}-\lim_{\zeta \rightarrow \pm \infty}\, (\chi(-x)-\chi(x))  e^{-i \zeta H_1} = 0.
$$

We  obtain (\ref{3.14}) for $ W_{-}(A)   $  and $ \varphi_{\v} $ from
(\ref{3.14}) for $ W_{-}(-A\circ L)  $, $ \varphi_{-\v}  $,  $
\chi \circ L $ instead of $ \chi $, and  $ B \circ L   $ instead of $ B $ using
equations (\ref{3.20}, \ref{3.20b}) and observing that   $ \bar{M}(\chi, A, \v)=
\bar{M}(\chi\circ L,-A \circ L  , -\v) $. For this purpose we use $ B \circ L   $ instead of $ B
$ in  the definition of $\eta $ in (\ref{n.12}).

\bull

\subsection{The Interacting Electron Wave Packet}
We first introduce an assumption that we use often.
\begin{assumption}
\label{ass}
Let $ \mu_{i}, \, i \in \{ 1, 2, 3  \} $  belong to $ \ere_{+} $.
Suppose that the following conditions hold.
\begin{enumerate}
\item
Either $ \mu_{i} \leqq \sigma_{0} , \,  i \in \{ 1, 2,
3  \}    $,  or $ \mu_{i} \geqq \sigma_{0} , \,  i \in \{ 1, 2, 3 \}
$.
\item
Either,  $ \mu_{i} \leqq \mu_{3} , \,  i \in \{ 1, 2  \}    $, or $
\mu_{i} \geqq \mu_{3} , \,  i \in \{ 1, 2  \}    $.
\end{enumerate}
We define $
\mu_{\max} := \max(\mu_{1}, \mu_{2})  $,  $ \mu_{\min} :=
\min(\mu_{1}, \mu_{2})  $,  and take  $ \nu = \mu_{\min}  $, if $ \mu_{i}
\leqq \mu_{3},   i \in \{ 1, 2  \}$ and    $ \nu = \mu_{\max}  $, if $
\mu_{i} \geqq \mu_{3},   i \in \{ 1, 2  \} $. We denote by $ Z := z(\mu_{\max}) $, if $ \mu_{i}
\leqq \sigma_{0},   i \in \{ 1, 2,
3  \} $; and $ Z := \max_{i \in \{ 1 ,2
\}}\{z_{\tilde{\omega}(\mu_{\max}),
  \mu_{i}}(h(\mu_{\max})) \}  $, if $ \mu_{i} \geqq \sigma_{0},   i \in \{ 1, 2,
3  \}  $. We suppose that $   Z
\geqq z_{\sqrt{\frac{2}{3}},\nu, \mu_{3}  }(h(\mu_{\max}))  $ and
 $ r_{1}
\rho(\mu_i,  z_{\sqrt{2},\nu, \mu_{3}  }(h(\mu_{\max})))\geq 1 $ for $ i \in \{ 1, 2  \} $.
\end{assumption}

\bull

The quantities $I_{ps}, I_{pp},I_{ss},$ and $ I_{sp}$ that we use below are
defined, respectively, in equations (\ref{a.31}), (\ref{a.33prima.1.menos1}),
(\ref{a.31.1.1.1}), and (\ref{a.31.1.1.1.1}) in Appendix A.

\begin{lemma}\label{interacting-wave-packet-lemma-1}
Suppose that Assumption \ref{ass} is satisfied and that $\sigma mv \geq 1$.
Then, for every gaussian wave function $
\varphi  $ with variance $ \sigma \in [\mu_{\min}, \mu_{\max}] $ and every $
\zeta \ \in \ere $ with
 $ | \zeta | \leq z(\sigma)  $,

\beq
\begin{array}{l}
\left\| \left( e^{i(z(\sigma)-\zeta) H_2}\, \chi(x)\, e^{- i (z(\sigma)-\zeta) H_1}
    -\chi(x)  e^{-i\int_0^{(z(\sigma)-\zeta)} \hat{\v}\cdot A(x + \tau \hat{\v} ) d
      \tau} \right)  e^{-i \zeta H_{1}} \varphi \right\| \leq \\\\
e^{-\frac{1}{2} \frac{1}{\tilde{\omega}(\sigma)^2}}(Z-\zeta)\left( \frac{1}{\sqrt{2}2m v} M_4 +
M_{5} \frac{ (\frac{\sigma m v}{2})^{1/2} +
  \frac{\sqrt{3}\pi^{1/4}}{2} }{\pi^{1/4}2 \sigma m v}  \right)
+C_{pp}(\sigma) I_{pp}(\mu_{1},\mu_{2} , \mu_{3}) +  \\\\
\frac{C_{ps}(\sigma)}{2} I_{ps}(\mu_{1},\mu_{2} ,\mu_{3}, \zeta)  +
C_{sp}(\sigma) I_{sp}(\mu_{1},\mu_{2} ,\mu_{3})  +
\frac{C_{ss}(\sigma)}{2} I_{ss}(\mu_{1},\mu_{2} ,\mu_{3}, \zeta),
\end{array}
\label{3.21}
\ene

\beq
\begin{array}{l}
\left\| \left( e^{iz(\sigma) H_2}\, \chi(x)\, e^{- i z(\sigma) H_1}
    -\chi(x)  e^{-i\int_0^{z(\sigma)} \hat{\v}\cdot A(x + \tau \hat{\v} ) d
      \tau} \right)  \varphi \right\| \leq \\\\
e^{-\frac{1}{2} \frac{1}{\tilde{\omega}(\sigma)^2}}(Z)\left( \frac{1}{\sqrt{2}2m v} M_4 +
M_{5} \frac{ (\frac{\sigma m v}{2})^{1/2} +
  \frac{\sqrt{3}\pi^{1/4}}{2} }{\pi^{1/4}2 \sigma m v}  \right)
+\frac{C_{pp}(\sigma)}{2} I_{pp}(\mu_{1},\mu_{2} , \mu_{3}) +  \\\\
\frac{C_{ps}(\sigma)}{2} I_{pp}(\mu_{1},\mu_{2} ,\mu_{3})  +
\frac{C_{sp}(\sigma)}{2} I_{sp}(\mu_{1},\mu_{2} ,\mu_{3})  +
\frac{C_{ss}(\sigma)}{2} I_{sp}(\mu_{1},\mu_{2} ,\mu_{3}).
\label{3.22}
\end{array}
\ene
\end{lemma}

\noindent {\it Proof:} As in the proof of Lemma 5.6 of \cite{bw} (see also \cite{w1})
we prove that,

\beq
\begin{array}{l}
\left( e^{i(z(\sigma)-\zeta) H_2}\, \chi(x)\, e^{- i (z(\sigma)-\zeta) H_1}
    -\chi(x)  e^{-i\int_0^{z(\sigma)-\zeta} \hat{\v}\cdot A(x + \tau \hat{\v} ) d
      \tau} \right)  e^{-i \zeta H_{1}} \varphi  =
\int_0^{z(\sigma)- \zeta}\, dz\, i e^{iz H_2}
 e^{-i\int_0^{z(\sigma)- \zeta -z} \hat{\v}\cdot A(x + \tau \hat{\v} ) d
      \tau}\\\\
\left[\sum_{i=1}^2
\left( f_i(x,z(\sigma)- \zeta -z)+ g_i(x,z(\sigma)- \zeta -z)\cdot \mo\right)\,\et+\right.
\left.
 f_3(x)\,\et\right] e^{-i\zeta H_{1}}\varphi,
 \end{array}
\label{3.23}
\ene
where,

\beq
\begin{array}{l}
f_1(x,\tau):= \frac{1}{2mv}\left[-\chi(x) (\mo \cdot A)(x+\tau \hv)+\chi(x) (A(x+ \tau
  \hv))^2-2 A(x+ \tau \hv)\cdot(\mo \chi)(x) +\right.\\\\
 \left. 2 \chi(x) A(x+ \tau  \hat{\v})
   \cdot  \eta (x, \tau \right)],
\end{array}
\label{3.24}
\ene
\beq
f_2(x,\tau):= \frac{1}{2mv}\left[ -\chi(x) (\mo \cdot \eta)(x,\tau)+ \chi(x) (\eta(x,\tau))^2- (\Delta \chi)(x)-2\eta(x,\tau)\cdot (\mo \chi)(x)
\right],
\label{3.25}
\ene
\beq
f_3(x):= (\mo\chi)(x)\cdot\hv,
\label{3.26}
\ene
\beq
g_1(x,\tau):= -\frac{1}{m v} \chi(x)\,A(x+\tau\hv),
\label{3.27}
\ene
\beq
g_2(x,\tau):= \frac{1}{m v}\left[-\chi(x)\, \eta(x, \tau)+ (\mo \chi)(x)\right].
\label{3.28}
\ene

It follows that,
\beq
\begin{array}{l}

\left\|  \left( e^{i(z(\sigma)-\zeta) H_2}\, \chi(x)\, e^{- i (z(\sigma)-\zeta) H_1}
    -\chi(x)  e^{-i\int_0^{z(\sigma)-\zeta} \hat{\v}\cdot A(x + \tau \hat{\v} ) d
      \tau} \right)  e^{-i \zeta H_{1}} \varphi  \right\| \leq
\\\\
\int_0^{z(\sigma) - \zeta }\,dz \,
\left\|f_1(x, z(\sigma)-\zeta - z )\et e^{-i\zeta H_{1}}  \,\varphi\right\|+
\int_0^{z(\sigma) -\zeta }\, dz
\left\|f_2(x, z(\sigma)- \zeta - z )\et\ e^{-i\zeta H_{1}}   \,\varphi    \right\|+
\\\\
\int_0^{z(\sigma) - \zeta }\,dz\, \left\|f_3(x)\et e^{-i \zeta H_{1}}
  \,\varphi\right\|+
\int_0^{z(\sigma)-
  z}\,dz\, \left\|g_1(x, z(\sigma) -\zeta -z)\cdot \mo\et e^{-i \zeta H_{1}} \,\varphi\right\| +
\\\\
\int_0^{z(\sigma)-\zeta}\,dz\,
\left\|g_2(x, z(\sigma) -\zeta -z)\cdot \mo\et  e^{-i \zeta H_{1}} \,\varphi\right\|.
\label{3.29}
\end{array}
\ene
We estimate the first integral in the right-hand side of (\ref{3.29}) using
equation (\ref{a.53}), the second using
(\ref{a.30prima}) and (\ref{a.33prima}), the third using (\ref{a.33prima}),  the fourth using (\ref{a.55}),
and the fifth using (\ref{a.33biprima}) and (\ref{a.52prima}). To use
(\ref{a.55}) note that  $\theta_{inv}(\sigma,z(\sigma))=-1/
\tilde{\omega}(\sigma)$. Then, as $ \sigma mv \geq 1$,
$\theta_{inv}(\sigma,z(\sigma))^2 \geq 1/2$.
After reordering terms we obtain  equation (\ref{3.21}). Equation (\ref{3.22}) is obtained in the
same way but using (\ref{a.33prima.1}) instead of (\ref{a.33prima}) and
(\ref{a.52prima.1}) instead of  (\ref{a.52prima}).

\bull

\begin{lemma}\label{interacting-wave-packet-lemma-2} For $ Z \geq h$,
\beq
\left\|\left( \chi(x)  e^{-i\int_{0}^{\infty} \hat{\v} \cdot
A(x + \tau \hat{\v}) d \tau  }-
\chi(x)  e^{-i\int_{0}^{Z- \zeta } \hat{\v} \cdot
A(x + \tau \hat{\v}) d \tau  }   \right) e^{-i \zeta H_{1} }  \varphi\right\|
\leq \frac{1}{2} \mathcal{R}(\zeta, Z).
\label{3.30}
\ene
\end{lemma}

\noindent {\it Proof:} By Duhamel's formula and (\ref{a.9.5}),
\beq
\begin{array}{l}
\left\|\left(\chi(x)  e^{-i \int_{0}^{\infty}\hat{\v} \cdot A( x +
\tau
      \hat{\v}) d \tau}- \chi(x)  e^{-i \int_{0}^{Z-\zeta}\hat{\v}
\cdot A( x + \tau
      \hv ) d \tau }  \right) e^{-i \zeta H_{1}} \varphi   \right\| \leq
\int_{Z-\zeta}^\infty \|\chi(x) \hv \cdot A(x+\tau \hv )
\,  e^{-i \zeta H_{1} } \varphi \| \, d\tau \\\\
\leq \frac{\| A \|_{\infty}}{\sqrt{2}}\int_{Z}^{\infty} e^{-\frac{1  }{2  } \
(h- \tau)^2 \frac{(\sigma m v)^2}{
\sigma^4m^2v^2 + \zeta^2  } } \, d \tau  = \frac{\|A\|_\infty}{\sqrt{2}}\,
 e^{-\frac{1}{2}(h-Z)^2 \frac{(\sigma m v )^{2}}{\sigma^4 m^2v^2 + \zeta^2} }
 \int_Z^\infty\, d\tau \,
\ds e^{ - \frac{1}{2} (\tau-Z)(\tau+Z-2h) \frac{(\sigma m v)^{2}}{\sigma^4 m^2
      v^4 + \zeta^2 } },
\label{3.31}
\end{array}
\ene
where we used that $(h-\tau)^2-(h-Z)^2= (\tau-Z)(\tau+Z-2 h)$. Finally since,
 $(\tau-Z)(\tau+Z-2 h) \geq (\tau-Z)^2$,

\beq
\begin{array}{l}
\left\|
 \left( \chi(x)  e^{-i
\int_{0}^{\infty}\hv
\cdot A( x + \tau
      \hv) d \tau }- \chi(x)  e^{-i \int_{0}^{Z-\zeta} \hv
\cdot A( x + \tau \hv) d \tau}  \right) e^{-i \zeta H_{1}} \varphi
\right\|
 \leq \\
 \frac{\|A\|_\infty}{\sqrt{2}}\,
 e^{-\frac{1}{2}(h-Z)^2 \frac{(\sigma m v )^{2}}{\sigma^4 m^2v^2 + \zeta^2} }
 \int_Z^\infty\, d\tau \,
\ds e^{ - \frac{1}{2} (\tau-Z)^2 \frac{(\sigma m v)^{2}}{\sigma^4
m^2
      v^4 + \zeta^2 } },
\label{3.32}
\end{array}
\ene
what proves the lemma.

\bull

In the Theorem below we estimate the exact electron wave packet $e^{-i\frac{\zeta}{v \hbar}H } W_{\pm}(A)  \varphi_{\v}$
for distances $\zeta$ such that, $ |\zeta | \leq z(\sigma)  $. As $z(\sigma) > h(\sigma)$ -see equation (\ref{n.7b})- this is the interaction region.

\begin{theorem} \label{interacting-wave-packet}
Suppose  that Assumption \ref{ass} is satisfied, and, furthermore that $\sigma m v \geq 1  $.
Let $ w= (w_1, \cdots , w_5) \in \ere^5 $ be such that $ w_{i} \geq
M_{i}(\chi,A,\v)  $ for $ i \in \{ 1, \cdots , 5  \} $. Then, for
every gaussian wave function $ \varphi  $ with variance $ \sigma \in
[\mu_{\min}, \mu_{\max}] $ and every $ \zeta \ \in \ere $ with $|\zeta| \leq z(\sigma)$,

\beq
\begin{array}{l}
 \| e^{-i\frac{\zeta}{v \hbar}H } W_{\pm}(A)  \varphi_{  \v} - \chi
 e^{-i\int_0^{\pm \infty}
\hat{\v}\cdot A(x + \tau \hat{\v} ) d\tau}
e^{- i\frac{\zeta}{v \hbar}H_{0}
  }\varphi_{ \v} \| \leq
e^{-\frac{1}{2 \tilde{\omega}(\sigma)^2}}
\ag^{0}_w(z(\sigma), \sigma) +
C_{pp}(\sigma) I_{pp}(\mu_{1},\mu_{2} , \mu_{3}) +\\\\
\frac{C_{ps}(\sigma)}{2} I_{ps}(\mu_{1},\mu_{2} ,\mu_{3}, \pm \zeta)  +
C_{sp}(\sigma) I_{sp}(\mu_{1},\mu_{2} ,\mu_{3})  +
\frac{C_{ss}(\sigma)}{2} I_{ss}(\mu_{1},\mu_{2} ,\mu_{3}, \pm \zeta) +
\frac{1}{2}\mathcal{R}( \pm \zeta, z(\sigma)),
 \label{3.33}
\end{array}
\ene

\beq
\begin{array}{l}
 \| W_{\pm}(A)  \varphi_{  \v} - \chi
 e^{-i\int_0^{\pm \infty}
\hat{\v}\cdot A(x + \tau \hat{\v} ) d\tau}
\varphi_{ \v} \| \leq
e^{-\frac{1}{2 \tilde{\omega}(\sigma)^2}}
(\ag^{-\infty}_w(z(\sigma), \sigma) + \frac{z(\sigma)}{2} \mathcal{A}(w)_4 +
\frac{z(\sigma)}{2 \sigma^2} \mathcal{A}(w)_5)  +\\\\
\frac{C_{pp}(\sigma)}{2} I_{pp}(\mu_{1},\mu_{2} , \mu_{3}) +
\frac{C_{ps}(\sigma)}{2} I_{pp}(\mu_{1},\mu_{2} ,\mu_{3})  +
\frac{C_{sp}(\sigma)}{2} I_{sp}(\mu_{1},\mu_{2} ,\mu_{3})  +\\\\
\frac{C_{ss}(\sigma)}{2} I_{sp}(\mu_{1},\mu_{2} ,\mu_{3}) +
\frac{1}{2}\mathcal{R}( 0, z(\sigma)). \label{f.20.1prima}
\end{array}
\label{3.34}
\ene
\end{theorem}

\noindent{\it Proof:}

We prove (\ref{3.33}) for $ W_{+}(A) $, the proof for $ W_{-}(A)  $ follows as in
(\ref{3.17}-\ref{3.20b}). Note that by the intertwining
relations of the wave operators (\ref{3.4}) and by (\ref{3.13}) we have that,

\beq
\begin{array}{l}\label{f.20.1}
\left\|\left( W_{+,\v}(A)- e^{i(z(\sigma)-\zeta)H_2}  \chi \,e^{-i(z(\sigma)-\zeta)H_1}
  \right) e^{-i \zeta H_1} \varphi \right\|=\left\| e^{-i\frac{(z(\sigma) - \zeta)}{v
        \hbar}H} W_+(A)   e^{-i \frac {\zeta}{ v \hbar } H_{0}} \varphi_{  \v} - \chi
e^{- i\frac{(z(\sigma)- \zeta) }{v \hbar}H_{0}} e^{-i \frac {\zeta}{ v \hbar } H_{0}} \varphi_{ \v} \right\| =\\\\
\left
\| e^{-i\frac{z(\sigma)}{v \hbar}H } W_{+}(A)  \varphi_{  \v} - \chi
e^{- i\frac{z(\sigma)}{v \hbar}H_{0}}\varphi_{ \v} \right\|.
\end{array}
\ene

We use the intertwining relations and (\ref{3.13}) again to obtain,
\beq
\label{3.35}
\begin{array}{l}

 \left\| e^{-i\frac{\zeta}{v \hbar}H } W_{+}(A)  \varphi_{  \v} - \chi
 e^{-i\int_0^{ \infty}
\hat{\v}\cdot A(x + \tau \hat{\v} ) d\tau}
e^{- i\frac{\zeta}{v \hbar}H_{0}
  }\varphi_{ \v} \right\|= \\\\
\|W_{+,\v}(A) e^{-i\zeta H_{1}} \varphi  -
\chi
 e^{-i  \int_0^{\pm \infty}
\hat{\v}\cdot A(x + \tau \hat{\v} ) d\tau}
e^{- i \zeta  H_{1}
  }\varphi \| \leq
\|(W_{+,\v}(A)  - e^{i(z(\sigma)- \zeta)H_{2}} \chi
e^{-i (z(\sigma)- \zeta) H_{1} })e^{-i\zeta H_{1}} \varphi \| + \\\\
\| ( e^{i(z(\sigma)-\zeta) H_2}\, \chi\, e^{- i (z(\sigma)-\zeta) H_1}
    -\chi  e^{-i\int_0^{z(\sigma)-\zeta} \hat{\v}\cdot A(x + \tau \hat{\v} ) d
      \tau} )  e^{-i \zeta H_{1}} \varphi \| +\\\\
\|( \chi  e^{-i\int_{0}^{\infty} \hat{\v} \cdot
A(x + \tau \hat{\v}) d \tau  }-
\chi  e^{-i\int_{0}^{z(\sigma)- \zeta } \hat{\v} \cdot
A(x + \tau \hat{\v}) d \tau  }  ) e^{-i \zeta H_{1} }  \varphi \|.
\end{array}
\ene

Equation (\ref{3.33}) is obtained by (\ref{f.20.1}), (\ref{3.35}), Theorem
\ref{incoming-wave-packet}, equation (\ref{3.21}) and Lemma
\ref{interacting-wave-packet-lemma-2}. The proof of (\ref{3.34}) is similar, but instead of (\ref{3.21}) we use (\ref{3.22}).

\subsection{Estimates for the Scattering Operator}
We first prove the following lemma.

\begin{lemma} \label{scattering-operator-lemma-1}
Suppose that  the conditions  of Theorem \ref{interacting-wave-packet} are satisfied. Then,
\beq
\begin{array}{l}

\left\|\left( W_{+,\v}^\ast\, e^{- i \int_0^{-\infty} \hv\cdot A(x+\tau
      \hv)\,d\tau}  - e^{ i \Phi }   \right) \chi(x)\,
\varphi\right\| \leq  3 e^{-\frac{1}{2} \frac{r_1^2}{ \sigma^2}} +

e^{-\frac{1}{2 \tilde{\omega}(\sigma)^2}}

(\ag^{-\infty}_w ( z(\sigma), \sigma ) + \frac{z(\sigma)}{2}\mathcal{A}(w)_4   +
\frac{z(\sigma)}{2\sigma^2}\mathcal{A}(w)_5  ) +

\\\\
\frac{C_{pp}(\sigma)}{2} I_{pp}(\mu_{1},\mu_{2} , \mu_{3}) +
\frac{C_{ps}(\sigma)}{2} I_{pp}(\mu_{1},\mu_{2} ,\mu_{3})  +
\frac{C_{sp}(\sigma)}{2} I_{sp}(\mu_{1},\mu_{2} ,\mu_{3})  +
\frac{C_{ss}(\sigma)}{2} I_{sp}(\mu_{1},\mu_{2} ,\mu_{3}) +
\\\\
\frac{1}{2}\mathcal{R}(0, z(\sigma)).
\label{3.36}
\end{array}
\ene
\end{lemma}

\noindent{\it Proof:}
As $   W_+^\ast \, W_+ = I  $,

\beq
\begin{array}{l}
\left\|\left( W_{+,\v}^\ast\, e^{- i \int_0^{-\infty} \hv\cdot
A(x+\tau
      \hv)\,d\tau     }  - e^{ i \Phi }   \right) \chi(x)\,
\varphi\right\|  = \left\| W_{+,\v}^\ast\, \left( e^{- i (
\int_0^{-\infty} \hv\cdot A(x+\tau
      \hv)\,d\tau  + \Phi ) } \chi  -  W_{+,\v} \chi     \right)  e^{ i \Phi }
\varphi\right\|  \leq
\\\\
\left\| \left( e^{- i  \int_0^{\infty} \hv\cdot A(x+\tau
      \hv)\,d\tau } \chi  -  W_{+,\v}     \right)
\varphi\right\| + \left\| (1-\chi) \varphi  \right\| + \left\| (e^{-
i( \int_0^{-\infty} \hv\cdot A(x+\tau
      \hv)\,d\tau   + \Phi ) } -e^{- i \int_0^{
      \infty} \hv\cdot
A(x+\tau
      \hv)\,d\tau     }  ) \chi \varphi \right\|.
\end{array}
\label{3.37}
\ene
Since $ \int_{-\infty}^\infty\, \hv\cdot A(x+\tau \hv)\,d\tau = \Phi $
for $ x $ in the cylinder $\{ x \in \ere^3 : x_1^2 + x_2^2 \leq
r_1^2\}$, (\ref{3.36}) follows from Theorem
\ref{interacting-wave-packet} and the following estimates,

 \beq
\|(1-\chi(x))\varphi\| \leq e^{-r_1^2 / 2 \sigma^2},\,\,\,
\left\| (e^{- i( \int_0^{-\infty} \hv\cdot A(x+\tau
 \hv)\,d\tau   + \Phi ) } -e^{- i \int_0^{\infty} \hv\cdot A(x+\tau \hv)\,d\tau
}  ) \chi \varphi \right\| \leq 2 e^{-r_1^2 / 2 \sigma^2}.
\label{3.38}
\ene
\bull
\\
In the theorem below  we approximate the scattering operator by its
high-velocity limit (see \cite{bw}).

\begin{theorem}
\label{scattering-operator} Suppose that Assumption \ref{ass} is satisfied. Let $ w= (w_1, \cdots , w_5) \in \ere^5 $ be such that $ w_{i} \geq
M_{i}(\chi,A,\v)  $ for $ i \in \{ 1, \cdots , 5  \} $. Then, for
every gaussian wave function $ \varphi  $ with variance $ \sigma \in
[\mu_{\min}, \mu_{\max}] $,

\beq
\begin{array}{l}
\left\|\left( S- e^{ i \Phi } \chi \right) \varphi_{\v} \right\|
\leq 3 e^{-\frac{1}{2} \frac{r_1^2}{ \sigma^2}} +
   e^{-\frac{1}{2
\tilde{\omega}(\sigma)^2}}(2
\ag_w^{-\infty}(z(\sigma), \sigma)+ z(\sigma)\mathcal{A}(w)_{4} +
\frac{z(\sigma)}{\sigma^{1/2}} \mathcal{A}(w)_{5} ) +
  C_{pp}(\sigma) I_{pp}(\mu_{1},\mu_{2} , \mu_{3}) +\\\\  C_{ps}(\sigma)
I_{pp}(\mu_{1},\mu_{2} ,\mu_{3})  + C_{sp}(\sigma)
I_{sp}(\mu_{1},\mu_{2} ,\mu_{3})  + C_{ss}(\sigma)
I_{sp}(\mu_{1},\mu_{2} ,\mu_{3})   +
\mathcal{R}(0, z(\sigma)).

\end{array}
\label{3.39}
\ene
\end{theorem}

\noindent {\it Proof:}
We denote,
\beq
S_{\v} := e^{-im \v\cdot x}\, S\,
e^{im\v\cdot x}.
\label{3.40}
\ene
We have that,

\beq
\begin{array}{l}
 \left\|\left( S- e^{ i \Phi} \chi\right)
\varphi_\v \right\|= \left\|\left( S_\v- e^{ i \Phi} \chi\right)
\varphi\right\|=\left\| W_{+,\v}^\ast\left( W_{-,\v}-\chi(x)\, e^{-i
\int_0^{-\infty} \hv \cdot A(x + \tau \hv) d \tau  }\right)\,
\varphi \right.+ \\\\
\left. \left( W_{+,\v}^\ast\, e^{      -i \int_0^{-\infty} \hv \cdot
A(x + \tau \hv) d \tau          }- e^{ i  \Phi }\right) \chi(x)\,
\varphi\right\|.
\label{3.41}
\end{array}
\ene
Equation (\ref{3.39}) follows from Theorem \ref{interacting-wave-packet},
Lemma \ref{scattering-operator-lemma-1} and (\ref{3.41}).

\subsection{The Outgoing  Electron Wave Packet}
In the following theorem we estimate the exact electron wave packet $e^{-i\frac{\zeta}{v \hbar}H } W_{-}(A)  \varphi_{  \v}$ for distances $\zeta$ in the
outgoing region, $\zeta \geq z(\sigma) > h(\sigma)$.

\begin{theorem}\label{out-going-wave-packet}
Suppose that Assumption \ref{ass} is satisfied. Let $ w= (w_1, \cdots , w_5) \in \ere^5 $ be such that $ w_{i} \geq
M_{i}(\chi,A,\v)  $ for $ i \in \{ 1, \cdots , 5  \} $. Then, for
every gaussian wave function $ \varphi  $ with variance $ \sigma \in
[\mu_{\min}, \mu_{\max}] $ and every $ \zeta \ \in \ere $ with $ \zeta
\geq z(\sigma)  $,
\beq
\begin{array}{l}
\| e^{-i\frac{\zeta}{v \hbar}H } W_{-}(A)  \varphi_{  \v} - \chi
 e^{i\Phi}
e^{- i\frac{\zeta}{v \hbar}H_{0}
  }\varphi_{ \v} \|
\leq 3 e^{-\frac{1}{2} \frac{r_1^2}{ \sigma^2}} +

 e^{-\frac{1}{2 \tilde{\omega}(\sigma)^2}}

\ag_w^\infty ( z(\sigma), \sigma) +
  C_{pp}(\sigma) I_{pp}(\mu_{1},\mu_{2} , \mu_{3}) +\\\\ C_{ps}(\sigma)
I_{pp}(\mu_{1},\mu_{2} ,\mu_{3})  + C_{sp}(\sigma)
I_{sp}(\mu_{1},\mu_{2} ,\mu_{3})  + C_{ss}(\sigma)
I_{sp}(\mu_{1},\mu_{2} ,\mu_{3})   +
\mathcal{R}(0, z(\sigma)).
\end{array}
\label{3.42}
\ene
\end{theorem}

\noindent{\it Proof:}

Using the definition of  $ S$  (see (\ref{3.5})) and the fact that $ e^{-i
  \frac{\zeta}{v \hbar}H} $ is unitary  we get,

\beq \label{3.43}
\begin{array}{l}
 \| e^{-i\frac{\zeta}{v \hbar}H } W_{-}(A)  \varphi_{  \v} - \chi
 e^{i\Phi}
e^{- i\frac{\zeta}{v \hbar}H_{0}
  }\varphi_{ \v} \|= \| W_{-}(A)  \varphi_{  \v} -  e^{i\frac{\zeta}{v \hbar}H} \chi
 e^{i\Phi}
e^{- i\frac{\zeta}{v \hbar}H_{0}
  }\varphi_{ \v} \|\leq \\\\
 \|e^{i\Phi}\left(  W_{+}(A)  \varphi_{  \v} - e^{i\frac{\zeta}{v \hbar}H } \chi
e^{- i\frac{\zeta}{v \hbar}H_{0}}  \right)\,\varphi_{ \v} \| + \| W_{-}(A)
\varphi_\v - W_+(A) e^{i \Phi} \varphi_\v \|.
\end{array}
\ene
Furthermore,
\beq
\| W_{-}(A)
\varphi_\v - W_+(A) e^{i \Phi} \varphi_\v \| \leq  \| W_{-}(A)
\varphi_\v - W_+(A) S \varphi_\v \|+ \| W_{+}(A)( S  -  e^{i \Phi}) \varphi_\v
\|.
\label{3.44}
\ene
Since the wave operators are asymptotically complete \cite{ku}, the operators
$W_{\pm}\,W_{\pm}^{\ast}$ are the orthogonal projector onto the common range of
$W_{\pm}$. Then,  $W_+\, W_+^\ast\, W_-= W_ -$, and we have that,
$$
 W_{-}(A)
\varphi_\v - W_+(A) S \varphi_\v =  W_{-}(A)\varphi_\v- W_+(A)\, W_+^{\ast}(A)\, W_-(A) \varphi_\v=0,
$$
 and by (\ref{3.43}, \ref{3.44})
\beq
 \| e^{-i\frac{\zeta}{v \hbar}H } W_{-}(A)  \varphi_{  \v} - \chi
 e^{i\Phi}
e^{- i\frac{\zeta}{v \hbar}H_{0}
  }\varphi_{ \v} \| \leq
 \|  W_{+}(A)  \varphi_{  \v} - e^{i\frac{\zeta}{v \hbar}H } \chi
e^{- i\frac{\zeta}{v \hbar}H_{0}
  }    \varphi_{ \v} \| + \|  S \varphi_\v - e^{i \Phi} \varphi_\v  \|.
\label{3.45}
\ene

The inequality  (\ref{3.42}) follows from Theorems \ref{incoming-wave-packet} and
\ref{scattering-operator}, and  from equation (\ref{3.45}).

\section{The Magnetic Field, the Magnetic Potential and the Cutoff
  Function}\label{section-the-magnetic-field-and-the-magnetic-potential-and-the-Cutoff-Fuction}
\sss

We have proven in Theorem 4.1 of \cite{bw} that the Hamiltonias (\ref{I.2.6}) with Dirichlet
boundary condition on $\partial \Lambda$ that correspond
to two different magnetic fields contained inside the magnet, and that have the
same flux $\Phi$  modulo $2\pi$ are unitarily equivalent. We have
also proven in \cite{bw} that   the scattering operator only
depends on the total flux $\Phi$ enclosed
inside the magnet, modulo $2\pi$. This implies that without losing generality
we can assume that

\beq
|\Phi| < 2 \pi,
\label{m.0}
\ene
what we do from now on. This also means that
we have a large freedom to choose the magnetic field, as long as it
is contained inside the magnet.  As mentioned in the introduction, we also
have a large freedom to choose the smooth cutoff function $\chi$. We use this freedom
to choose the magnetic field, the magnetic potential and the smooth cutoff function
that is convenient for the computation of the error bounds.
Below  we construct a magnetic field inspired in the experimental results of
Tonomura et. al. \cite{to1}. We also choose a magnetic potential and a cutoff
function, and we provide bounds for them.

\subsection{Mollifiers}

We denote for $ z \in \ere$,
\beq
\psi(z):= \frac{1}{\iota}\left\{\begin{array}{cc}
 e^{-1/(1-z^2)},& \,|z| \leq 1,\\\\
0,\, &|z| \geq 1,
\end{array}
\right.
\label{m.1.2}
\ene
where,
\beq
\iota:= \int_{-1}^{1} \,  e^{-1/(1-z^2)}\, dz.
\label{m.2}
\ene
For $\varepsilon >0$ we define,
\beq
\psi_\varepsilon(z):= \frac{1}{\varepsilon} \psi(z/\varepsilon),
\label{m.3}
\ene
and for every  $a,b \in \ere$, with $ a < b $ and every  $  \varepsilon \in
\ere_+ $ with   $ \varepsilon  < \frac{1}{2} (b-a)$, we take,
\beq
\psi_{a,b,\varepsilon}(z):= \int_a^b \, dy\, \psi_\varepsilon(z-y)=\left\{
\begin{array}{l}1, z \in [a+\varepsilon, b-\varepsilon],
\\\\
0, z \notin [a-\varepsilon, b+\varepsilon].
\end{array} \right.
\label{m.4}
\ene
Then,
\beq
\left\| \psi_{a,b,\varepsilon} \right\|_\infty =1,
\label{m.5}
\ene

\beq
\left\| \psi_{a,b,\varepsilon}' \right\|_\infty\leq \frac{1}{\iota e  \,\varepsilon},
\label{m.6}
\ene
\beq
\left\| \psi_{a,b,\varepsilon}^{''} \right\|_\infty \leq \frac{2 N}{\iota\,
  \varepsilon^2}, \,\hbox{where}\,
N:= 2e^{-(3/2 + \sqrt{3/4})}(3/2 + \sqrt{3/4})^2 (1-(3/2 + \sqrt{3/4})^{-1})^{1/2}.
\label{m.7}
\ene

\subsection{The Magnetic Field}

Recall that the magnet is the set, \beq
\tilde{K}:=\left\{(x_1,x_2,x_3)\in \ere^3: 0 < \tilde{r}_1 \leq
(x_1^2+x_2^2)^{1/2}\leq \tilde{r}_2, |x_3| \leq \tilde{h} \right\}.
\label{m.8} \ene

We use
cylindrical coordinates: for $ (x_1, x_2, x_3) \in \ere^3   $, we
take  $ r:= (x_1^2+x_2^2)^{1/2}, \,  0 \leq \theta < 2\pi, \, x_3 $.
For $ \te <  \frac{ \tilde{r}_2- \tilde{r}_1}{4}, \td < \frac{\tilde{h}}{2}$, we
define,

\beq B= B(x, \te,\td):=\frac{ \Phi}{ C_{\te,\td} }\,
\psi_{\tilde{r}_1+\te,\tilde{r}_2-\te,\te}(r)\,
\psi_{-\tilde{h}+\td, \tilde{h}-\td,\td}(x_3) (-\sin\theta,
\cos\theta,0),
\label{m.11}
\ene

where for a transverse section of
$\tilde{K}, \mathrm{TS} $,
\beq
C_{\te,\td}:= \int_{\mathrm TS }\,
\psi_{\tilde{r}_1+\te,\tilde{r}_2-\te,\te}(r)\,
\psi_{-\tilde{h}+\td, \tilde{h}-\td,\td} (x_3) \geq 2 (\tilde{h}-2\td)\,
(\tilde{r}_2-\tilde{r}_1- 4 \te).
\label{m.12}
\ene
Then, $\nabla \cdot B=0$ and the flux of $B$ over any transverse section of $\tilde{K}$ is $\Phi$.\\
This choice of $B$, that is approximately constant along any transverse section
of $\tilde{K}$ and is directed along the unit vector $ (-\sin(\theta), \cos(\theta), 0)$
is inspired by the experimental results of Tonomura et al. \cite{to1}:
in Figure 4 (a) of \cite{to1}, the fringes on the
shadow of the magnet suggest that the component of the magnetic
field that is orthogonal to a transverse section of the magnet is constant
over this transverse section.

By (\ref{m.5}, \ref{m.6}, \ref{m.12}),

\beq
\|B\|_\infty \leq \frac{\pi}{  (\tilde{h}-2\td)\,
(\tilde{r}_2-\tilde{r}_1- 4 \te)},
\label{m.13}
\ene
\beq
\left\|\frac{\partial}{\partial x_j} B \right\|_\infty  \leq
\frac{\pi}{
   (\tilde{h}-2\td)\, (\tilde{r}_2-\tilde{r}_1- 4 \te)}
(\frac{1}{\iota e \te} +  \frac{1}{ \tilde{r}_1}  ), \, j=1,2,
\label{m.14}
\ene

\beq
\left\|\frac{\partial}{\partial x_3} B\right\|_\infty \leq \frac{\pi}{
  (\tilde{h}-2\td)\,
 (\tilde{r}_2-\tilde{r}_1- 4 \te)} \frac{1}{\iota e \td}.
\label{m.15}
\ene
With this choice of $B$ we have that (see  (\ref{n.12})).

\beq \|\eta(x,\tau)\|_\infty \leq 2\tilde{h} \,\frac{\pi}{
(\tilde{h}-2\td)\, (\tilde{r}_2-\tilde{r}_1- 4 \te)},
\label{m.16}
\ene

\beq \|\mo \cdot \eta(x,\tau)\|_\infty \leq  2\tilde{h}
\frac{\pi}{(\tilde{h}-2\td)\, (\tilde{r}_2-\tilde{r}_1- 4 \te)}\,
\left( \frac{1}{\iota e \te}+ \frac{1}{\tilde{r}_1} \right).
\label{m.17}
\ene

\subsection{The Magnetic Potential}

The potential $ A = A (x,\te, \td) $ associated to the field
$ B = B(x, \te, \td)  $  satisfies the differential equation
 $ \nabla \times A =B$. As  $ B $ has no vertical component, we can take $ A $
 parallel to the vertical axis.
\beq
A=A(x, \te,\td):= \frac{-\Phi}{C_{\te,\td}}\,
\psi_{-\tilde{h}+\td, \tilde{h}-\td,\td}(x_3)\, \left(0,0,
\int_{(y_1,y_2)}^{(x_1,x_2)} \psi_{\tilde{r}_1+\te,
\tilde{r}_2-\te,\te}(r)\, (\cos\theta, \sin\theta)\right),
\label{m.19}
\ene
where $ (y_1,y_2)$ is any point with $|(y_1,y_2)|
\geq \tilde{r}_2$ and the line integral is over any curve in $
\ere^2 $ that connects the point $ (y_{1}, y_{2}) $ with $ (x_{1},
x_{2} ) $. The value of $A $ is independent of the curve chosen. The
potential $A$ has support in the convex hull of $\tilde{K}$,
that we denoted by $\tilde{D} $. Moreover, by (\ref{m.5}, \ref{m.6}),

\beq
\|A\|_\infty \leq \frac{\pi}{  (\tilde{h}-2\td)\,
(\tilde{r}_2-\tilde{r}_1- 4 \te)} (\tilde{r}_2-\tilde{r}_1),
\label{m.20} \ene \beq \left\|\frac{\partial}{\partial x_j} A
\right\|_\infty \leq \frac{\pi}{    (\tilde{h}-2\td)\,
(\tilde{r}_2-\tilde{r}_1- 4 \te)}, j=1,2, \label{m.21} \ene

\beq
\left\|\frac{\partial}{\partial x_3} A\right\|_\infty \leq \frac{\pi}{
  (\tilde{h}-2\td)\, (\tilde{r}_2-\tilde{r}_1- 4 \te)} \frac{1}{ \iota e \td}
\, (\tilde{r}_2-\tilde{r}_1).
\label{m.22}
\ene

\subsection{The Cutoff Function}\label{cutoff-function}

We use the freedom that we have in the choice of the cutoff function $\chi(x)$ to select it in a convenient way.
Take $0 <  \varepsilon < \tilde{r}_1 ,\delta >0$. We define (see (\ref{n.4})),
\beq
r_1:= \tilde{r}_1-\varepsilon >0,\, \,  r_2:= \tilde{r}_2+ \varepsilon, \, \,  h:= \tilde{h}+\delta.
\label{m.23}
\ene
We define
\beq
\chi(x):= 1- \psi_{r_1+\varepsilon/2, r_2-\varepsilon/2, \varepsilon/2}(r)\,
\psi_{-h+\delta/2, h-\delta/2, \delta/ 2}(x_3).
\label{m.24}
\ene
Then (see (\ref{n.3})),
\beq
\chi(x)=
\left\{
\begin{array}{cc}
 0,& x \in \tilde{K}, \\\\
1,& x \in \ere^3 \setminus K.
\end{array}
\right.
\label{m.25}
\ene

Moreover, by (\ref{m.5}, \ref{m.6}, \ref{m.7}),
\beq
\left\| \chi     \right\|_\infty =1,
\label{m.27}
\ene
\beq
\left\| \frac{\partial}{\partial x_j}   \chi  \right\|_\infty \leq \frac{2 }{ \iota  e  \varepsilon}, j=1,2,
\label{m.28}
\ene
\beq
\left\| \frac{\partial}{\partial x_3}   \chi  \right\|_\infty \leq \frac{2 }{\iota  e  \delta},
\label{m.29}
\ene

\beq
\left\|\mo^2 \chi \right\|_\infty \leq  \frac{8\, N}{ \iota \varepsilon^2}+
\frac{2}{e r_1 \iota \varepsilon}+\frac{8\, N}{ \iota \delta^2}.
\label{m.30}
\ene

We denote by
\beq
\begin{array}{l}
I :=\frac{1}{\pi}  (\tilde{h}- 2 \td )( \tilde{r}_2 -\tilde{r}_{1} - 4 \te  ),
\\\\
J :=  \frac{\tilde{r}_2- \tilde{r}_{1}}{I}.
\end{array}
\ene

We designate by $ \bar{m}(\chi)= \bar{m}:= (m_{1}(\chi), \cdots,  m_{5}(\chi))  \in \ere^5 $ the
vector with the following components,

\beq
\begin{array}{l}\label{n-b.1}
m_1(\chi)=  m_1 : = \frac{8N}{\iota \varepsilon^2 } + \frac{2}{\iota \varepsilon r_1 e} +
\frac{8N}{\iota \delta^2 } + \left(2+ (\tilde{r}_2- \tilde{r}_1 )
 \ds \frac{1}{\iota \tilde{\delta} e}  \right)  I^{-1} + \frac{4}{\iota \delta e} J + J^2,
\\\\
m_2(\chi)= m_2 : = 2 \left( \frac{4}{\iota \varepsilon e}  + \frac{2}{\iota \delta
e} \right) + 2 J,
\\\\
m_3(\chi)= m_3 : =\ds \frac{2}{\iota \delta e} + J,
\\\\
m_4(\chi)= m_4 : = \left(2 + (\tilde{r}_2 -\tilde{r}_1)\ds \frac{1}{\iota \tilde{\delta}
e}\right)I^{-1} + J^{2} + \frac{4}{\iota \delta e}J,
\\\\
m_5(\chi)= m_5 : = 2 J.
\end{array}
\ene
 Now we define the following quantities,

\beq
\begin{array}{l}
c_{pp}(\sigma) := \frac{1}{\pi^{1/4} m v }\left(\ds  \frac{8 N}{\iota
\varepsilon^2}  + \frac{2}{\iota \varepsilon r_1 e}  +\ds \frac{8
N}{\iota \delta^2}  + \ds \frac{4\tilde{h}}{I} \frac{4}{\iota
\varepsilon e} \right) + \ds\frac{4}{\pi^{1/4} \iota \delta
 e},
\\\\
c_{ps}(\sigma) : = \frac{1}{\pi^{1/4} m v}\left(  \frac{2\tilde{h}
}{I}\left(\ds \frac{1}{\iota \tilde{\varepsilon} e }  +
\frac{1}{\tilde{r}_1} \right) + \left(\frac{2\tilde{h}}{I}\right)^{2} \right),
\\\\
c_{sp}(\sigma) : =  \frac{1}{\pi^{1/4} \sigma m v }\left( \frac{8}{\iota
\varepsilon e} +\ds \frac{4}{ \iota \delta e } \right),\\\\
c_{ss}(\sigma) : =  \frac{1}{\pi^{1/4} \sigma m v } \frac{4
\tilde{h}}{I},
\\\\
R(\zeta, Z  ) =  R(\zeta) : = \ds\frac{m_5}{2} \frac{(\sigma^4 m^2 v^2 +
\zeta^2)^{1/2}}{ \sigma m v} \pi^{1/2} e^{- \frac{1}{2} \frac{(h -
Z)^{2} (\sigma m v )^2}{ \sigma^4 m^2 v^2 + \zeta^2 } }.
\end{array}
\label{4.30}
\ene

\begin{remark}\label{notation-and-bounds}
For the field, the potential and cutoff function
constructed in this section we have that,

\beq
\begin{array}{l}
M_{i} \leq m_{i}, \, \, \, i \in \{ 1, \cdots, 5 \},
\\\\
C_{pp}(\sigma) \leq c_{pp}(\sigma), \, \,C_{ps}(\sigma) \leq
c_{ps}(\sigma), \, \,C_{ss}(\sigma) \leq c_{ss}(\sigma), \,
\,C_{sp}(\sigma) \leq c_{sp}(\sigma),
\\\\
\mathcal{R}(\zeta, Z)  \leq R(\zeta, Z).
\end{array}
\ene
\end{remark}

\noindent {\it Proof:} the Remark follows from explicit computation.

\bull

We introduce some notation that we use below. We define the vectors  $  \av^j(v,\bar{m})=  \av^j : =( \ap^j_1 , \ap^j_{1/2},
 \ap^j_0, \ap^j_{-1/2}, \ap^j_{-1} )  $, for $ j \in \{ -\infty, 0, \infty
   \}  $:

\beq
\begin{array}{l}
\label{av}
\av^{-\infty}(v,\bar{m})  =\av^{-\infty}: = ( mv r_1 \frac{\mathcal{A}(\bar{m})_1 }{2} + mvr_2
(\frac{2\tilde{h}}{r_1})^{1/2}\frac{1}{( 1- 5 \times 10^{-10}  )^{1/2}} \frac{\mathcal{A}(\bar{m})_2}{2}  , mv r_1
\frac{\mathcal{A}(\bar{m})_3}{2}, - \\\\134.99 \tilde{h}
\frac{\mathcal{A}(\bar{m})_1 }{2}, - 134.99 \tilde{h}
\frac{\mathcal{A}(\bar{m})_3 }{2}, 0),
\\\\
\av^{-0}(v,\bar{m})  =\av^{-0} : = ( \ap_1^{-\infty}, \ap_{1/2}^{-\infty}, \ap_0^{-\infty} + 135.91 \tilde{h}
\mathcal{A}(\bar{m})_4, \ap_{-1/2}^{-\infty} + \\\\135.91 \tilde{h}
\mathcal{A}(\bar{m})_5, \frac{\sqrt{\pi}}{2} \frac{m_5}{2} (1 + 1.11 \times 10^{-6})^{1/2}
 \frac{136.82}{mv}\tilde{h} ),
\\\\
\av^{\infty}(v,\bar{m})  =\av^{\infty} : = ( 3\ap_1^{-\infty},3 \ap_{1/2}^{-\infty}, 3\ap_0^{-\infty} + 138 \tilde{h}
\mathcal{A}(\bar{m})_4 , 3\ap_{-1/2}^{-\infty} + 138 \tilde{h}
\mathcal{A}(\bar{m})_5 , 0).

\end{array}
\ene
Finally, for  $  j \in \{ - \infty, 0, \infty  \} $ we denote,

\beq
\label{ap}
\ap^j(\sigma, v, \bar{m}) =  \ap^j(\sigma) :=    \sum_{i \in \{ 1, 1/2, 0, -1/2, -1  \}}
\ap^j_i \sigma^i.
\ene

\section{Tonomura et al. Experiments. Continued}\label{tonomura-experiments}
\sss
\subsection{Experimental Data}\label{tonomura-experiments-experimental-data}

We consider the 2 different magnets with their dimensions given in table I of \cite{to3}.
We denote them by $ \{ \tilde{K}_{j}  \}_{j \in \{ 1, 2   \} } $,
\beq \label{E.1}
\tilde{K}_{j}:= \{  x = (x_1,x_2,x_3) \in \ere^{3} :  \tilde{r}_{1,j} \leq \sqrt{x_1^2+x_2^2}
\leq \tilde{r}_{2, j}, | x_3 | \leq \tilde{h}  \}. \ene
We use the notation
\beq
\label{E.1.1}
\chi_j , \, j \in \{ 1, 2  \}
\ene
for the corresponding cutoff function constructed in Section \ref{cutoff-function}.

The height $ \tilde{h} $ is $  10^{-6} cm $ for both magnets and
$$
\tilde{r}_{1,1}= 1.5 \times 10^{-4} cm,
$$
$$
\tilde{r}_{2,1}= 2.5 \times 10^{-4} cm,
$$

$$
\tilde{r}_{1,2}= 1.75 \times 10^{-4} cm,
$$
$$
\tilde{r}_{2,2}= 2.75 \times 10^{-4} cm.
$$
In the Tonomura et al. experiments \cite{to2} the electron has an energy of $150\, keV$. In this experiments
they consider impenetrable magnets as we do in this paper. In the experiments  \cite{to1} they consider
penetrable magnets and energies of $80 \,keV$, $100\, keV$ and $125\, keV$. Since our method  applies also in the case
of penetrable magnets, we will consider in our estimates below the two extreme energies and an
 intermediate energy, although the most important one is the one of $150\, keV$ that is the one used for the
case of impenetrable magnets. Thus we consider the following energies.

$$
E_{1}= 150 \, keV,
$$
$$
E_{2}= 100 \, keV,
$$
$$
E_{3}= 80 \, keV.
$$

They used an electron wave packet that might be represented at the time of
emission ( $ t \to -\infty  $)  by the gaussian wave function,

\beq
\label{E.2}
 \left(\frac{1}{\alpha_z^2
\pi}\right)^{1/4}\left(\frac{1}{\alpha_r^2 \pi}\right)^{2/4} e^{ -i \frac{t}{\hbar} H_{0}} e^{i
\frac{M}{\hbar}\v \cdot x } e^{ -\frac{x_1^{2} + x_2^2}{ 2\alpha_r^2}}\,\,
e^{ -\frac{x_3^{2}}{\ds 2\alpha_z^2}}   .
\ene

The transverse variance of the wave function $ \alpha_r $ is several
times the radius of the torus ($ r_{2,j}, j=1,2 $), so the electron wave packet
covers the magnet.

The part of the wave packet that goes through the hole of the torus has
a different behavior than the one that  goes outside the hole. There
appears to be  no interference between those two parts of the wave packet, because a
clear figure of the shadow of magnet is formed behind the torus.
This was pointed out by  Tonomura et al. \cite{to1}, \cite{to2}. We can,
therefore, model only the part of the electron wave packet that goes
trough the hole of the magnet. Hence, we take the transverse variance $
\alpha_r $ smaller than the inner radius of the magnet. The
anisotropy of the variance ($ \alpha_z \ne \alpha_r  $) does not
introduce new ideas to the analysis and all the proofs that we do
assuming that $ \alpha_z = \alpha_r  $ can be done in the same way
if $ \alpha_z \ne \alpha_r $. We obtain similar results in both
situations. Taking $ \alpha_z \ne \alpha_r $ complicates the
notations and, therefore, for simplicity, we will assume that $
\alpha_z = \alpha_r = \sigma  $. So, when emitted, the electron that
goes trough the hole is represented by,

\beq
\label{E.3}
\psi_{ \v,0 }(x,t) := \frac{1}{(\sigma^2 \pi)^{3/4}}
e^{-i\frac{t}{\hbar} H_{0} } e^{i \frac{M}{\hbar}\v \cdot x }
e^{-\frac{x^{2}}{2\sigma^2}},
\ene
with the variance $ \sigma $ smaller than the inner radius of the magnet.

The real electron wave packet, under the experiment conditions, that behaves as
(\ref{E.3}) when the time goes to $  - \infty $ is given by the wave
function (see  (\ref{I.2.9})),

\beq \label{E.4}
\psi_{\v}(x,t):=e^{- i \frac{t}{\hbar} H}\, W_{-}\varphi_{\v}=  e^{-i \frac{\zeta}{\hbar v} H}\,W_{-}\varphi_{\v}.
\ene
Remember that we take $\v=(0,0,v)$ and that $\zeta:=v t$  is the classical position of the electron, in the vertical direction, at time $t$.

The energy for the free wave packet (or of the perturbed wave packet at $ -
\infty $) is given by
\beq
\langle \frac{1}{2M}{\mathbf P}^2
\varphi_\v   ,  \varphi_\v \rangle = \frac{1}{2}M v^2 + \frac{3}{4}
\frac{\hbar^2}{M \sigma^2}  \approx  \frac{1}{2M} v^2.
\ene
 When $ \sigma$ is big  ( $ \sigma m v >> 1 $ ) the second factor
 is much smaller  than the first. If we take for example $ \sigma m v \geq \sqrt{15}
 $ the second factor is less that $ 1/10  $ times the first. Therefore,
 when $  \sigma m v >> $ 1, we can suppose that the energy is given
 by  the classical energy, $ \frac{1}{2M} v^2  $. With this
 assumption we can calculate the velocities, and the velocities times $m $   corresponding to the
 energies $  E_1, E_2, E_3 $:
$$
v_{1}= 2.2971 \times 10^{10} cm/s, \quad mv_1 = 1.9842 \times
10^{10} cm^{-1},
$$
$$
v_{2}=  1.8755 \times 10^{10} cm/s, \quad mv_2 = 1.6201 \times
10^{10} cm^{-1},
$$
$$
v_{3}=  1.6775 \times 10^{10} cm/s, \quad mv_3 = 1.4491 \times
10^{10} cm^{-1}.
$$
For now on we suppose that the obstacle $ \tilde{K} $ is either $ \tilde{K}_1 $ or $
\tilde{K}_2 $ and that the velocity $ v $ is either $ v_1, v_2$ or $ v_3  $.

\subsection{ Selection of the Parameters }\label{parameters-election}

We have obtained rigorous upper bounds for the difference between the exact solution to
the Schr\"odinger equation and the Aharonov-Bohm Ansatz, and for the difference
between the scattering operator and its high-velocity limit. These bounds hold for any
choice of the parameters  $ S_1, \tilde{\delta},
\tilde{\varepsilon},  \delta $ and $ \varepsilon $. We use this
freedom to choose these  parameters in a convenient way. From now on, we choose the
parameter $S_1 > 0$  such that
\beq
 r_1 \rho(S_1)=1.
\ene
 This choice is made to optimize the error bound in Theorem
 \ref{incoming-wave-packet}. This theorem was proven using Lemmata
 \ref{lem-a.3}, \ref{lemm-a.5}. For example, for the convergence of the integral on the
 left-hand side of equation (\ref{a.12}) we need the decay  of
 $\rho(\sigma,z)$ for large $ z$, but for $z$ small this factor is
 very large. For this reason we split this integral in two regions (where we use
 different estimates) introducing the parameter $S_1$.

Furthermore.,
\beq
\begin{array}{l}
\tilde{\varepsilon} : = \frac{\tilde{r}_2 - \tilde{r}_1}{200},
\\\\
 \tilde{\delta} : = \frac{\tilde{h}}{100},
\\\\
 \delta : = \max(10 \,\sigma, \tilde{h}),
\\\\
 \varepsilon : = \frac{\tilde{r}_1}{50}.
\end{array}
 \ene
This selection was obtained using numerical estimates to optimize  the error
bound for the time evolution of the electron wave packet.

\section{The Time Evolution of the Electron Wave Packet. Continued}\label{time-continued}
\sss
\begin{lemma}\label{time-evolution-angle}
For the  data used  in the Tonomura et al. experiments, $ v \in
\{  v_1, v_2, v_3  \}  $ and $  \tilde{K}  \in \{ \tilde{K}_{1}, \tilde{K}_{2} \} $, suppose that
$ \sigma \in [\frac{4.5}{mv}, \tilde{r}_1 / 2 ]  $ and $ \zeta \in   \ere $.  Then,

\beq
\label{te-tee.m1}
\begin{array}{l}
e^{-\frac{1}{2 \tilde{\omega}(\sigma)^2}} \ag^{-\infty}_{\bar{m}} ( z(\sigma), \sigma ) \leq
e^{- \frac{33}{34} \frac{(\sigma mv)^2}{2 }} \ap^{-\infty}(\sigma) + 10^{-420},
\\\\
e^{-\frac{1}{2 \tilde{\omega}(\sigma)^2}}\ag^{0}_{\bar{m}} ( z(\sigma), \sigma ) +\frac{1}{2}  R(\zeta,
z(\sigma)) \leq e^{- \frac{33}{34} \frac{(\sigma mv)^2}{2 }}  \ap^{0}(\sigma) + 10^{-420},
\\\\
e^{-\frac{1}{2 \tilde{\omega}(\sigma)^2}}\ag^{\infty}_{\bar{m}} ( z(\sigma), \sigma ) + R(0,
z(\sigma)) \leq e^{- \frac{33}{34} \frac{(\sigma mv)^2}{2 }}    \ap^{\infty}(\sigma) + 10^{-420}.
\end{array}
\ene

\end{lemma}

\noindent{\it Proof:}

\begin{itemize}

\item First case, $ \sigma \in [\sigma_0, \frac{\tilde{r}_1}{2}]  $.

As $ \tilde{\omega}(\sigma)^{-1}  \leq  \sqrt{\frac{33}{34}} \sigma m v  $, we have
that

\beq
\label{te-tee.0}
1 \leq  \frac{ (\sigma m v)^2  }{ (\sigma m v)^2 - \tilde{\omega}(\sigma)^{-2} }  \leq  34.
\ene

For these values of $ \sigma $,   $ \tilde{\omega}(\sigma)^{-1} = \sqrt{2000}$. Then,
using (\ref{a.28}) and the experimental values we get,

\beq
\label{te-tee.1}
2.1023 \times 10^{-6} \leq z(\sigma) \leq .0673.
\ene

We also have,

\beq
\label{te-tee.2}
.0042  \leq S_1 \leq 303.8306.
\ene

Using (\ref{te-tee.1}) and (\ref{te-tee.2}) we get,

\beq
\label{te-tee.3}
(h r_2^2 \sigma^3m^3v^3)^{1/2}\, \left(\max(z(\sigma), S_1)\right)^{-1/2} \leq 2.9127 \times 10^5,
\ene
and

\beq
\label{te-tee.4}
\frac{(\sigma^4 m^2 v^2 +\zeta^2)^{1/2}}{ \sigma m v  }\leq (\sigma^2  +
\frac{33 z(\sigma)^2 }{34\times 2000} )^{1/2} \leq 0.0015.
\ene
Now we note that  (see the definition of  $ R(\zeta, z(\sigma))$ in (\ref{4.30})).

\beq \label{te-tee.4.1}
 R(\zeta, z(\sigma) ) \leq  \frac{m_5}{2} \frac{(\sigma^4 m^2 v^2 +
z(\sigma)^2)^{1/2}}{ \sigma m v} \pi^{1/2} e^{- \frac{1}{2 \tilde{\omega}(\sigma)^2}}, \,\,
 \,\,\,\,\,\,\,  R(0, z(\sigma)) \leq
\frac{m_{5}}{2} \pi^{1/2} \sigma  e^{- \frac{(h-z(\sigma))^2}{2 \sigma^2}}.
\ene

We bound the quantities $  \ag^{-j}, \,  j \in
\{-\infty, 0, \infty \}  $ uniformly  for $ \sigma \in [\sigma_0,
\frac{\tilde{r}_1}{2}]  $ and for the experimental energies and magnets, using
(\ref{te-tee.1}, \ref{te-tee.2}, \ref{te-tee.3}, \ref{te-tee.4.1})
 and  the smaller experimental values of
$ \tilde{r_1}, (\tilde{r}_2-\tilde{r}_1), \tilde{h}  $ and $ m v  $ to determine
the components of $ \bar{m}  $. We  use the fact that for the values of sigma that we
consider, $ e^{-\frac{1}{2\tilde{\omega}(\sigma)^2}} \leq e^{- 1000}  $ to obtain,

\beq
\begin{array}{l}
e^{-\frac{1}{2 \tilde{\omega}(\sigma)^2}} \ag^{-\infty} ( z(\sigma), \sigma ) \leq
 10^{-420},
\\\\
e^{-\frac{1}{2 \tilde{\omega}(\sigma)^2}}\ag^{0} ( z(\sigma), \sigma ) +\frac{1}{2}  R(\zeta,
z(\sigma)) \leq  10^{-420},
\\\\
e^{-\frac{1}{2 \tilde{\omega}(\sigma)^2}}\ag^{\infty} ( z(\sigma), \sigma ) + R(0,
z(\sigma)) \leq  10^{-420}.
\end{array}
\ene

\item Second case, $ \sigma \in [\frac{4.5}{mv}  ,\sigma_0]  $. For these values of
 $ \sigma $, $ \frac{ (\sigma mv)^2 }{     (\sigma mv)^2 + \tilde{\omega}(\sigma)^{-2}     } = 34 $, then
by (\ref{a.28}), $ 34 \,h+\sqrt{34}\, \sqrt{33} \, h \leq z(\sigma) \leq 34 \,h  + \sqrt{34}\,\sqrt{
  \frac{33}{34} \sigma^4 m^2 v^2  + 33 \,h^2 }  $ and by triangle inequality
$ z(\sigma) \leq 34 h + \sqrt{33}\,\sigma^2 m v + 34h  $ and then, we have that,

\beq
\label{te-tee.5}
134.99 \,\tilde{h} \leq z(\sigma) \leq 136.82\,\tilde{h}.
\ene

It can be verified that,

\beq
\label{te-tee.6}
\begin{array}{l}
\max(z(\sigma), S_1) = S_1 \leq \sigma mv r_1,
\\\\
\max(z(\sigma), S_1)^{-1/2} (h r_2^2 \sigma^3m^3v^3)^{1/2} \leq \sigma m v r_2
(\frac{h}{r_1})^{1/2} \frac{1}{(1-5 \times 10^{-10})^{1/2}},
\\\\
\frac{(\sigma^4 m^2 v^2 + z(\sigma)^2)^{1/2}  }{\sigma m v} \leq  (1.11 \times
10^{-6} + 1)^{1/2} \frac{136.82 \tilde{h}}{ \sigma m v },
\end{array}
\ene
where in the last inequality we used (\ref{te-tee.5}).
Using  (\ref{te-tee.5}) again we get

\beq \label{te-tee.7}
R(0,z(\sigma)) \leq \frac{m_5}{2}\pi^{1/2} \sigma
e^{-\frac{1}{2}(\frac{133.99\tilde{h}}{ \sigma  })^2  } \leq 10^{-10^8}.
\ene
Finally we obtain  (\ref{te-tee.m1}) using (\ref{n.18}), (\ref{av}), (\ref{te-tee.4.1}),
(\ref{te-tee.5}), (\ref{te-tee.6}), (\ref{te-tee.7}) and the fact that $ \mathcal{A}_{4}(\bar{m})
\leq \mathcal{A}_{1}(\bar{m})  $ and  $ \mathcal{A}_{5}(\bar{m})
\leq \mathcal{A}_{3}(\bar{m})$ (note that in this case $ e^{-\frac{1}{2\tilde{\omega}(\sigma)^2}} = e^{-\frac{33}{34} \frac{(\sigma m v)^2}{2}}$).

\end{itemize}

\begin{remark}\label{time-evolution-angle-remark}
For  $ j \in \{- \infty, 0, \infty   \}  $,  $ e^{-\frac{33}{34}\frac{(\sigma m v)^2}{2}} \ap^{j}(\sigma) $ is decreasing
on the interval $
 [ \frac{4.5}{mv}, \infty) $.
\end{remark}
\noindent{\it  Proof:}
Calculating the numbers $ \ap^j_i  $ we find that $ \ap^j_i \geq 0  $ for $ i
\in \{ 1, 1/2 , -1 \}  $, and also $ \ap^0_{-1/2} \geq 0  $. The other
components of the vectors $ \ap^j  $ are negative. We suppose that $ j \in \{
-\infty, \infty  \} $, the case $ j = 0 $ can be done in the same way (the term
$ \ap^0_{-1/2}  $ is manipulated as the term $ \ap^0_{-1}  $). Since $ \ap^{j}(\sigma) \geq 0   $ and $ \sigma mv \geq 4.5 $, we have
that $ \frac{d}{d\sigma}   e^{-\frac{33}{34}\frac{(\sigma m v)^2}{2}}       \ap^j  \leq e^{- \frac{33}{34}\frac{(\sigma mv)^2}{2}  }
(-b_1(\sigma) + b_2(\sigma) ) $, where $ b_{1}(\sigma) = \frac{33}{34} 4.5 mv (
\ap^j_1\sigma + \ap^j_{1/2} \sigma^{1/2}  ) \geq 0  $ and $ b_{2} =- \frac{33}{34}4.5 mv
( \ap^j_0 + \ap^j_{-1/2} \sigma^{-1/2}   ) + \sum_{i \in \{ 1, 1/2, 0, -1/2
  \}} i \ap^j_i \sigma^{i-1} \geq 0 $. As $ b_1  $ is increasing and $ b_2 $
decreasing, $ -b_1 + b_2  $ is decreasing, as $ -b_1(\frac{4.5}{ mv} )  +
b_2(\frac{4.5}{ mv}) \leq 0 $, we have that  $  \frac{d}{d\sigma}
e^{-\frac{1}{2 \tilde{\omega}(\sigma)^2}}
\ap^j  \leq 0   $ for $ \sigma \in  [ \frac{4.5}{mv}, \sigma_0 ] $.

\bull

Below we introduce a partition of an interval that is adapted to the order of magnitude.

\begin{definition}\label{time-evolution-definition}
{\rm For any  number $ a > 0  $ we designate by $ O_a \in \ZETA $ the order of a, (i.e. $ O_a $ is
such that  $ 10^{O_a} \leq a < 10^{O_a + 1}     $).
For an interval $ [a, b],  a > 0  $ and a positive number $ N_0 $  we define
the partition
$  \mathcal{P}(a,b,N_0): = \{ p_i \}_{i = 1}^{k}  $ ($ p_{i} < p_{i+1} \forall
i \in \{1, \cdots, k-1   \}  $) as follows:
\begin{itemize}
\item case 1: $ b \leq 10^{O_a+1}  $.
If $b-a \leq N_{0} 10^{O_a}$ we take $ k=2, \, p_1 = a, \, p_2 = b $. If $ b-a
>N_0  10^{ O_a}  $ we take $ k \geq 3, $  $ p_1 = a,\, p_k = b  $ and $ p_{i} , i \in \{ 2, \cdots k-1  \}
$ such that $ p_i < p_{i+1}  $,   $ p_{i+1}- p_{i} = N_0 10^{O_a}  $ for $ i
\in \{ 1,\cdots, k-2 \}  $  and
 $ p_k - p_{k-1} \leq N_0 10^{0_a}  $.

\item case 2: $ b > 10^{O_a+1}  $.
For every $  j \in \{0,\cdots, O_b - O_a \}  $ we define a set $  \mathcal{P}^j
$ as follows.  We take $ \mathcal{P}^0  $ as in the case 1 but taking   $
10^{O_a+1}   $ instead of $ b  $. $  \mathcal{P}^{O_b-O_a}    $ is taken as in
the case 1 taking  $  10^{O_b}  $ instead of a. If $ O_b - O_a  \geq 2  $,  for $
j \in \{ 1, \cdots, O_b-O_a -1    \}  $ we define $ \mathcal{P}^{j}  $ as in
the case 1 taking $ 10^{O_a + j}  $ instead of $ a $ and $ 10^{O_a + j + 1}  $
instead of $ b  $.
Now we define $ \mathcal{P}(a,b,N_0) = \cup_{j \in \{ 0, \cdots, O_b -O_a  \}}
\mathcal{P}^j  $.
\end{itemize}}
\end{definition}

\begin{definition}\label{time-evolution-definition-cambio}
We denote by $ \{ \Sigma_{j} \}_{j = 1}^{11}  $ the following sets:\\
$ \Sigma_1 :=  \mathcal{P}(\frac{r_1}{\log(10)250},\frac{r_1}{\log(10)197},.0003) $,
$ \Sigma_2 :=  \mathcal{P}(\frac{r_1}{\log(10)197},\frac{r_1}{\log(10)150},.0005) $,
$ \Sigma_3 :=  \mathcal{P}(\frac{r_1}{\log(10)150},10^{-5},.0008) $,
$ \Sigma_4 :=  \mathcal{P}(10^{-5},1.1 \times 10^{-5},.0001) $,
$ \Sigma_5 :=  \mathcal{P}(1.1 \times 10^{-5},1.3 \times 10^{-5},.0002) $,
$ \Sigma_6 :=  \mathcal{P}(1.3 \times 10^{-5},1.7 \times 10^{-5},.0004) $,
$ \Sigma_7 :=  \mathcal{P}(1.7 \times 10^{-5},2\times 10^{-5},.0008) $,
$ \Sigma_8 :=  \mathcal{P}(2 \times 10^{-5},\frac{\tilde{r}_1}{2},.0015) $,
$ \Sigma_9 :=  \mathcal{P}(10^{-6},  \frac{r_1}{\log(10)250}  ,1000) $,
$ \Sigma_{10} :=  \mathcal{P}(\sigma_0,  10^{-6}  ,1000) $,
$ \Sigma_{11} :=  \mathcal{P}(\frac{4.5}{mv}, \sigma_0, .1) $.

\end{definition}

\begin{lemma}\label{time-evolution-integrals} 
Suppose that the energies and magnets are the ones used on Tonomura et
al. experiments.
Let $ \mu_{i} \in \ere_+  , i \in \{ 1,2, 3 \}  $. Suppose that $ \{ \mu_{i}  \}_{i =
  1}^{2}  $ is contained in one of the sets $ \Sigma_{j}  $ for $ j \in \{ 1, \cdots, 11 \}  $. We take $ \mu_3 = 10^{-6}
 $ if $  \{ \mu_{i}  \}_{i =1}^{2}     $ is contained in $  \Sigma_j  $ for $ j \in \{  1, \cdots, 10  \}$ and we take $ \mu_3 = \sigma_0 $
if  $ \{ \mu_{i}  \}_{i = 1}^{2}  $ is contained in the last set. We suppose furthermore, that $ \mu_1  $ and $
\mu_2 $ are consecutive numbers in the set where they  belong and $\mu_1 <
\mu_2  $. Then, for every $
\sigma \in  [\mu_1 , \mu_2]  $ and every $ \zeta \in \ere $  with  $ |\zeta |  \leq
z(\sigma) $ we have that,
\beq
\begin{array}{l}\label{te-tee.8}
c_{pp}(\sigma) I_{pp}(\mu_{1},\mu_{2} , \mu_{3}) +
\frac{c_{ps}(\sigma)}{2} I_{ps}(\mu_{1},\mu_{2} ,\mu_{3},  \zeta)  +
c_{sp}(\sigma) I_{sp}(\mu_{1},\mu_{2} ,\mu_{3})  +
\frac{c_{ss}(\sigma)}{2} I_{ss}(\mu_{1},\mu_{2} ,\mu_{3},  \zeta) \leq
\\\\
4 e^{-\frac{r_1^2}{2 \sigma^2}}+ 10^{-3} e^{-\frac{33}{34} \frac{(\sigma mv)^2}{2}} \ap^0(\sigma)
+ 10^{-101}, 
\\\\
c_{pp}(\sigma) I_{pp}(\mu_{1},\mu_{2} , \mu_{3}) + c_{ps}(\sigma)
I_{pp}(\mu_{1},\mu_{2} ,\mu_{3})  + c_{sp}(\sigma)
I_{sp}(\mu_{1},\mu_{2} ,\mu_{3})  + c_{ss}(\sigma)
I_{sp}(\mu_{1},\mu_{2} ,\mu_{3}) \leq
\\\\
4 e^{-\frac{r_1^2}{2 \sigma^2}} +  10^{-7}  e^{-\frac{33}{34} \frac{(\sigma mv)^2}{2}}
\ap^\infty(\sigma) + 10^{-101},
\end{array}
\ene
where the functions $I_{pp}, I_{ps}, I_{sp}$, and $I_{ss}$ are evaluated at $Z:=z(\mu_2)$ if $\mu_j \leq \sigma_0$ and at
$Z:= \max_{j\in\{1,2\} }\{z_{\omega(\mu_2),\mu_j}(h(\mu_2))\}$, if $\mu_j \geq \sigma_0$.
\end{lemma}

\noindent{\it Proof:}
 We use a computer to calculate   $ r_{1}
\rho(\mu_i, Z) $ for $ i \in \{ 1, 2 ,3 \} $ and  we prove that these
quantities are bigger than $1$. As $z(\sigma) \leq Z$ (see {\ref{a.31biprima}),  $ r_{1}
\rho(\mu_i, Z) \geq 1$  for $ i \in \{ 1, 2 ,3 \} $ implies that $|\zeta| \leq  r_{\mu_1,\mu_2}$ and that
$r_{\nu,\mu_3}\geq Z$, what simplifies  $I_{ss}$ (see equation (\ref{a.31.1.1.1})).
We estimate the integrals as it is shown in the appendix using a computer, taking $  \delta_0 = 1  $
if $  \mu_1 mv  > 10 $  and $  \delta_0 = \frac{1}{10}  $ if $  \mu_1 mv  \leq 10 $.
We use the computer again to show that (\ref{te-tee.8}) is valid with
 ($ 4 e^{-\frac{r_1^2}{2 \mu_1^2}}+ 10^{-3}  e^{-\frac{33}{34} \frac{(\mu_2 mv)^2}{2}} \ap^0(\mu_2)$) instead of ($4 e^{-\frac{r_1^2}{2
     \sigma^2}}+ 10^{-3}  e^{-\frac{33}{34} \frac{(\sigma mv)^2}{2}}   \ap^0(\sigma)$),
    ($4 e^{-\frac{r_1^2}{2 \mu_1^2}} +  10^{-7}  e^{-\frac{33}{34} \frac{(\mu_2  mv)^2}{2}}\ap^\infty(\mu_2)$)
 instead of ($ 4 e^{-\frac{r_1^2}{2 \sigma^2}} +
   10^{-7}  e^{-\frac{33}{34} \frac{(\sigma mv)^2}{2}} \ap^\infty(\sigma)  $), $ -Z   $
   instead of $ \zeta  $ and $
   c_{T}(\mu_1)  $ instead of  $ c_{T}(\sigma)  $ (for $ T \in\{ pp, ps, sp, ss  \} $). Finally by
   Remark \ref{time-evolution-angle-remark} and the fact that $ c_{T} (\sigma)
   \leq c_{T}(\mu_1), \, I_{T}(\mu_1, \mu_2, \mu_3, \zeta) \leq   I_{T}(\mu_1,
   \mu_2, \mu_3, -Z),  \, \,  T \in  \{ pp, ps, sp, ss   \}   $ (see (\ref{a.31biprima})) , we obtain (\ref{te-tee.8}).

\bull

\subsection{The Incoming Electron Wave Packet. Continued}
\begin{theorem}\label{tonomura-experiments-incoming-wave-packet}
Suppose that the magnets and energies are the ones of the experiments of
Tonomura et al.. Then for every gaussian wave function with variance $ \sigma
\in [\frac{4.5}{mv}, \frac{\tilde{r}_1}{2}]  $
and every $ \zeta \in \ere    $ with $ \zeta \leq -z(\sigma)   $ we have,

\beq \label{te-tee.9}
\begin{array}{l}
\| e^{- i \frac{\zeta}{v \hbar}H}W_{-}(A)\varphi_{\v}- \chi e^{-i
  \frac{\zeta}{v \hbar}H_0  } \varphi_{\v}
\| \leq e^{- \frac{33}{34} \frac{(\sigma m v)^2}{2} } \sum_{i \in \{1, 1/2, 0,
  -1/2, -1  \}} \ap_{i}^{-\infty} \sigma^{i}  + 10^{-420},
\end{array}
\ene
where the quantities $ \ap^{-\infty}_{i}  $ are explicit numbers that depend
only on the magnet and the energy  that we take (see (\ref{av}))

\end{theorem}

\noindent{\it Proof:}
Equation (\ref{te-tee.9}) is a consequence of Theorem
\ref{incoming-wave-packet}, Remark \ref{notation-and-bounds} and Lemma \ref{time-evolution-angle}.

\subsection{The Interacting Electron Wave Packet. Continued}
\begin{theorem}\label{tonomura-experiments-interacting-wave-packet}
Suppose that the magnets and energies are the ones of the experiments of
Tonomura et al.. Then for every gaussian wave function with variance $ \sigma
\in [\frac{4.5}{mv},
\frac{\tilde{r}_1}{2}]  $
and every $ \zeta \in \ere    $ with $ |\zeta| \leq z(\sigma)   $ we have,

\beq \label{te-tee.10}
\begin{array}{l}
\| e^{- i \frac{\zeta}{v \hbar}H}W_{-}(A) \varphi_{\v} - \chi e^{-i \int_{0}^{-\infty} \hv
  \cdot  A(x+ \tau \v ) d\tau }  e^{-i \frac{\zeta}{v \hbar}H_0  } \varphi_{\v}
\| \leq \\\\  4 e^{- \frac{r_{1}^2}{2 \sigma^2}}   +   e^{- \frac{33}{34}
  \frac{(\sigma m v)^2}{2} }
\sum_{i \in \{1, 1/2, 0,
  -1/2, -1  \}} (1+ 10^{-3})  \ap_{i}^{0} \sigma^{i}  + 10^{-101} + 10^{-420}, 
\end{array}
\ene
where the quantities $ \ap^{0}_{i}  $ are explicit numbers that depend
only on the magnet and the energy  that we take (see (\ref{av}))

\end{theorem}

\noindent{\it Proof:}
Let $ \sigma  \in [\frac{4.5}{mv}, \frac{\tilde{r}_1}{2}]  $, then there are $
\mu_1, \mu_2 $ and $ \mu_3 $ such that $ \mu_{1}, \mu_{2}, \mu_{3}  $ and $ \sigma $
satisfies the hypothesis of  Lemma  \ref{time-evolution-integrals}. We prove
using a computer that they satisfy also the hypothesis of the Theorem
\ref{interacting-wave-packet}. We obtain (\ref{te-tee.10}) from Theorem
\ref{interacting-wave-packet}, Remark \ref{notation-and-bounds} and Lemmata
\ref{time-evolution-angle}, \ref{time-evolution-integrals}.

\bull

\subsection{Outgoing Electron Wave Packet and Scattering Operator. Continued }

\begin{theorem}\label{tonomura-experiments-outgoing-wave-packet}
Suppose that the magnets and energies are the ones of the experiments of
Tonomura et al.. Then, for every gaussian wave function with variance $ \sigma
\in [\frac{4.5}{mv},
\frac{\tilde{r}_1}{2}]  $
and every $ \zeta \in \ere    $ with $\zeta \geq z(\sigma)   $ we have,

\beq \label{te-tee.10.2}
\begin{array}{l}
\| e^{- i \frac{\zeta}{v \hbar}H}W_{-}(A) \varphi_{\v} -  e^{i \Phi} \chi
 e^{-i \frac{\zeta}{v \hbar}H_0  } \varphi_{\v}
\| \leq \\\\  7 e^{- \frac{r_{1}^2}{2 \sigma^2}}   +   e^{- \frac{33}{34}
  \frac{(\sigma m v)^2}{2} }
\sum_{i \in \{1, 1/2, 0,
  -1/2, -1  \}}(1+10^{-7}) \ap_{i}^{\infty} \sigma^{i}  + 10^{-101} + 10^{-420}, 
\end{array}
\ene

\beq \label{te-tee.10.1}
\begin{array}{l}
\| S \varphi_{\v} - e^{i \Phi} \chi  \varphi_{\v}
\| \leq \\\\  7 e^{- \frac{r_{1}^2}{2 \sigma^2}}   +   e^{- \frac{33}{34}
  \frac{(\sigma m v)^2}{2} }
\sum_{i \in \{1, 1/2, 0,
  -1/2, -1  \}}(1+10^{-7}) \ap_{i}^{\infty} \sigma^{i}  + 10^{-101} + 10^{-420}, 
\end{array}
\ene
where the quantities $ \ap^{\infty}_{i}  $ are explicit numbers that depend
only on the magnet and the energy  that we take (see (\ref{av})).

\end{theorem}

\noindent{\it Proof:}
Let $ \sigma  \in [\frac{4.5}{mv}, \frac{\tilde{r}_1}{2}]  $. Then, there are $
\mu_1, \mu_2 $ and $ \mu_3 $ such that $ \mu_{1}, \mu_{2}, \mu_{3}  $ and $ \sigma $
satisfies the hypothesis of  Lemma  \ref{time-evolution-integrals}. We prove
using a computer that they satisfy also the hypothesis of the Theorem
\ref{out-going-wave-packet}. We obtain (\ref{te-tee.10}) from Theorem
\ref{out-going-wave-packet}, Remark \ref{notation-and-bounds} and Lemmata
\ref{time-evolution-angle}, \ref{time-evolution-integrals}.
 To get equation (\ref{te-tee.10.1}) we remember that to obtain the error bound in Theorem
\ref{out-going-wave-packet} we used the error bound for the scattering operator of Theorem
\ref{scattering-operator}.  Then, the error bound that we get for the outgoing
wave function in Theorem \ref{out-going-wave-packet} bounds the error bound for the scattering
operator.

\section{Aharonov-Bohm Ansatz. Discontinuous Change of Gauge Formula from the Zero  Vector Potential}
\label{aharonov-bohm-ansatz}
\sss
In this section we denote by $ A $ the vector potential constructed in
Section \ref{section-the-magnetic-field-and-the-magnetic-potential-and-the-Cutoff-Fuction}.
 We take also the parameters, magnets and
energies introduced in Section \ref{tonomura-experiments}.

\subsection{Statement of the Aharonov-Bohm Ansatz}\label{aharonov-bohm-ansatz-statement}

Let $ A_1  $ and $ A_2  $ be two differentiable  magnetic potentials defined in $\ere^{3}
\setminus \tilde{K}  $ with curl zero and that have the same flux $\Phi$. Suppose, furthermore, that

\beq
|A_i(x)| \leq C \frac{1}{1+|x|}, \,\, a_i(r):=\hbox{\rm max}_{x \in \ere^{3}
\setminus \tilde{K}, |x| \geq r}\,
\{ |A_i(x)\cdot \hat{x}| \} \in L^1(0,\infty).
\label{a-b-anzats-h.0}
\ene
 Choose any  point $ x_{0} \in \ere \setminus \tilde{K}  $.  We define
\beq
\label{a-b-anzats-h.1}
\lambda_{A_2, A_1}(x):= \int_{ x_{0} }^{x} (A_2 - A_1),
\ene
where the integral is over any curve in $ \ere^3 \setminus \tilde{K}  $
that connects $ x_0  $ with $ x $. This integral does not depends on the curve
because both potentials have curl zero, and both have the same flux $ \Phi
$. If this  last condition is not true we can not define $\lambda_{A_2, A_1}$. Then,

\beq
\label{a-b-anzats-h.2}
A_{2} = A_{1} + \nabla \lambda_{A_2, A_1}.
\ene
The solution to the Schr\"odinger equation with magnetic potential $ A_{2} $
and initial condition given when the time is zero by the estate $ \psi  $,
 is obtained in terms of the corresponding one for
the magnetic potential $A_{1}  $, by the change of gauge formula,
\beq \label{a-b-anzats-h.3}
e^{-i\frac{t}{\hbar}H(A_2)}\psi = e^{i \lambda_{A_2, A_1}} e^{-i \frac{t}{\hbar}
  H(A_1)} e^{-i \lambda_{A_2,
    A_1} } \psi.
\ene
The solution to the Schr\"odinger equation for the vector potential $ A_1 $
that behaves as
\beq
\label{a-b-anzats-h.4}
e^{-i\frac{t}{\hbar} H_0 } \psi
\ene
when the time goes to minus infinity is given by the formula (see equation
\ref{I.2.9}),
\beq
\label{a-b-anzats-h.5}
e^{-i\frac{t}{\hbar} H(A_1)} W_{-}(A_1)  \psi.
\ene
In other words, (\ref{a-b-anzats-h.5} ) is the solution to the Schr\"odinger
equation  when the initial conditions are taken at time minus infinity by
(\ref{a-b-anzats-h.4}). Now we give the change of gauge formula for the
Schr\"odinger equation with initial conditions taken at time minus infinity:

\beq\label{a-b-anzats-h.6}
e^{-i\frac{t}{\hbar} H(A_2)} W_{-}(A_2)  \psi =
 e^{i \lambda_{A_2, A_1} (x) }  e^{-i\frac{t}{\hbar} H(A_1)} W_{-}(A_1) e^{-i
   \lambda_{A_2, A_1, \infty}(-\mo)}  \psi,
\ene
where $ \lambda_{A_2, A_1, \infty}(x) := \lim_{r \to \infty} \lambda_{A_2, A_1}(rx)  $. (see equation (5.8) in \cite{bw}).

Although the magnetic potential, $A$, constructed in Section
\ref{section-the-magnetic-field-and-the-magnetic-potential-and-the-Cutoff-Fuction}
has curl equal zero, it has non zero  flux.  Therefore, there is
no change of gauge between the vector potential zero and $ A $. Suppose now
that for every time the electron is practically localized in a region, $  \mathcal{D} $, that has no
holes (that is simply connected) or, in other words, in a region where $ \lambda_{A,0}  $ can be defined by
equation (\ref{a-b-anzats-h.1}) if we take curves that connects $ x_{0} $ with
 $ x  $ lying on this
region. On this region $A$ is  gauge equivalent to the vector
potential zero and the change of gauge formulae (\ref{a-b-anzats-h.3}) should follow approximately (although not exactly,
because there is not a real change of gauge between $A$ and the zero potential). The error will depend on how
much of the electron lies in the complement of $ \mathcal{D} $. This is the
Ansatz of Aharonov and Bohm \cite{ab}. Let us be more specific.
In our case we take,

\beq
\label{a-b-anzats-h.8}
\mathcal{D} := (\ere^3 \setminus \tilde{K}) \setminus \mathcal S,
\ene
where
\beq \label{a-b-anzats-h.9}
\mathcal S := \{ (x_1, x_2, 0)  \in  \ere^3 : \sqrt{x_1^2 + x_2^2  } > \tilde{r}_2  \}.
\ene

For two vector potentials $A_1 $ and $A_2 $  whose curl is zero (and
that do not necessarily have the same flux) we define the function given in
(\ref{a-b-anzats-h.1}) in the simply connected region $
\mathcal{D}$: given $ x_{0} = (x_{0,1}, x_{0,2}, x_{0,3} ) \in \mathcal{D}  $
with $ x_{0,3} < - \tilde{h}   $ and $ x $ in $ \mathcal{D}$ we define,
\beq
\label{a-b-anzats-h.10}
\lambda_{A_2,A_1}(x):= \int_{ x_{0} }^{x} (A_2 - A_1),
\ene
where the integral is over any curve in $ \mathcal{D} $ connecting $ x_0  $
with $ x $. Note  that for an electron to cross from the
negative vertical axis to the positive one over $ \mathcal{D}  $, it has to
go through the hole of the magnet.

Then, we have that,
\beq
\label{a-b-anzats-h.11}
A_2(x) = A_1(x) + \nabla \lambda_{A_2, A_1}(x), \quad  x \in \mathcal{D}.
\ene
We extend  $ \lambda_{A_2, A_1}$ to
$\ere^{3} \setminus \tilde{K} $ by zero without
changing notation, i.e., $\lambda_{A_2, A_1}(x) = 0 $, for $ x \in \mathcal{S} $. Note that $ \lambda_{A_2, A_1}  $
is discontinuous on $\mathcal S$.

The Ansatz of Aharonov and Bohm can be stated in the following way.

\begin{definition}{Aharonov-Bohm Ansatz with  Initial  Condition  at Zero} \label{aharonov-bohm-anzats}\\
{\rm Let $ A_1 $ be a magnetic potential defined in $ \ere^3 \setminus \tilde{K}  $
such curl $A_1 = 0$, and with flux not necessarily zero. Let $\psi$ the initial
data at time zero of a solution  to the Schr\"odinger equation that
stays in $ \mathcal{D} $ for all times. Then, the change of gauge formula (\cite{ab}, page 487),
\beq
e^{-i\frac{t}{\hbar} H(A_1) } \psi \approx  \psi_{AB}(x,t):=
e^{i \lambda_{A_1, 0}(x)}e^{-i\frac{t}{\hbar}H_0} e^{-i
  \lambda_{A_1,0}(x)} \psi
\label{a-b-anzats-h.9.1}
\ene
holds.}
\end{definition}
\noindent  Note that if the initial state at $t=0$ is taken as
$ e^{-i\lambda_{A_1, 0}(x) } \,\psi$ the Aharonov-Bohm Ansatz is the
multiplication of the free solution by the Dirac magnetic factor
$e^{i \lambda_{A_1,0}(x) }$ \cite{di}.

Equation (\ref{a-b-anzats-h.9.1}) is formulated when the initial conditions are
taken at time zero. Now we reformulate it taking initial conditions
when the time is minus infinity and for the high velocity state $ \varphi_{\v}
$. For the high-velocity state $ \varphi_{\v}  $ and for big $ v $, we
have that,
\beq
\label{a-b-anzats-h.7}
 e^{-i \lambda_{A_2, A_1, \infty}(-\mo)} \varphi_\v \approx
  e^{-i \lambda_{A_2, A_1, \infty}(-\hv)}  \varphi_\v.
\label{7.12}
\ene
For this statement see the proof of Theorem 5.7 of \cite{bw}. Formula (\ref{a-b-anzats-h.6}) with $W_-(0)=I$, and equation (\ref{7.12}) suggest the
following formulation  of the Aharonov-Bohm Ansatz, with initial condition at time minus infinity and for high-velocity states.

\begin{definition}{Aharonov-Bohm Ansatz with Initial condition at $-\infty$. General Potentials} \label{aharonov-bohm-anzats-incoming-general}
\\{\rm Let $ A_1 $ be a magnetic potential defined in $ \ere^3 \setminus \tilde{K}  $
such curl$A_1 = 0 $, and with flux not
necessarily zero.
Let $ \psi_{\v}(A_1)(x,t)$,
$$
\psi_{\v}(A_1)(x,t):= e^{-i\frac{t}{\hbar} H(A_1)} W_{-}(A_1) \,\varphi_{\v}
$$
be the solution to the Schr\"odinger equation that behaves
as
\beq
\label{a-b-anzats-h.9.2}
e^{-i\frac{t}{\hbar} H_{0}}\varphi_{\v}
\ene
when the time goes to minus infinity.
We suppose that $ \psi_{\v}(A_1)(x,t) $ is approximately localized in $ \mathcal{D} $
for every time. Then, the following change of gauge formula follows,
\beq
\label{a-b-anzats-h.9.3}
\psi_{\v}(A_1)(x,t) \approx e^{i\lambda_{A_1,0}(x) } e^{-i\frac{t}{\hbar} H_0} e^{-i \lambda_{A_1,0,
    \infty}(-\hv)} \varphi_\v,
\ene
where $  \lambda_{A_1,0, \infty} (x ) = lim_{r \to \infty} \lambda_{A_1,0}(rx)  $.}
\end{definition}

\bull

Let us show that formula  (\ref{a-b-anzats-h.9.3}) can formally be derived from (\ref{a-b-anzats-h.9.1}).
We take $ \psi = e^{i\lambda_{A_1,0}(x) }  e^{-i \lambda_{A_1,0,
    \infty}(-\hv)} \varphi_{\v} $ in (\ref{a-b-anzats-h.9.1}).
Then, we have  that $ e^{-i\frac{t}{\hbar}H(A_1)} \psi \approx e^{i \lambda_{A_1,0}(x) }  e^{-i\frac{t}{\hbar}H_0 }
e^{-i \lambda_{A_1,0, \infty}(-\hv)} \varphi_\v  $. For big velocities, the
time evolution $ e^{-i\frac{t}{\hbar} H_0} \varphi_\v $ is localized near the classical
position $ \v t  $ \cite{ew} . Therefore,

$$
e^{i \lambda_{A_1,0}(x) }  e^{-i\frac{t}{\hbar}H_0 }
e^{-i \lambda_{A_1,0, \infty}(-\hv)} \varphi_\v \approx  e^{i
  \lambda_{A_1,0}(\v t) }  e^{-i\frac{t}{\hbar}H_0 }
e^{-i \lambda_{A_1,0, \infty}(-\hv)} \varphi_\v ,
$$
and thus, $ e^{-i\frac{t}{\hbar} H(A_1)}
\psi  $ behaves as (\ref{a-b-anzats-h.9.2}) when the time goes to minus infinity.
Then,
$$ \psi_\v(A_1)(x,t) \approx e^{-i \frac{t}{\hbar} H(A_1) } \psi \approx e^{i \lambda_{A_1,0}(x) }  e^{-i\frac{t}{\hbar}H_0 }
e^{-i \lambda_{A_1,0, \infty}(-\hv)} \varphi_\v
$$
and (\ref{a-b-anzats-h.9.3})
follows.

For a general $C^1 $ vector potential  $ A_{1} $ with curl equal zero and flux $\Phi $,
 there is a real change of gauge (given by formula (\ref{a-b-anzats-h.1})) between this potential
and the vector potential $ A $  with support in the convex hull of $\tilde{K}$ constructed in Section
\ref{section-the-magnetic-field-and-the-magnetic-potential-and-the-Cutoff-Fuction}. As the vector potentials $ A $ and $A_1
$ are gauge equivalent, they define the same physics and, therefore, we can
always chose the vector potential $A $. For this potential, $   \lambda_{A,0,
    \infty}(-\hv) = 0  $, and then, the  Aharonov-Bohm Ansatz for initial
conditions at minus infinity  and  the potential $A$ is as follows.

\begin{definition}{Aharonov-Bohm Ansatz}
\label{aharonov-bohm-anzats-incoming}

Let $ A $ be the magnetic potential  constructed in Section
(\ref{section-the-magnetic-field-and-the-magnetic-potential-and-the-Cutoff-Fuction}). \\
Let $ \psi_{\v}(x,t):= e^{-i\frac{t}{\hbar} H(A)} W_{-}(A) \,\varphi_{\v} $ be the solution to the Schr\"odinger equation that behaves
like
\beq
\label{a-b-anzats-h.9.4}
\psi_{\v,0}:=e^{-i\frac{t}{\hbar} H_{0}}\varphi_{\v}
\ene
when  time goes to minus infinity.
We suppose that $ \psi_{\v} $ is approximately localized in $ \mathcal{D} $
for all times. Then, the following change of gauge formula holds,
\beq
\label{a-b-anzats-h.9.5}
\psi_{\mathbf v} \approx  \psi_{AB, \mathbf v}(x,t):=
e^{i\lambda_{A,0}(x) } e^{-i\frac{t}{\hbar} H_0} \varphi_\mathbf v.
\ene
\end{definition}

\bull

\noindent Observe that  the Aharonov-Bohm Ansatz is the
multiplication of the free solution by the Dirac magnetic factor
$e^{i \lambda_{A,0}(x) }$ \cite{di}.

Note that as we noticed before, the electron -when emitted, would follow the free evolution
 $ e^{-i \frac{t}{\hbar}H_0} \varphi_\v  $ under the assumption that we take a representation where the magnetic potential
($A$) vanishes at this time. If we take a representation given by a general vector potential ($A_1$) with flux $\Phi$,
 we should change the initial conditions at minus infinity by   $  e^{i \lambda_{A_1,0,\infty}(-\hv)}
 e^{-i \frac{t}{\hbar}H_0} \varphi_\v$ (notice that $\lambda_{A_1,0,\infty}(-\hv)  =  \lambda_{A_1,A,\infty}(-\hv) $ ).

In the following sections we give a rigorous proof that
(\ref{a-b-anzats-h.9.5}) holds and  we obtain error bounds for the difference
between the exact solution and the Aharonov-Bohm Ansatz. We also provide  a physical interpretation of the error bound and we
relate it to the probability for the electron  to be outside the region $ \mathcal{D} $.

\section{The Time Evolution of the electron Wave Packet. Final Estimates}
\label{final} \sss
In this Section we use the same symbol,  $   e^{-i \frac{\zeta}{v \hbar}H _{0} }  $, for the restriction of the free evolution to $ \Lambda $ and,
 moreover, we designate by $  \|  \cdot  \| $ the norm in $  L^2(\Lambda)  $.
\subsection{Incoming Electron Wave Packet. Final Estimates}\label{ab-anzats-section-incoming-wave-packet}

\begin{lemma} \label{ab-anzats-ic-w-p-lemma}
For every gaussian wave function, $ \varphi $, with variance $ \sigma  $ and
for every $ \zeta  \in \ere  $ with $ \zeta \leq -z(\sigma)  $, the following
estimate holds.
\beq \label{ab-anzats-ic-w-p.1}
\| \chi e^{-i \frac{\zeta}{v \hbar}H _{0} } \varphi_{\v} - e^{i \lambda_{A,0} }
e^{-i \frac{\zeta}{v \hbar} H_0}  \varphi_\v  \| \leq  \sqrt{2} e^{-\frac{33}{34}
  \frac{(\sigma m v)^{2}}{2}  } + 10^{-420}.
\ene

\end{lemma}

\noindent{\it Proof:}
Let $\mathcal{D}_{-h} $ be the set $ \{ (x_1, x_2, x_3)  \in \ere^{3} : x_3 \leq -
h  \} $. We have that, $ \lambda_{A,0} (x) = 0  $ and $ \chi(x)=1  $ for  $ x \in \mathcal{D}_{-h}
$. Using polar coordinates we
obtain (see (\ref{3.13}), (\ref{a.4}) and Remark \ref{rem-a.1}).
\beq \label{ab-anzats-ic-w-p.2}
\| \chi e^{-i \frac{\zeta}{v \hbar}H _{0} } \varphi_{\v} - e^{i \lambda_{A,0} }
e^{-i \frac{\zeta}{v \hbar} H_0}  \varphi_\v     \|^{2} \leq \frac{4}{\pi^{3/2}}
\int_{( \ere^3 \setminus \mathcal{D}_{-h} - \hv \zeta)\rho(\sigma, \zeta)} e^{-x^2} dx \leq 2 e^{-
  \theta_{inv}(\sigma, z(\sigma))^2 }.
\ene
Finally we notice that $ \sqrt {2} \, e^{-
\frac{  \theta_{inv}(\sigma, z(\sigma))^2 }{2}} \leq 10^{-434} $ for $
\sigma \geq \sigma_0 $.

\bull

Using Theorem \ref{tonomura-experiments-incoming-wave-packet} and Lemma \ref{ab-anzats-ic-w-p-lemma}
we prove that,

\begin{theorem}{Aharonov-Bohm Ansatz. Incoming Wave Packet}\label{ab-anzats-ic-w-p}\\
Suppose that the magnets and energies are the ones of the experiments of
Tonomura et al.. Then for every gaussian wave function with variance $ \sigma
\in [\frac{4.5}{mv}, \frac{\tilde{r}_1}{2}]  $
and every $ \zeta \in \ere    $ with $ \zeta \leq -z(\sigma)   $, the solution
to the Schr\"odinger equation that behaves as  (\ref{a-b-anzats-h.9.4}) when the
time goes to minus infinity, $e^{- i \frac{\zeta}{v \hbar}H}W_{-}(A)\varphi_{\v}$, is given at the time $ t = \frac{\zeta}{v}  $ (
$\zeta$ being the vertical coordinate) by the Aharonov-Bohm Ansatz,
\beq\label{ab-anzats-ic-w-p.2.1}
 e^{i\lambda_{A,0}(x) } e^{-i\frac{t}{\hbar} H_0} \varphi_\v,
\ene
up to an error bound of the form:

\beq \label{ab-anzats-ic-w-p.3}
\begin{array}{l}
\| e^{- i \frac{\zeta}{v \hbar}H}W_{-}(A)\varphi_{\v}- e^{i \lambda_{A,0} }
e^{-i \frac{\zeta}{v \hbar} H_0}  \varphi_\v
\| \leq e^{- \frac{33}{34} \frac{(\sigma m v)^2}{2} } (\sum_{i \in \{1, 1/2, 0,
  -1/2, -1  \}} \ap_{i}^{-\infty} \sigma^{i} + \sqrt{2})  + 10^{-419},
\end{array}
\ene
where the quantities $ \ap^{-\infty}_{i}  $ are explicit numbers that depend
only on the magnet and the energy  that we take (see (\ref{av})).
\end{theorem}

\subsection{Interacting Electron Wave Packet. Final Estimates}\label{ab-anzats-section-interacting-wave-packet}
\begin{lemma} \label{ab-anzats-i-w-p-lemma}
For every gaussian wave function, $ \varphi $, with variance $ \sigma \in[4.5/mv, \tilde{r}_1/2] $ and
for every $ \zeta  \in \ere  $ with $ |\zeta| \leq z (\sigma)  $, the following
estimate holds.
\beq \label{ab-anzats-i-w-p.1}
\begin{array}{l}
\|\chi e^{-i \int_0^{-\infty} \hv \cdot A (x + \tau \hv)d \tau  } e^{-i \frac{\zeta}{v
    \hbar}H _{0} } \varphi_{\v} -
 e^{i \lambda_{A,0} }e^{-i \frac{\zeta}{v \hbar} H_0}  \varphi_{\v}    \|
\leq  2 e^{-\frac{1}{2}  r_{1}^2 \rho (\sigma, \zeta)^{2} } \leq  2.0031\,\,
e^{-\frac{1}{2} \frac{r_1^2}{\sigma^2}}   +\\\\
2 e^{-\frac{33}{34}  \frac{(\sigma m v)^{2}}{2}  } + 10^{-456}.
\end{array}
\ene
\end{lemma}
\noindent{\it Proof:}
We denote by  $ {\mathcal HM} :=  \{ (x_1, x_2, x_3) \in \ere^3 : \sqrt{x_1^2
  + x_2^2} \leq r_1  \}$. For $ x \in {\mathcal HM}$,  $- \int_0^{-\infty} \hv \cdot A (x + \tau \hv)d \tau    = \lambda_{A,0}^{h}(x)$ and
  $\chi(x) = 1$.
 Using polar coordinates we obtain (see (\ref{3.13}, \ref{a.4})),
\beq
\label{ab-anzats-i-w-p.2}
\| \chi e^{-i \int_0^{-\infty} \hv A (x + \tau \hv)d \tau  } e^{-i \frac{\zeta}{v
    \hbar}H _{0} } \varphi_{\v} -
 e^{i \lambda_{A,0} }e^{-i \frac{\zeta}{v \hbar} H_0}  \varphi_\v     \|^2  \leq \frac{4}{\pi^{3/2}}
\int_{( \ere^3 \setminus \mathcal{HM} - \hv \zeta)\rho(\sigma, \zeta)} e^{-x^2} dx \leq
4 e^{-
 r_{1}^2 \rho(\sigma, \zeta)^{2} }.
\ene

The second inequality in (\ref{ab-anzats-i-w-p.1}) is proved in three  cases:

\begin{itemize}
\item{$\sigma \in [\frac{4.5}{mv}, \sigma_0]  $.}

By (\ref{te-tee.5}), see also Sections \ref{notation} and
\ref{parameters-election}, \beq  e^{- r_1^2 \rho(\sigma, \zeta)^2}
\leq e^{-\frac{r_1^2}{ (z(\sigma) - h)^2 }
 \frac{1}{\tilde{\omega}(\sigma)^2}} \leq e^{-\frac{r_1^2}{(134.82 \tilde{h})^2}
 \frac{33}{34} (\sigma mv)^2  }  \leq    e^{-\frac{33}{34} (\sigma mv)^2  }.
\ene

\item {$ \sigma \in [\sigma_{0}, 3.2 \times 10^{-6}   ].$}
For these values of $ \sigma $ we have that $ \tilde{\omega} (\sigma) = 2000^{-1/2}
$. We use  (\ref{te-tee.0}),  (\ref{a.28}) and the triangle inequality for the
 square-root term to obtain,

\beq
\label{ab-anzats-i-w-p.4}
\frac{z(\sigma)}{\sigma^2 mv } \leq \frac{1}{mv}\left(\frac{68 h}{\sigma^2}  +
  \frac{\sqrt{2000\times 34}}{\sigma}  \right).
\ene
Then,
\beq
\label{ab-anzats-i-w-p.5}
e^{- \frac{1}{2} \rho(\sigma, \zeta)^2 r_1^2} \leq {\mathrm exp}\left[-\frac{r_1^2}{2 \sigma^2}
  \frac{1}{1+\frac{1}{(mv)^2}(\frac{68 h}{\sigma^2}  +
    \frac{\sqrt{2000 \times 34}}{\sigma}  )^2 }  \right] = {\mathrm exp}\left[-\frac{r_1^2}{2 }
  \frac{1}{\sigma^2+\frac{1}{(mv)^2}(\frac{68 h}{\sigma}  +
    \sqrt{2000\times 34}  )^2} \right].
\ene
The function $ f(\sigma) = 1/ \left(\sigma^2+\frac{1}{(mv)^2}(\frac{68 h}{\sigma}  +
    \sqrt{2000\times 34}  )^2\right) $ restricted to the interval $[\sigma_0 , 10^{-7}  ]
  $ has  derivative equal to zero on the positive
  axis only at the unique point of intersection of the function   $
  \sigma^4 $ and the line  $ \frac{68 h}{(mv)^2}  (68 h + \sqrt{2000 \times 34}\sigma  )
  $, see Sections \ref{notation} and \ref{parameters-election}. For the interval $ [10^{-7}, 3.2 \times 10^{-6} ] $
 the derivative of $ f
  $ is zero over the positive axis in the unique solution of the equation  $
  \sigma^4 = \frac{68\tilde{h}}{(mv)^2}  (68 \tilde{h} + (\sqrt{2000 \times 34} + 680
  )\sigma)  $, see Sections \ref{notation} and \ref{parameters-election}. Then, it follows that,

\beq
\label{ab-anzats-i-w-p.6}
\begin{array}{l}
 {\mathrm \exp}\,\left[-\frac{r_1^2}{2 }
  \frac{1}{\sigma^2+\frac{1}{(mv)^2}(\frac{68 h}{\sigma}  +
    \sqrt{2000\times 34}  )^2 }  \right] \leq \\\\
    \max_{\nu \in  \{ \sigma_0,\, 10^{-7} , \,
  3.2 \times 10^{-6}  \}}
 {\mathrm exp}\,\left[-\frac{r_1^2}{2 }
  \frac{1}{\nu^2+\frac{1}{(mv)^2}(\frac{68 h}{\nu}  +
    \sqrt{2000\times 34}  )^2 }\right].
\end{array}
\ene
Evaluating (\ref{ab-anzats-i-w-p.6}) using the experimental energies and
magnets, we find that,

\beq
\label{ab-anzats-i-w-p.7}
e^{- \frac{1}{2} \rho(\sigma, \zeta)^2} \leq 10^{-458}.
\ene

\item {$\sigma \in [ 3.2 \times 10^{-6}, \frac{\tilde{r}_1}{2}].  $}

Now we use that
\beq
\label{ab-anzats-i-w-p.8}
 e^{-\ r_{1}^2 \rho(\sigma, \zeta)^{2} } \leq
 e^{-\frac{r_1^2}{\sigma^2}}
e^{- \left( r_{1}^2 \rho(\sigma, z(\sigma)\right)^{2}-  \frac{r_1^2}{\sigma^2}
  ) } = e^{-\frac{r_1^2}{\sigma^2}}\, {\mathrm exp}\left[   \frac{r_1^2}{\sigma^2}    \frac{
  ( \frac{z(\sigma)}{ \sigma^{2} m v } )^2}{ 1  + ( \frac{z(\sigma)}{ \sigma^{2} m v } )^2  }\right].
\ene

By (\ref{a.28}) $ \frac{z(\sigma)}{ \sigma^{2} m v }  $ is decreasing as a
function of $ \sigma  $ (see Sections \ref{notation}, and \ref{parameters-election} and
notice that $ \frac{(\sigma mv )^2}{ (\sigma mv)^2 - \tilde{\omega}^2 }  $ is
decreasing on $\sigma$) and then,
we have that,

\beq
\label{ab-anzats-i-w-p.9}
 \sqrt{2} \,{\mathrm exp}\left[   \frac{r_1^2}{2\sigma^2}    \frac{
  ( \frac{z(\sigma)}{ \sigma^{2} m v } )^2      }{     1
 +      ( \frac{z(\sigma)}{ \sigma^{2} m v } )^2  }\right] \leq
\sqrt{2}\, {\mathrm exp} \left[   \frac{r_1^2}{2(3.2 \times 10^{-6})^2}    \frac{
  ( \frac{z(3.2 \times 10^{-6})}{ (3.2 \times 10^{-6})^{2} m v } )^2      }{
  1  +    ( \frac{z(3.2 \times 10^{-6})}{ (3.2 \times 10^{-6})^{2} m v } )^2  }
\right] \leq 1.4171.
\ene
\end{itemize}

\begin{remark}\label{ab-anzats-i-w-p-remark}
{\rm The term appearing in the middle inequality of equation
(\ref{ab-anzats-i-w-p.1}) is two times  the square root of the probability for
the free particle to be outside the hole of the magnet ($ \mathcal{H}M  $) when the
electron is classically at the position  $  (0, 0, \zeta) $:

\beq \label{ab-anzats-i-w-p.9.2}
\int_{\ere^3 \setminus \mathcal{H}M} |(e^{-i \frac{\zeta}{v \hbar} H_0} \varphi_\v)(x) | ^2
dx = e^{- r_{1}^2 \rho(\sigma, \zeta)^{2} }.
\ene
Recall that $ \mathcal{H}M  $ is defined in the proof of Lemma
\ref{ab-anzats-i-w-p-lemma}.
Equation (\ref{ab-anzats-i-w-p.9.2}) is  a
measure of the part of the electron that hits the magnet when the
classical electron (the electron under classical mechanics rules)
lies within a distance less than $ z(\sigma)  $ from the center of
the magnet. By the  second inequality in (\ref{ab-anzats-i-w-p.1} )
we can see that the probability of the electron to be outside the
hole of the magnet at time $ \zeta/v $ splits in two terms: one,
$e^{-\frac{1}{2} \frac{r_1^2}{\sigma^2}}$, is  due to the
probability of the free electron to be outside the hole when $ \zeta
= 0 $ (see formula (\ref{ab-anzats-i-w-p.9.2})).
 This factor provides us an idea of the influence of the magnet over the
 electron given by the size of the wave
packet (i.e., how much does
the electron hits the magnet -see Section \ref{radius-wave-packet}),  and the
other, $e^{-\frac{33}{34}
  \frac{(\sigma m v)^{2}}{2}  }$, is related with the spreading of the electron
as  time increases - see Section \ref{spreading-angle}. This factor is important when   $\sigma   $ is small, because
by Heisenberg uncertainly principle when the electron is localized in a small
region, its momentum is not localized and therefore the electron spreads. Those
two factors are essentially the causes of all the error bounds that we have in
this paper. The error bounds are mainly produced by the
probability of the electron to hit the magnet when it is classically at
the position $ (0,0, \zeta)  $, with $ |\zeta| \leq z(\sigma)  $. In Section
\ref{physical-interpretation-error-bounds} we provide an analysis of these terms and we give precise
definitions of the size of the electron wave packet and of the opening  angle,
that is due to the spreading.}
\end{remark}

Using Theorem \ref{tonomura-experiments-interacting-wave-packet} and Lemma \ref{ab-anzats-i-w-p-lemma}
we prove,
\begin{theorem}{Aharonov-Bohm Ansatz. Interacting Electron Wave Packet}
\label{ab-anzats-i-w-p}\\
Suppose that the magnets and energies are the ones of the experiments of
Tonomura et al.. Then, for every gaussian wave function with variance $ \sigma
\in [\frac{4.5}{mv}, \frac{\tilde{r}_1}{2}]  $
and every $ \zeta \in \ere    $ with $ |\zeta| \leq z(\sigma)   $   the solution
to the Schr\"odinger  equation, $e^{- i \frac{\zeta}{v \hbar}H}W_{-}(A)\varphi_{\v}$, that behaves as  (\ref{a-b-anzats-h.9.4}) when
time goes to minus infinity is given at the time $ t = \frac{\zeta}{v}  $ (
$\zeta$ being the vertical coordinate)  by the Aharonov-Bohm Ansatz,
\beq
\label{ab-anzats-i-w-p.2.1}
 e^{i\lambda_{A,0} } e^{-i\frac{t}{\hbar} H_0} \varphi_\v,
\ene
up to an error bound of the form:

\beq
\begin{array}{l}
\| e^{- i\frac{\zeta}{v\hbar} H}W_{-}(A)\varphi_{\v}- e^{i \lambda_{A,0} }
e^{-i \frac{\zeta}{v \hbar} H_0}  \varphi_\v
\| \leq
\\\\
6.0031 e^{- \frac{r_{1}^2}{2 \sigma^2}}   +   e^{- \frac{33}{34}
  \frac{(\sigma m v)^2}{2} }
(\sum_{i \in \{1, 1/2, 0,
  -1/2, -1  \}} (1+10^{-3})  \ap_{i}^{0} \sigma^{i} +2  )  + 10^{-101} + 10^{-420} + 10^{-456} ,
\end{array}
\ene 
where the quantities $ \ap^{0}_{i} $ are explicit numbers that depend only on the magnet and the energy  that we take (see (\ref{av})).

\end{theorem}

\subsection{Outgoing Electron Wave Packet and Scattering Operator. Final Estimates}\label{ab-anzats-section-outgoing-wave-packet}

\begin{lemma} \label{ab-anzats-o-w-p-lemma}
For every gaussian wave function, $ \varphi $, with variance $ \sigma  $ and
for every $ \zeta  \in \ere  $ with $ \zeta \geq z(\sigma)  $, the following
estimate holds.
\beq
\label{a-b-anzats-o-w-p.1}
\| \chi e^{i \Phi }e^{-i \frac{\zeta}{v \hbar}H _{0} } \varphi_{\v} - e^{i \lambda_{A,0} }
e^{-i \frac{\zeta}{v \hbar} H_0}  \varphi_\v     \|  \leq  \sqrt{2} \,  e^{-\frac{33}{34}
  \frac{(\sigma m v)^{2}}{2}  } + 10^{-420}.
\ene

\end{lemma}

\noindent{\it Proof:}
Let $\mathcal{D}_h $ be the set $ \{ (x_1, x_2, x_3)  \in \ere^{3} : x_3 \geq
h  \}  $, note that $ \lambda_{A,0} (x) = \Phi $ and $ \chi(x) = 1 $ for $ x \in \mathcal{D}_h
$. The proof follows in the same way as the proof of Lemma \ref{ab-anzats-ic-w-p-lemma}.

\bull

 Theorem \ref{tonomura-experiments-outgoing-wave-packet} and Lemma \ref{ab-anzats-o-w-p-lemma}
imply the following theorem.

\begin{theorem}{Aharonov-Bohm Ansatz. Outgoing Electron Wave Packet}\label{ab-anzats-o-w-p}\\
Suppose that the magnets and energies are the ones of the experiments of
Tonomura et al.. Then for every gaussian wave function with variance $ \sigma
\in [\frac{4.5}{mv}, \frac{\tilde{r}_1}{2}]  $
and every $ \zeta \in \ere    $ with $ \zeta \geq z(\sigma)   $   the solution
to the Schr\"odinger equation, $e^{- i \frac{\zeta}{v \hbar}H}W_{-}(A)\varphi_{\v}$, that behaves as  (\ref{a-b-anzats-h.9.4}) when the
time goes to minus infinity is given  at the time $ t = \frac{\zeta}{v}  $ ($\zeta$ being the vertical coordinate) by the Aharonov-Bohm Ansatz,
\beq\label{a-b-anzats-o-w-p.1.1}
 e^{i\lambda_{A,0} } e^{-i\frac{t}{\hbar} H_0} \varphi_\v,
\ene
up to an error bound of the form:

\beq
\label{a-b-anzats-o-w-p.2}
\begin{array}{l}
\| e^{- i \frac{\zeta}{v \hbar}H}W_{-}(A)\varphi_{\v}- e^{i \lambda_{A,0} }
e^{-i \frac{\zeta}{v \hbar} H_0}  \varphi_\v
\| \leq \\\\
  7 e^{- \frac{r_{1}^2}{2 \sigma^2}}   +   e^{- \frac{33}{34}
  \frac{(\sigma m v)^2}{2} }
(\sum_{i \in \{1, 1/2, 0,
  -1/2, -1  \}}(1+10^{-7}) \ap_{i}^{\infty} \sigma^{i} + \sqrt{2})  +  10^{-101} + 2
\times 10^{-420},
\end{array} 
\ene
and, furthermore, the scattering operator satisfies,
\beq
\|  S(A)\varphi_\v - e^{i\Phi}\varphi_\v \| \leq 7 e^{- \frac{r_{1}^2}{2 \sigma^2}}   +   e^{- \frac{33}{34}
  \frac{(\sigma m v)^2}{2} }
(\sum_{i \in \{1, 1/2, 0,
  -1/2, -1  \}}(1+10^{-7}) \ap_{i}^{\infty} \sigma^{i} + \sqrt{2})  + 10^{-101} +
2\times 10^{-420},
\ene 
where the quantities $ \ap^{\infty}_{i}  $ are explicit numbers that depend only on the magnet and the energy  that we take (see (\ref{av})).
\end{theorem}

\subsection{Uniform in Time Estimates for the Electron Wave Packet}\label{ab-anzats-section-uniform-estimations}

\begin{remark}\label{ab-anzats-u-e-remark}
{\rm The error bound of Theorem \ref{ab-anzats-ic-w-p} is smaller
that the one of Theorem  \ref{ab-anzats-i-w-p} and this last one is
bounded by the error bound of Theorem \ref{ab-anzats-o-w-p}. This
is physically reasonable, because for an electron to be an interacting
electron, it has to be first incoming electron and for an electron
to be outgoing electron it has to be before an interacting electron,
so the error should be accumulative. Let us prove this.
That the error of Theorem \ref{ab-anzats-ic-w-p}  is smaller than the one of
the Theorem  \ref{ab-anzats-i-w-p} follows directly from the definitions
(\ref{av}). To prove that the error in Theorem \ref{ab-anzats-o-w-p} bounds the
one of Theorem \ref{ab-anzats-i-w-p} we use again (\ref{av}) and that  (remember that
 $ \sigma m v \geq 4.5$),
\beq
\begin{array}{l}
(1+ 10^{-3}) \frac{ \ap^{0}_{-1}}{\sigma} + (2-\sqrt{2} ) =  (1+ 10^{-3})  \left[ \frac{\sqrt{\pi}}{2} \frac{m_5}{2} (1 + 1.11 \times 10^{-6})^{1/2}
 \frac{136.82}{\sigma mv}\tilde{h}\right] +  (2-\sqrt{2}) \leq \\\\
  (2- 10^{-3} ) (
 150(1 - \frac{1}{50}) \sigma m v -134.99    ) \frac{\tilde{h}}{2} \frac{\frac{1}{\sqrt{2}}+ \frac{\sqrt{3} \pi^{1/4}  }{2}
 }{\sqrt{\sigma m v} \pi^{1/4} } m_5  $  $    \leq  (2 -  10^{-3}  ) (\sigma^{1/2} mv r_1 - \sigma^{-1/2} 134.99 \tilde{h} )
\frac{\mathcal{A}_3(\bar{m})}{2}    \leq \\\\
(2-  10^{-3}  ) (
 \sigma^{1/2} \ap^{-\infty}_{1/2} + \sigma^{-1/2} \ap^{-\infty}_{-1/2} ).
\end{array}
\ene 
}
\end{remark}

\bull

This gives us the following theorem.

\begin{theorem}{Aharonov-Bohm Ansatz. Time-Uniform Estimates}\label{ab-anzats-o-w-p-u-t}\\
Suppose that the magnets and energies are the ones of the experiments of
Tonomura et al.. Then, for every gaussian wave function with variance $ \sigma
\in [\frac{4.5}{mv}, \frac{\tilde{r}_1}{2}]  $
and every $ \zeta \in \ere    $   the solution
to the Schr\"odinger equation, $e^{- i \frac{\zeta}{v \hbar}H}W_{-}(A)\varphi_{\v}$, that behaves as  (\ref{a-b-anzats-h.9.4}) when the
time goes to minus infinity is given at the time $ t = \frac{\zeta}{v}  $ (
$\zeta$ being the vertical coordinate) by the Aharonov-Bohm Ansatz,
\beq\label{ab-anzats-u-e.1}
 e^{i\lambda_{A,0}(x) } e^{-i\frac{t}{\hbar} H_0} \varphi_\v,
\ene
up to an error bound of the form:

\beq \label{ab-anzats-u-e.2}
\begin{array}{l}
\| e^{- i \frac{\zeta}{v \hbar}H}W_{-}(A)\varphi_{\v}- e^{i \lambda_{A,0} }
e^{-i \frac{\zeta}{v \hbar} H_0}  \varphi_\v
\| \leq \\\\
  7 e^{- \frac{r_{1}^2}{2 \sigma^2}}   +   e^{- \frac{33}{34}
  \frac{(\sigma m v)^2}{2} }
(\sum_{i \in \{1, 1/2, 0,
  -1/2, -1  \}}(1+10^{-7}) \ap_{i}^{\infty} \sigma^{i} + \sqrt{2})  + 10^{-101} +
2\times 10^{-420}.
\end{array}
\ene
Moreover, the scattering operator satisfies,
\beq
\|  S\varphi_\v - e^{i\Phi}\varphi_\v \| \leq  7 e^{- \frac{r_{1}^2}{2 \sigma^2}}   +   e^{- \frac{33}{34}
  \frac{(\sigma m v)^2}{2} }
(\sum_{i \in \{1, 1/2, 0,
  -1/2, -1  \}}(1+10^{-7}) \ap_{i}^{\infty} \sigma^{i} + \sqrt{2})  + 10^{-101} +
2\times 10^{-420}.
\label{ab-anzats-u-e.2.b}
\ene
The quantities $ \ap^{\infty}_{i}  $   are explicit numbers that depend only on the magnet and the energy  that we take (see (\ref{av})).
\end{theorem}

\bull

By (\ref{n-b.1}) and (\ref{E.1.1}) $ m_i(\chi_1) \geq m_{i}(\chi_2), \, i \in
\{1, \cdots, 5  \}  $, and as $ \sigma^{1/2} mv r_1 \geq \frac{134.99 \, \tilde{h}}{\sigma^{1/2}}  $
  ($ \sigma mvr_1 \geq 134.99 \, \tilde{h}      $, remember that $ \sigma mv \geq 4.5 $), we have that $ \ap^{j}(\sigma, v,
\bar{m}(\chi_1))  \geq  \ap^{j}(\sigma, v,
\bar{m}(\chi_2))  $ (see (\ref{ap})).   We have also (see (\ref{av}) and (\ref{ap})) that $ \ap^{\infty}
(\sigma,v_{i},\bar{m}(\chi_{1}) ) \leq  \ap^{\infty}
(\sigma,v_{1},\bar{m}(\chi_{1}) )   $ for $ i \in \{1,
  2, 3\}  $ and $  j \in \{1, 2  \} $ (notice that $ \mathcal{A}(\bar{m})_1  \geq \mathcal{A}(\bar{m})_4  $ and
$  \mathcal{A}(\bar{m})_3  \geq \mathcal{A}(\bar{m})_5  $). So if we write $\ap_{i}^{\infty}(v_1,
  \bar{m}(\chi_1))  $ in (\ref{ab-anzats-u-e.2}, \ref{ab-anzats-u-e.2.b}) instead of $ \ap_i^\infty  $
  we obtain also error bounds, but now the coefficients $ \ap^{\infty}_{i} $ are fixed for all the magnets and velocities. Taking this into
  consideration  we calculate the values of $ \av^\infty(v_1, \bar{m}(\chi_1))
  $ and we obtain the following theorem.

\begin{theorem}{Aharonov-Bohm Ansatz and Tonomura et al. Experiments}
\label{th6.13.viejo}
\\
Suppose that the magnets and energies are the ones of the experiments of
Tonomura et al.. Then, for every gaussian wave function with variance $ \sigma
\in [\frac{4.5}{mv}, \frac{\tilde{r}_1}{2}]  $
and every $ \zeta \in \ere    $, the solution
to the Schr\"odinger equation, $e^{- i \frac{\zeta}{v \hbar}H}W_{-}(A)\varphi_{\v}$, that behaves as  (\ref{a-b-anzats-h.9.4}) when the
time goes to minus infinity is given at the time $ t = \frac{\zeta}{v}  $ (
$\zeta$ being the vertical coordinate) by the Aharonov-Bohm Ansatz,
\beq\label{ab-anzats-u-w-e.3.viejo}
 e^{i\lambda_{A,0}(x) } e^{-i\frac{t}{\hbar} H_0} \varphi_\v,
\ene
up to an error bound of the form:
\beq \label{ab-anzats-o-w-p.2.viejo}
\begin{array}{l}
\| e^{- i \frac{\zeta}{v \hbar}H}W_{-}(A)\varphi_{\v}- e^{i \lambda_{A,0} }
e^{-i \frac{\zeta}{v \hbar} H_0}  \varphi_\v
\| \leq \\\\
  7 e^{- \frac{r_{1}^2}{2 \sigma^2}}   +   e^{- \frac{33}{34}
  \frac{(\sigma m v)^2}{2} }
(1.04 \times 10^{14}\sigma + 3.91 \times 10^{8} \sigma^{1/2} - 1.41 \times
10^{3} - 1.14 \times 10^{-2} \frac{1}{\sigma^{1/2} }  )  + 10^{-101} +
2\times 10^{-420}.
\end{array} 
\ene
Furthermore, the scattering operator satisfies,
\beq
\begin{array}{l}
 \|  S\varphi_\v - e^{i\Phi}\varphi_\v \| \leq  7 e^{- \frac{r_{1}^2}{2 \sigma^2}}   +   e^{- \frac{33}{34}
  \frac{(\sigma m v)^2}{2} }
(1.04\times 10^{14}\sigma + 3.91 \times 10^{8} \sigma^{1/2} - 1.41
\times 10^{3} - 1.14 \times 10^{-2} \frac{1}{\sigma^{1/2} }  )
+\\\\
 10^{-101} +
2\times 10^{-420}.
\end{array}
\label{ab-anzats-o-w-p.2.b.viejo}
\ene
\end{theorem}

\bull

We now bound the right hand side of (\ref{ab-anzats-o-w-p.2.viejo}) by  $ 7 e^{- \frac{r_{1}^2}{2 \sigma^2}}  + \mathcal{F} (\sigma, mv)  $, where
  $  \mathcal{F} (\sigma, mv) :=   e^{- \frac{33}{34}
  \frac{(\sigma m v)^2}{2} }
(1.04 \times 10^{14}\sigma + 3.91 \times 10^{8} \sigma^{1/2})  + 10^{-101} + 2\times10^{-420} $.   We notice that
$  \mathcal{F} $ is decreasing for $mv$ fixed and $  \sigma mv \geq 4.5 $.
We compute $ \mathcal{F}(15.5 \times 10^{-10}, mv_3)  $
and show that this quantity is less than  $10^{-100}$, it follows that  $ \mathcal{F}(\sigma,mv) \leq 10^{-100} $
 for $  \sigma \geq 15.5\times 10^{-10} $ and the experimental velocities. Then
  $  \mathcal{F} (\sigma, mv)   \leq   e^{- \frac{33}{34}
  \frac{(\sigma m v)^2}{2} } (1.04 \times 10^{14}(  15.5 \times 10^{-10}  ) + 3.91 \times 10^{8} (15.5 \times 10^{-10})^{1/2})  + 10^{-100}  $
$ =      177 \times 10^{3}        e^{- \frac{33}{34}
  \frac{(\sigma m v)^2}{2} }   + 10^{-100}  $. We obtain the following theorem, that is our main result, and that is quoted as
Theorem \ref{th1.1}   in the introduction.

\begin{theorem}{Aharonov-Bohm Ansatz and Tonomura et al. Experiments. Final Estimates}
\label{th6.13}
\\
Suppose that the magnets and energies are the ones of the experiments of
Tonomura et al.. Then, for every gaussian wave function with variance $ \sigma
\in [\frac{4.5}{mv}, \frac{\tilde{r}_1}{2}]  $
and every $ \zeta \in \ere    $, the solution
to the Schr\"odinger equation, $e^{- i \frac{\zeta}{v \hbar}H}W_{-}(A)\varphi_{\v}$, that behaves as  (\ref{a-b-anzats-h.9.4}) when the
time goes to minus infinity is given at the time $ t = \frac{\zeta}{v}  $ ($\zeta$
being the vertical coordinate) by the Aharonov-Bohm Ansatz,
\beq\label{ab-anzats-u-w-e.3}
 e^{i\lambda_{A,0}(x) } e^{-i\frac{t}{\hbar} H_0} \varphi_\v,
\ene
up to an error bound of the form:
\beq \label{ab-anzats-o-w-p.2}
\begin{array}{l}
\| e^{- i \frac{\zeta}{v \hbar}H}W_{-}(A)\varphi_{\v}- e^{i \lambda_{A,0} }
e^{-i \frac{\zeta}{v \hbar} H_0}  \varphi_\v
\| \leq \\\\
  7 e^{- \frac{r_{1}^2}{2 \sigma^2}}   +   177 \times 10^{3} e^{- \frac{33}{34}
  \frac{(\sigma m v)^2}{2} }  + 10^{-100}.
\end{array}
\ene
Furthermore, the scattering operator satisfies,
\beq
 \|  S\varphi_\v - e^{i\Phi}\varphi_\v \| \leq  7 e^{- \frac{r_{1}^2}{2 \sigma^2}}   +  177 \times 10^{3}  e^{- \frac{33}{34}
  \frac{(\sigma m v)^2}{2} }
+ 10^{-100}.
\label{ab-anzats-o-w-p.2.b}
\ene
\end{theorem}

\begin{remark}\label{probabilidad_interaccion}
{\rm

In the experiments of Tonomura et al. \cite{to2},  they send an electron wave packet that partially hits the magnet. The part of the electron
wave packet that hits the magnet does not go behind the magnet  because we can see the black shadow of the magnet behind it. In other words,
 this part of the electron wave packet will be in the region   $ \{ (x_1, x_2, x_3)  \in \Lambda : x_3 \leq
h  \}  $. We can bound,
therefore, the probability of interaction of the electron with the magnet by the probability for the electron to not be behind the
 magnet for large time. We denote, as in the proof of Lemma \ref{ab-anzats-o-w-p-lemma}, by
 $\mathcal{D}_h $ the set $ \{ (x_1, x_2, x_3)  \in \ere^{3} : x_3 \geq
h  \}  $. Actually    $\mathcal{D}_h $ is the region behind the magnet. Then the probability of interaction of the electron
 with the magnet is bounded by,
\beq \label{p_i.1}
\| \chi_{\Lambda \setminus \mathcal{D}_h} e^{- i \frac{t}{ \hbar}H}W_{-}(A)\varphi_{\v}  \|^2
\ene
when the time goes to $ \infty  $, where $ \chi_{\Lambda \setminus \mathcal{D}_h}     $ is the characteristic function of the set
 $  \Lambda \setminus \mathcal{D}_h   $.  We take as before $ \zeta = v t  $, then we have,

\beq \label{p_i.2}
\begin{array}{l}
\| \chi_{\Lambda \setminus \mathcal{D}_h} e^{- i \frac{t}{\hbar}H}W_{-}(A)\varphi_{\v}  \|^2
\leq (   \| e^{- i \frac{\zeta}{v \hbar}H}W_{-}(A)\varphi_{\v}- e^{i \lambda_{A,0} }
e^{-i \frac{\zeta}{v \hbar} H_0}  \varphi_\v
\|    +    \|   \chi_{\Lambda \setminus  \mathcal{D}_h} e^{-i \frac{\zeta}{ v \hbar} H_0}  \varphi_\v \|    )^2.
\end{array}
\ene

We take $ \hat{\omega}(\sigma) := \frac{1}{ \sqrt{\frac{33}{34}}  \sigma mv }  $ and $ \hat{z} (\sigma):= z_{\hat{\omega}(\sigma), \sigma}(h(\sigma))   $,
see Section
\ref{notation}. Using polar coordinates we
obtain for $ \zeta \geq \hat{z}(\sigma)  $ (see Section \ref{notation}, (\ref{3.13}), (\ref{a.4}) and Remark \ref{rem-a.1}),

\beq \label{p_i.2.1}
\|   \chi_{\Lambda \setminus  \mathcal{D}_h} e^{-i \frac{\zeta}{ v \hbar} H_0}  \varphi_\v \|^{2}
\leq \frac{1}{\pi^{3/2}}
\int_{( \ere^3 \setminus \mathcal{D}_{h} - \hv \zeta)\rho(\sigma, \zeta)} e^{-x^2} dx \leq \frac{1}{2}  e^{-
  \theta_{inv}(\sigma, \hat{z}(\sigma))^2 } = \frac{1}{2} e^{-\frac{33}{34} (\sigma mv)^2}.
\ene

Letting the time go to $  \infty $ in (\ref{p_i.1}) and using Theorem \ref{th6.13}, (\ref{p_i.2}) and (\ref{p_i.2.1})  we obtain that
 the probability of interaction of the electron
 with the magnet is bounded by,
\beq \label{p_i.3}
\begin{array}{l}
(  7 e^{- \frac{r_{1}^2}{2 \sigma^2}}   +  177001  e^{- \frac{33}{34}
  \frac{(\sigma m v)^2}{2} }
+ 10^{-100})^2.
\end{array}
\ene




 }
\end{remark}

\section{Physical Interpretation of the Error Bounds}\label{physical-interpretation-error-bounds}
\sss
We analyze the error bounds given in equation (\ref{ab-anzats-o-w-p.2}, \ref{ab-anzats-o-w-p.2.b}).
The error bounds appearing in the whole paper are produced by the same
factors. Equations  (\ref{ab-anzats-o-w-p.2}, \ref{ab-anzats-o-w-p.2.b}) provide uniform in time error bounds
that apply  to all experimental magnets and energies. The behaviour of the error
bound is the same for the three energies and the two magnets, so there is no
loss of generality if we select a magnet and an energy in our analysis. We will use
the biggest energy
$(E_{1})$ and the second magnet ($ K_2 $)  to provide numbers and
graphics. So, for now on we take the magnet $K_2 $ and the energy $ E_1 $.

\noindent The main factors that produce the error bound in equation
(\ref{ab-anzats-o-w-p.2}, \ref{ab-anzats-o-w-p.2.b}) are the terms,

\begin{enumerate}

\item Size of electron wave packet factor.
\beq
\label{p-i-e-b.1}
 e^{- \frac{r_{1}^2}{2 \sigma^2}}.
\ene

\item Opening angle of the electron wave packet factor.
\beq
\label{p-i-e-b.2}
 e^{- \frac{33}{34}
\frac{(\sigma m v)^2}{2} }.
\ene
\end{enumerate}
When the variance $\sigma $  is close to the radius of the magnet,
(\ref{p-i-e-b.1}) is close to $ 1 $  and (\ref{p-i-e-b.2}) is extremely small,
because in this case $ \sigma m v  $ is big. Then, when the electron is big compared to the
inner radius, (\ref{p-i-e-b.1}) is the important term, which justifies our name.
When the  variance is very small -such that $ \sigma m v  $ is close to $ 1 $-
the factor (\ref{p-i-e-b.2}) is close to one and (\ref{p-i-e-b.1}) is
extremely small ( $\frac{r_1}{\sigma}  $ is big), and then, the important factor is
(\ref{p-i-e-b.2}). But when the variance in position ($ \sigma $) is small,
by Heisenberg uncertainly principle the variance in momentum is big, and then,
the  component of the momentum transversal to the axis of the magnet is
large. In consequence, the opening angle of the
electron wave packet is large, and the electron spreads fast as it propagates. This justifies the name of (\ref{p-i-e-b.2}).

By the previous discussion, we divide the analysis of the error bounds in
 (\ref{ab-anzats-o-w-p.2}, \ref{ab-anzats-o-w-p.2.b}) in three sections: big sigma ($ \sigma  $ close to
 the inner radius of the magnet), small sigma ($ \sigma m v   $ close to 1) and
 intermediate sigma (sigma neither big, nor small).

\subsection{Big Sigma, $ \sigma \in  [ \frac{ r_{1}}{22},  \frac{\tilde{r}_1}{2}]
  $} 
Remember that $ r_1 = \tilde{r}_1 - \varepsilon $ and that
$\varepsilon  $ is defined in Section \ref{parameters-election}. Here $
\tilde{r}_1 = 1.75 \times 10^{-4} cm $ (see Section
\ref{tonomura-experiments-experimental-data}).  Then, in terms of absolute values,
 big sigma ranges over the interval $[ 7.7955 \times 10^{-6},  8.7500 \times
10^{-5} ]$.
In Figure \ref{big-sigma-figure} we show the graphic of the error bound in  (\ref{ab-anzats-o-w-p.2}) as a
function of $ \frac{\sigma}{r_1}  $, for big sigma, and in the table below we give some representative values.

\begin{tabular}{|l|r|}
   \hline
    \multicolumn{2}{|l|}{Error Bound as a Function of} \\
    \multicolumn{2}{|l|}{Sigma Over $r_1$ for Big Sigma.} \\
   \hline
     Sigma Over $r_1$      &   \textbf{\em Error Bound}\\
\hline
   .34305 & $ 10^{-1}$ \\
 \hline
   .27626  & $10^{-2}$ \\
   \hline
   .23764  & $10^{-3}$ \\
   \hline
   .21170  & $10^{-4}$ \\
   \hline
   .19274  & $10^{-5}$ \\
   \hline
   .17811  & $10^{-6}$ \\
   \hline
   .16637  & $10^{-7}$ \\
   \hline
   .15668  & $10^{-8}$ \\
   \hline
   .14851  & $10^{-9}$ \\
   \hline
   .14150  & $10^{-10}$ \\
   \hline
\end{tabular}

\subsection{Intermediate Sigma, $ \sigma \in [
  6.7591 \times 10^{-6} r_1,  \frac{r_1}{22} ]  $, or  $ \sigma \in [ \frac{23}{mv}, \frac{154678}{mv} ]  $} \label{intermediate_sigma}
Remember that $ mv = 1.9842 \times 10^{10} $  (see Section
\ref{tonomura-experiments-experimental-data}). Therefore, in terms of absolute
values,  intermediate sigma
ranges over the interval $ [ 1.1592 \times 10^{-9}, 7.7955 \times 10^{-6}  ]
$. For these values of sigma, $ \frac{r_1}{\sigma} \geq 22 $ and $  \sigma mv
\geq 23  $, and therefore, the error bound in  (\ref{ab-anzats-o-w-p.2}) is
less than $ 10^{-99}  $. \\
For intermediate sigma the probability of interaction of the electron with the magnet is less than $ 10^{-199} $
 (see Remark \ref{probabilidad_interaccion} ).

\subsection{Small Sigma, $ \sigma \in [ 1.3224 \times 10^{-6} r_1 ,
6.7591  \times 10^{-6} r_1 ]  $, or $\sigma  \in [ \frac{4.5}{mv},  \frac{23}{mv}    ] $}
In terms of absolute values we have that $ \sigma \in [2.2679 \times 10^{-10},
1.1592 \times 10^{-9}]  $. In Figure \ref{small-sigma-figure-prima} we show the graphic of the error bound in  (\ref{ab-anzats-o-w-p.2}) as a
function of $ \frac{\sigma}{r_1}  $, for small sigma, and in the table below we give some representative values.

\begin{tabular}{|l|r|}
   \hline
    \multicolumn{2}{|l|}{Error Bound as a Function of} \\
    \multicolumn{2}{|l|}{Sigma Over $r_1$ for Small Sigma.} \\
   \hline
     Sigma Over $r_1$      &   \textbf{\em Error Bound}\\
\hline
   1.6001 $\times 10^{-6}$ & $10^{-1}$ \\
 \hline
   1.7234  $\times 10^{-6}$  & $ 10^{-2}$ \\
   \hline
   1.8384  $\times 10^{-6}$  & $10^{-3}$ \\
   \hline
   1.9467   $\times 10^{-6}$   & $10^{-4}$ \\
   \hline
   2.0492   $\times 10^{-6}$   & $10^{-5}$ \\
   \hline
   2.1469   $\times 10^{-6}$  & $10^{-6}$ \\
   \hline
   2.2403   $\times 10^{-6}$  & $10^{-7}$ \\
   \hline
   2.3299e   $\times 10^{-6}$  & $10^{-8}$ \\
   \hline
   2.4162    $\times 10^{-6}$  & $10^{-9}$ \\
   \hline
   2.4996   $\times 10^{-6}$   & $10^{-10}$ \\
   \hline
\end{tabular}

\subsection{The Radius of the Electron Wave Packet}\label{radius-wave-packet}
As before, we denote  by $ \mathcal{H}M $ the cylinder $ \{  (x,y,z) \in \ere^3 :
\sqrt{ x^2 + y^2   } \leq r_{1}  \}   $.  $ \mathcal{H}M $ is
basically the hole of the magnet. The factor $  e^{- \frac{
r_{1}^{2}}{2 \sigma^{2}}}  $ is practically  the square
root of the probability for the free particle at time zero  to be
outside the hole of the magnet:

\beq
\label{r-e.9}
e^{- \frac{r_{1}^{2}}{2 \sigma^{2}}} = \left\| \chi_{ \ere \setminus \mathcal{H}M }
  (\frac{1}{\sigma^2 \pi})^{3/4} e^{-\frac{x^2}{2\sigma^{2}}}   \right\|.
\ene
This factor
represents the part of the electron wave packet that hits the magnet or goes outside
(the square root appears because our estimations are in norm and not in probability).
It is natural to have this factor in the error bound because we are only modeling
 the particles that go trough the hole. This factor is significant only
when the variance is close to the inner radius of the magnet.
As  the proximity of the electron to the magnet increases the
error in equations (\ref{ab-anzats-o-w-p.2}, \ref{ab-anzats-o-w-p.2.b}), it is important to define intuitively what is the
meaning of this closeness or, in other words, what is the size of the
electron wave packet. We agree  that the free electron is actually localized in
configuration space in a ball   centered in the classical position $ \v t
$ and with radius  chosen in such a way that the probability of finding the
electron on this ball is $ 99 \% $. We measure the radius of the wave packet at
the time $ t= 0  $ - when the free particle is  in the center of the magnet - and
denote it by $ R(\sigma)  $. Then, we have:

$$
\label{E.10} R:=  R(\sigma) = 2.382 \,\,\sigma.
$$

The error due to the part of the electron that hits the magnet (\ref{r-e.9}) is
practically  zero (smaller than  $ 10^{-99}$) when $ R \leq .1082 r_{1} ( R \leq 1.8556 \times 10^{-5})  $.
In Figure \ref{radius-figure} we show the error bound of equation
(\ref{ab-anzats-o-w-p.2}) as a function of the radius of the wave packet over $r_1 $ for big
sigma, $ \sigma \in  [ \frac{ r_{1}}{22},  \frac{\tilde{r}_1}{2}] \,\,( .1082 \, r_1  \leq R \leq .5102 \, r_1 )$.

Even when the size of
the wave packet is comparable to the inner radius of the magnet we have error
bounds extremely small. We give some data to show this behavior:

\begin{tabular}{|l|r|}
   \hline
    \multicolumn{2}{|l|}{Error Bound as a Function of the Radius} \\
    \multicolumn{2}{|l|}{of the Wave Packet Over $r_1$ for Big Sigma.} \\
   \hline
   Radius of the Wave Packet over $r_1$    &   \textbf{\em Error Bound}\\
\hline
   .81716  & $10^{-1}$ \\
 \hline
   .65806  & $10^{-2}$ \\
   \hline
   .56606 & $10^{-3}$ \\
   \hline
   .50427 & $10^{-4}$ \\
   \hline
   .45911 & $10^{-5}$ \\
   \hline
   .42425 & $10^{-6}$ \\
   \hline
   .39629 & $10^{-7}$ \\
   \hline
   .37322 & $10^{-8}$ \\
   \hline
   .35376 & $10^{-9}$ \\
   \hline
   .33703 & $10^{-10}$ \\
   \hline
\end{tabular}

\subsection{The Opening  Angle of the Electron Wave Packet}
\label{spreading-angle}
Although it is impossible to define an opening  angle of the electron,
because it is everywhere,  we agree to say that the free electron (in
momentum representation) is actually in a ball, $B_P(M\v)$ with center the
classical momentum  $ (M \v ) $ and radius $P$  such that there is a   $ 99 \%  $
probability for the electron to have its momentum within this ball.
We  define the opening angle, $ \omega (\sigma)$, in the obvious way (see
Figure \ref{spreading-angle-draw}),

$$ \label{e.10}
\sin (\frac{\omega(\sigma)}{2}) := \frac{P}{M v}= \frac{2.382}{\sigma m v}.
$$




When sigma is big,  the opening angle is very small and when sigma is
small, the opening angle is big, this is nothing more than Heisenberg
uncertainty principle.

The factor $ e^{-\frac{33}{34}\frac{(\sigma mv)^{2}}{2} }  $ of the error bound
 (\ref{ab-anzats-o-w-p.2}, \ref{ab-anzats-o-w-p.2.b}) has the following interpretation in terms of the opening angle:

$$
\label{E.11}
e^{-\frac{33}{34}\frac{(\sigma mv)^{2}}{2} } =
e^{-2.7535(\frac{1}{\sin(\omega(\sigma)/2)})^2 }.
$$

This factor is practically zero   (smaller than $ 10^{-100}$)
when $  \omega \leq 11.8  $ degrees  ($\sigma \geq 1.1592 \times 10^{-9}  $  or
$ \sigma m v \geq 23  $),
 and then, it begins to increase as $\omega$ increases ( $\sigma$
decreases).
In Figure \ref{spreading-angle-figure} we show the error bound in equation  (\ref{ab-anzats-o-w-p.2}) as a function of
the opening angle for small sigma,  $ \sigma \in [ 1.3224 \times 10^{-6} r_1 ,
6.7591  \times 10^{-6} r_1 ]  $, and in the table below we give some representative values.

\begin{tabular}{|l|r|}
   \hline
    \multicolumn{2}{|l|}{Error Bound as a Function of} \\
    \multicolumn{2}{|l|}{the Opening Angle for Small Sigma.} \\
   \hline
     Opening Angle (degrees)    &   \textbf{\em Error Bound}\\
\hline
   51.8407 & $10^{-1}$ \\
 \hline
   47.8885  & $10^{-2}$ \\
   \hline
   44.7231 & $10^{-3}$ \\
   \hline
   42.1135  & $10^{-4}$ \\
   \hline
   39.9137  & $10^{-5}$ \\
   \hline
   38.0265  & $10^{-6}$ \\
   \hline
   36.3842  & $10^{-7}$ \\
   \hline
   34.9380  & $10^{-8}$ \\
   \hline
   33.6517  & $10^{-9}$ \\
   \hline
   32.4979  & $10^{-10}$ \\
   \hline
\end{tabular}

\section{Conclusions}\label{conclusions}
\sss
In Theorems \ref{ab-anzats-ic-w-p},  \ref{ab-anzats-i-w-p},
\ref{ab-anzats-o-w-p}, \ref{ab-anzats-o-w-p-u-t}, \ref{th6.13.viejo} and \ref{th6.13}
we found the time evolution of the electron up to an error bound that we provide
explicitly. The approximate wave function of the electron that we give is  the
one given by the Aharonov-Bohm Ansatz.  It coincides also with the part of the electron wave packet that goes
through the hole of the magnet in Tonomura et al. experiments \cite{to2}.
As we noticed before (see Section
\ref{aharonov-bohm-ansatz-statement}) the Aharonov-Bohm Ansatz  is valid if the
evolution of the exact wave packet is localized at every time in
a simply connected region, with no holes,  (for example in (\ref{a-b-anzats-h.8})).
The main factors that produce the error bounds are the size of the wave packet
(see (\ref{p-i-e-b.1})) and
the opening angle (see (\ref{p-i-e-b.2})). These factors can be understood
also in terms of the part of the wave packet that hits the magnet when
the electron crosses the hole of the magnet (see Remark
\ref{ab-anzats-i-w-p-remark}) and, therefore, they are related with the part of
the electron not localized in a simple connected region (see
(\ref{a-b-anzats-h.8})) at every time.
In Section \ref{physical-interpretation-error-bounds} we analyzed the error
bounds and we have shown that our estimates for the time evolution are valid for a
rather big interval that starts when the opening angle is close to $ 55  $
degrees ( $\sigma \approx  1.3224 \times 10^{-6} r_1$ ) and ends when the size of the wave packet is
close to the inner radius of the magnet (close to $r_1 $).  We have shown also that
the error bounds decrease very fast -exponentially- as the variance gets away
from the extremes of the interval.
 For intermediate sigma ($ \sigma \in [
  6.7591 \times 10^{-6} r_1,  \frac{r_1}{22} ]  $), the time evolution given by
the Aharonov-Bohm Ansatz (\ref{ab-anzats-u-w-e.3})
  differs from the exact one only by a number less than  $ 10^{-99}  $ in norm.
As it is shown in Remark \ref{probabilidad_interaccion} and Section \ref{intermediate_sigma}, for intermediate
sigma, the probability that  the electron wave
packet interacts with the magnet is smaller than $10^{-199}$
 and so,
 there are no fields in the trajectory of the electron. Nevertheless, the
 solution is the one given by the Aharonov-Bohm Ansatz (\ref{ab-anzats-u-w-e.3})
 and it is affected by the vector potential $ A $ by a  phase
 factor  $ e^{i \lambda_{A,0} } $. This phase factor is the one that appears
 in Tonomura et al. experiments \cite{to2}. Although in the experiments of Tonomura et
al. \cite{to2} there is no interaction with the magnetic field, there is an interaction
with the impenetrable magnet. Tonomura et al. \cite{to2} argued that it is not necessary to consider the part of
 the electron wave packet  that hits the magnet -they used a rather big one- because the shadow of the
 magnet was clearly seen in the hologram. Our results show that it
 would be quite interesting to perform an experiment with a medium size electron
 wave packet with an  intermediate sigma. One could use, as  well,  a bigger magnet. Our results show that quantum mechanics predicts in this
 case the interference patterns observed by Tonomura et al. \cite{to2}  with extraordinary
 precision.

In the Aharonov-Bohm Ansatz the electron  is not
accelerated, it propagates following the free evolution, with the
wave function multiplied by a phase. As we
prove that the Aharonov-Bohm Ansatz approximates the exact solution
with an error bound uniform in time that can be smaller that
$10^{-99}$ in norm, we rigorously prove that quantum mechanics
predicts that no force acts on the electron, in agreement with the
experimental results of Caprez et al. \cite{cap}.

Summing up, the experiments of Tonomura et al. \cite{to3,to1,to2} give a strong  evidence of the existence of the interference fringes predicted by
Franz \cite{f} and by Aharonov and Bohm \cite{ab}. The experiment of Caprez et al. \cite{cap} verifies that the interference fringes are not due to a
force acting on the electron, and the results of this paper rigorously prove that quantum mechanics theoretically predicts the observations of these
experiments  in a extremely precise way. This gives a firm experimental and theoretical basis to the existence of the Aharonov-Bohm effect \cite{ab},
namely, that magnetic fields act at a distance on charged particles, even if they are identically zero in the space accessible to the particles, and that
this action at a distance is carried by the circulation of the magnetic potential, what gives magnetic potentials a real physical significance.

\section{Appendix A. Estimates for the Free Evolution of gaussian States}
 \sss
In this appendix we prove estimates for the solutions to the boosted
free Schr\"odinger equation,
 \beq i\frac{\partial }{\partial z}
\varphi(x,z) = H_1 \varphi (x,z), \, \, \,  \varphi (x,0)=\varphi(x), \label{a.1}
\ene

where the boosted free Hamiltonian $ H_{1} $ is defined in (\ref{f.1}).

Recall that under the change of variable $ t:= z/v$, the solutions
of (\ref{a.1})
are solutions of the boosted free Schr\"odinger equation with
Hamiltonian $ e^{-im \v\cdot x} H_0 \, e^{im\v\cdot
  x}$. Classically, a particle that starts at the origin with velocity $\v
= (0,0,v)$, will be located at time $ t$  at the position $
(0,0,z) $. At the high-velocity limit, the quantum evolution follows
the classical one and the parameter $ z  $ can be taken as the
position in the $ z-
$direction of the particle.
We consider the case where the initial state is gaussian,
\beq \varphi(x):= \frac{1}{(\sigma^2 \pi)^{3/4}} \,e^{-x^2/ 2
\sigma^2},
\ene with variance $\sigma$. The solution to (\ref{a.1}) is given
by,

\beq
 e^{-iz H_1}\varphi = e^{-iz m v /2} \,\frac{\sigma^{3/2}}{\pi^{3/4}}
 \frac{1}{(\sigma^2+iz/ mv)^{3/2}} \,
 e^{-(x-z \hv)^2/ 2(\sigma^2+i z/mv )}.
 \label{a.4}
 \ene
we will often use the following simple result.

\begin{remark}\label{rem-a.1}{\it
Suppose that  $C_3\leq C_2 \leq C_1 \leq 0$. Then,
\begin{enumerate}
\item
\beq \int_{C_3}^{C_2} e^{-z^2} dz \leq e^{-C_1^2}
\int_{C_3-C_1}^{C_2-C_1} e^{-z^2} dz \leq e^{-C_1^2} \int_0^\infty
e^{-z^2} dz. \label{a.5} \ene
\item
\beq \int_{C_3}^{C_2} z^2 e^{-z^2} dz \leq -\frac{C_2}{2}
e^{-C_2^2}+ \frac{C_3}{2} e^{-C_3^2}+ \frac{1}{2}
e^{-C_1^2}\int_{C_3-C_1}^{C_2-C_1} e^{-z^2} dz \leq
e^{-C_1^2}(-\frac{C_2}{2}+\frac{1}{2} \int_0^\infty e^{-z^2} dz).
\label{a.6} \ene
\end{enumerate}}
\end{remark}
\noindent{\it Proof:}
$$
\int_{C_3}^{C_2}e^{-z^2} dz \leq   e^{-C_1^2 }\int_{C_3}^{C_2}
e^{-(z^2- C_1^2)}\leq e^{-C_1^2} \int_{C_3}^{C_2} e^{(z-C_1)^2} dz,
$$
where we used that, $ z^2-C_1^2 \geq (z-C_1)^2$. This proves 1.
Furthermore, 2 follows from 1 and the following equation.

$$
\int_{C_3}^{C_2}  \frac{z}{2} \,\, 2  z \, e^{-z^2} d z=
-\frac{z}{2}  e^{-z^2} \big|^{C_2}_{C_3} +\int_{C_3}^{C_2}
\frac{1}{2} e^{-z^2}dz.
$$

.
\begin{lemma}\label{lem-a.2}
Let $f$ be a bounded complex valued function with  support contained
in $D$. Then, for $ z \geq h$ and $ d \geq h - z  $,
\begin{enumerate}
\item
\beq \left\| f(x) \et \varphi \right\|\leq
\frac{\|f\|_\infty}{\sqrt{2}} \ds \,e^{- \theta_{inv}(\sigma,
z)^{2}/2}, \label{a.9} \ene

\item
\beq \left\| f(x + d \hat{\v}) \et \varphi \right\|\leq
\frac{\|f\|_\infty}{\sqrt{2}} \ds \,e^{- \theta_{inv}(\sigma,
z, z+d,h(\sigma))^{2}/2}, \label{a.9.5} \ene

\item
\beq \left\| f(x)\et \varphi\right\| \leq \frac{\|f\|_\infty
}{\pi^{1/4}}
 \, e^{- \theta_{inv}(\sigma,z)^2 /2 } \sqrt{2h} \, r_2\, \rho(z)^{3/2}.
\label{a.10} \ene
\end{enumerate}

\end{lemma}
\noindent {\it Proof:}
We use the function $ \rho(z) $ defined in (\ref{n.8}),
\begin{eqnarray}
\left\| f(x) \et \varphi \right\|^2 \leq
\frac{\|f\|^2_\infty}{\pi^{3/2}} \int_{(D-\hv z)\rho(z)} \, e^{-x^2}
\,dx \leq \frac{\|f\|^2_\infty}{\pi^{1/2}}\, \int_{(-h-
z)\rho(z)}^{(h-z)\rho(z)} \, d\mu \,e^{-\mu^2} \left( 1- \ds
e^{-r^2_2 \rho(z)^{2} }\right) \leq \nonumber\\\nonumber \\
\frac{\|f\|^2_\infty}{\pi^{1/2}}\, {\ds
e^{-\theta_{inv}(\sigma,z)^{2}}} \int_{-2h \rho(z)}^{0}
 \, dz \,e^{-z^2} \left( 1- \ds e^{-r^2_2 \rho(z)^{2} }\right),
\label{a.11}
 \end{eqnarray}
where in the last inequality we used (\ref{a.5}). Equation
(\ref{a.9}) follows from (\ref{a.11}). Equation (\ref{a.9.5})
is obtained similarly. Equation (\ref{a.10})
follows from (\ref{a.11}) and the estimate,
$$
 \int_{-2h \rho(z)}^{0}
 \, dz e^{-z^2} \left( 1- \ds e^{-r^2_2 \rho(z)^{2} }\right) \leq 2h r_2^2 \rho(z)^{3}.
$$

\begin{lemma}\label{lem-a.3}
Let $f$ be a bounded complex valued function with  support contained
in $D$. Then, for $Z \geq h, s \geq 0$, \beq
\begin{array}{l}
\int_{Z}^{\infty}
\left\| f(x) \et \varphi \right\|\leq   \frac{\|f\|_\infty}{\sqrt{2}} {\ds
  e^{-\theta_{inv}(\sigma,Z)^{2}/2}}
(\max(Z,s)-Z)+ \\ \\
\frac{\|f\|_{\infty}}{\pi^{1/4}} {\ds e^{- \theta_{inv}( \sigma,
\max(Z,s))^{2}/2}} \sqrt{2h} r_2 (\sigma m v)^{3/2}
\int_{\max(Z,s)}^\infty \frac{1} {(\sigma^4 m^2 v^2+\zeta^2)^{3/4}}
d\zeta. \label{a.12}
\end{array}
\ene
\end{lemma}

\noindent {\it Proof:} We prove the lemma writing     the integral
in the left hand side of (\ref{a.12}) as follows

$$
\int_{Z}^{\infty}\left\| f(x) \et \varphi \right\|=
\int_{Z}^{\max(Z,s)}\left\| f(x) \et \varphi
\right\|+\int_{\max(Z,s)}^{\infty}\left\| f(x) \et \varphi \right\|
$$
and using (\ref{a.9}) in the first integral, (\ref{a.10}) in the
second, and the fact that $\theta_{inv}(\sigma ,z)^{2}$ is
increasing in $z$ for $ z \geq h $.

\bull

\begin{lemma} \label{lem-a.4}
Let $g: \ere^3 \rightarrow \CE^3$ be  bounded  and with support
contained in $D$ and let $z \geq h$. Then,
\begin{enumerate}

\item
\beq \left\| g(x)\cdot \mo \, \et \varphi \right\| \leq \frac{
\|g\|_\infty }{\pi^{1/4} \sigma} e^{ -\theta_{inv}(\sigma,z)^{2} /2}
\left[ \frac{-\theta_{inv}(\sigma,z)}{2}+ \frac{3\sqrt{\pi}}{4}
\right]^{1/2}. \label{a.15} \ene
\item
\beq \left\| g(x)\cdot \mo \,\et \varphi \right\|\leq  \frac{
\|g\|_\infty }{\pi^{1/4} \sigma} e^{ -\theta_{inv}(\sigma,z)^{2} /2}
\left[4(\sigma m v)^2+2\right]^{1/2} \sqrt{h} \, r_2\,
\rho(z)^{3/2}. \label{a.16} \ene

\end{enumerate}

\end{lemma}
\noindent {\it Proof:}
\beq \left\| g(x)\cdot \mo \,\et \varphi
\right\|^2 \leq \frac{ \|g\|_\infty^2 }{\pi^{3/2} \sigma^2}
\int_{(D-\hv z)\rho(z)} \, x^2 \, e^{-x^2} \,dx \leq  \frac{
\|g\|_\infty^2 }{\pi^{1/2} \sigma^2}
[\Upsilon(\sigma,z)+\Theta(\sigma,z)]\left( 1- e^{- r_2^2
\rho(z)^{2}}\right),
\label{a.17}
\ene
where $\Theta $ and $\Upsilon $ are defined in Section \ref{notation}.
Equation (\ref{a.15}) follows from (\ref{a.17}) applying the last
inequality in (\ref{a.5}) to $\Upsilon(\sigma,z)$ and the last
inequality in (\ref{a.6}) to $ \Theta(\sigma,z)$. Furthermore, using
the middle  inequality in (\ref{a.6}) we obtain that,

\beq
\Theta(\sigma, z) \leq\frac{\rho(z)}{2}[ (z-h)e^{- (z-h)^2
\rho^{2}(z)}- (z+h) e^{- (z+h)^2 \rho^{2}(z)}] +\frac{1}{2}
e^{-\theta_{inv}(\sigma,z)^{2}} \int_{-2h \rho(z)}^0\, e^{-z^2}\, dz.
\label{a.18}
\ene
Note that,
\beq
 e^{- (z-h)^2 \rho(z)^{2}}-  e^{- (z+h)^2 \rho(z)^{2}}\leq e^{-\theta_{inv}(\sigma,z)^{2}} 4 z h \rho(z)^{2},
 \label{a.19}
 \ene
\beq \int_{-2h\rho(z)}^{0}\, e^{-z^2}\, dz \leq  2h \rho(z).
\label{a.20}
\ene
Writing $ z-h= z+h-2h$  in (\ref{a.18}) we obtain that,
\beq
\Theta(\sigma, z) \leq\frac{\rho(z)}{2} (z+h) e^{-\theta_{inv}(\sigma,z)^{2}} 4 z h \rho(z)^{2}
\leq  e^{-\theta_{inv}(\sigma,z)^{2}}   4h \rho(z) (\sigma mv)^2.
\label{a.21}
\ene
Moreover, applying the middle inequality in (\ref{a.5}) to
$\Upsilon(\sigma,z)$ we prove that,
\beq \Upsilon(\sigma,z)\leq
e^{-\theta_{inv}(\sigma, z)^{2}}\, 2h\, \rho(z).
\label{a.22}
\ene
Equation (\ref{a.16}) follows from (\ref{a.17}, \ref{a.21},
\ref{a.22}).

\begin{lemma}\label{lemm-a.5}
Let $g: \ere^3 \rightarrow \CE^3$ be  bounded  and with support
contained in $D$. Then, for any  $Z \geq h$ with
$\theta_{inv}(\sigma,Z)\geq 1$, $ s \geq 0$,

\beq
\begin{array} {c}
\int_Z^\infty \left\| g(x)\cdot \mo \, \et \varphi \right\| \leq
\frac{ \|g\|_\infty }{\pi^{1/4} \sigma} e^{
-\theta_{inv}(\sigma,Z)^{2} /2} \left(\frac{|\theta_{inv}(\sigma,
Z)|^{1/2}}{\sqrt{2}}+ \frac{\sqrt{3}\,\,\pi^{1/4}}{2}\right)
(\max(Z,s)-Z)+
\\\\
\frac{ \|g\|_\infty }{\pi^{1/4} \sigma}
e^{-\theta_{inv}(\sigma,\max(Z,s))^{2}/2}\left(2 \sigma m v + \sqrt{2}\right)  \sqrt{h}\,\, r_2
(\sigma m v)^{3/2} \int_{\max(Z,s)}^\infty (\sigma^4 m^2 v^2+
\zeta^2)^{-3/4}.
\end{array}
\label{a.23} \ene
\end{lemma}

\noindent {\it Proof:} We split  the integral in the left hand side
of (\ref{a.23}) as follows

$$
\int_{Z}^{\infty}\left\| g(x) \cdot \mo \,\et \varphi \right\|=
\int_{Z}^{\max(Z,s)}\left\| g(x)\cdot \mo \,\et \varphi \right\|
+\int_{\max(Z,s)}^{\infty}\left\| g(x)\cdot\mo \,\et \varphi \right\|
$$
and using (\ref{a.15}) in the first integral, (\ref{a.16}) in the
second, and the fact that the functions $e^{-x/2}\,\sqrt{x},
e^{-x/2}\,  x^{1/4}$ are decreasing for $ x \geq 1$ (notice also that  $\theta_{inv}(\sigma ,z)^{2}$ is
increasing in $z$ for $ z \geq h $).

\begin{remark}\label{rem-a.6}{\it
Suppose that $ z, \, \zeta \in \ere^+$, $  s  $ and $ b  $ are real numbers such that  $
z \geq \zeta, \, s \geq z-2\zeta, \,  b >0$.
Then,

\begin{enumerate}

\item
In any interval $I:= [\sigma_1, \sigma_2]$ such that $\forall \sigma
\in I, - \theta_{inv}(\sigma,z,s,\zeta)\geq \sqrt{1/2}$,

\beq \Upsilon(\sigma,z,s, \zeta) e^{-b \rho(\sigma,z)^{2}}\leq
\max[\Upsilon(\sigma_1,z,s,\zeta) e^{-b \rho(\sigma_1,z)^{2}},
\Upsilon(\sigma_2,z,s,\zeta) e^{-b \rho(\sigma_2,z)^{2}}].
\label{a.24} \ene
\item
In any interval $I:= [\sigma_1, \sigma_2]$ such that $\forall \sigma
\in I, -\theta_{inv}(\sigma,z,s,\zeta) \geq \sqrt{3/2}$,

\beq \Theta(\sigma,z,s,\zeta)e^{-b \rho(\sigma,z)^{2}}\leq
\max[\Theta(\sigma_1,z,s,\zeta)e^{- b \rho(\sigma_1,z)^{2}},
\Theta(\sigma_2,z,s,\zeta)e^{-b \rho(\sigma_2,z)^{2}}].
\label{a.25}
\ene

\end{enumerate}}
\end{remark}
\noindent {\it Proof:} We give the proof of 1. The proof of 2 is
similar. We have that

\beq
\begin{array}{c}
\frac{\partial}{\partial \sigma} \Upsilon(\sigma,z,s,\zeta)\,e^{-b
\rho(\sigma,z)^{2}} = \frac{1}{\sigma}\frac{m^2 v^2}{(\sigma^4m^2v^2+z^2) }
\left(\left(\frac{z}{mv}\right)^2- \sigma^4\right)
 e^{-b \rho(\sigma, z)^{2}} \left[ e^{-(\zeta+s)
^2\rho(\sigma,z)^{2}}  (\zeta+s) \rho(\sigma, z)\,  -\right. \\\\\left.
e^{-(z-\zeta)^2 \rho(\sigma,z)^{2}} (z-\zeta) \rho(\sigma, z)\, \, - 2 b\,
\rho(\sigma,z)^2\,  \Upsilon(\sigma,z,s, \zeta) \right].
\end{array}
\label{a.26}
\ene

As the function $e^{-x^2}x$ is decreasing for $ x \geq 1/ \sqrt{2}$,
the term in the square brackets is  ({\ref{a.26}) is negative. Then,
the left-hand side of (\ref{a.26}) is different from zero for
$\sigma \in I$ if $\sqrt{z/mv}\notin I$ and otherwise, it is
negative for $\sigma <\sqrt{z/mv}$ and it is positive for $\sigma
>\sqrt{z/mv}$. This proves 1.

\bull

Remember that
 $z_{\tilde{\omega},\sigma}(h)$ is defined in Section \ref{notation}.  It is
 given by,

\beq
z_{\tilde{\omega},\sigma}(h)= \frac{h(\sigma m v)^2}{(\sigma m v)^2- \tilde{\omega}^{-2}} +
\frac{\sigma m v}{((\sigma m v)^2- \tilde{\omega}^{-2})^{1/2}} \left( \tilde{\omega}^{-2}
\sigma^2+ h^2 \left(\frac{(\sigma m v)^2}{(\sigma m v)^2-
\tilde{\omega}^{-2}}-1\right) \right)^{1/2}.
\label{a.28}
\ene

\begin{remark}\label{rem-a.7}
{\it Suppose that $\sigma_2 \leq \sigma \leq \sigma_1$. Then,
\beq
 z_{\tilde{\omega},\sigma}(\zeta)\leq \max (z_{\tilde{\omega},\sigma_1}(\zeta),z_{\tilde{\omega},\sigma_2}(\zeta)).
 \label{a.29}
 \ene}
\end{remark}
\noindent{\it Proof:}
Note that as a function of $\sigma$,  $ \rho(\sigma,z) $ is
increasing  for $ \sigma \leq \sqrt{z/mv}$ and that it is decreasing
for $\sigma > \sqrt{z/mv}$. Suppose that $ z_{\tilde{\omega},\sigma_1}(\zeta)
\leq z_{\tilde{\omega},\sigma}(\zeta)$. Then, $ \sigma \leq \sqrt{z/mv}$, because
if $\sigma > \sqrt{z/mv}$,

$$
\tilde{\omega}^{-1}= -\theta_{inv}(\sigma_{1}, z_{\tilde{\omega},\sigma_1}(\zeta),  z_{\tilde{\omega},\sigma_1}(\zeta)  , \zeta) <-
\theta_{inv}(\sigma, z_{\tilde{\omega},\sigma_1}(\zeta),  z_{\tilde{\omega},\sigma_1}(\zeta)  , \zeta),
$$
since, $- \theta_{inv}(\sigma,z,z,\zeta)= (z-\zeta)\rho(\sigma,z)$   and as $ -\theta_{inv}$ is increasing in $z \geq 0$, this implies that
$z_{\tilde{\omega},\sigma}(\zeta) < z_{\tilde{\omega},\sigma_1}(\zeta)$. Then,
 $\sigma_2 < \sigma  \leq \sqrt{z/mv}$, and it follows that,

 $$
 \tilde{\omega}^{-1} =- \theta_{inv} (\sigma, z_{\tilde{\omega},\sigma}(\zeta), z_{\tilde{\omega},\sigma}(\zeta) , \zeta) \geq -
 \theta_{inv}( \sigma_2 ,z_{\tilde{\omega},\sigma}(\zeta), z_{\tilde{\omega},\sigma}(\zeta) , \zeta).
 $$
 But as also,

 $$
 \tilde{\omega}^{-1} = - \theta_{inv}(\sigma_2, z_{\tilde{\omega},\sigma_{2}}(\zeta), z_{\tilde{\omega},\sigma_2}( \zeta), \zeta),
 $$
and $-\theta_{inv}$ is increasing in $z \geq 0$, we have that $z_{\tilde{\omega},\sigma}(\zeta)
\leq z_{\tilde{\omega},\sigma_2}(\zeta)$.

\begin{lemma}\label{lemm-a.8}
Let $ \mu_{i}, \, i \in \{ 1, 2, 3  \} $  belong to $ \ere_{+} $.
Suppose that the following conditions are satisfied,
\begin{enumerate}
\item
Either $ \mu_{i} \leq \sigma_{0} , \,  i \in \{ 1, 2,
3  \}    $,  or $ \mu_{i} \geq \sigma_{0} , \,  i \in \{ 1, 2, 3 \}$.
\item
  $ \mu_{i} \leq \mu_{3} , \,  i \in \{ 1, 2  \}    $, or $
\mu_{i} \geq \mu_{3} , \,  i \in \{ 1, 2  \}    $.
\end{enumerate}
 We define $
\mu_{\max} := \max(\mu_{1}, \mu_{2})  $,  $ \mu_{\min} :=
\min(\mu_{1}, \mu_{2})  $ and take $ \nu = \mu_{\max}  $, if $
\mu_{i} \geq \mu_{3}, i \in \{1,2\} $  and  $ \nu = \mu_{\min}  $, if $ \mu_{i}
\leq \mu_{3}, i \in \{1,2\} $. We denote by $ Z := z(\mu_{\max}) $, if $ \mu_{i}
\leq \sigma_{0} $; and $ Z := \max_{i \in \{ 1 ,2
\}}\{z_{\tilde{\omega}(\mu_{\max}),
  \mu_{i}}(h(\mu_{\max})) \}  $, if $ \mu_{i} \geq \sigma_{0}  $. We suppose that $   Z
\geq z_{\sqrt{2},\nu, \mu_{3}  }(h(\mu_{\max}))  $.
Let $ f : \ere^{3} \times \ere_{+} \times \ere \to \mathbb{C} $  be a
complex valued function and we take  $f_{\sigma, z}(x) : = f(x, \sigma,
z) $. Suppose the support  of  $f_{\sigma, z} $  is  contained in
$K-[0, (z(\sigma) - \zeta -z)]\hv$ for some $ \zeta \in \ere $, every $ \sigma \in \ere_{+} $ and every $ z \in \ere   $
with $ z + \zeta \leq z(\sigma)  $. Then, for every
gaussian wave function $ \varphi  $  with variance $ \sigma \in
[\mu_{\min}, \mu_{\max}]  $,
\beq \label{a.30prima}
\int_0^{z(\sigma)-\zeta} \left\| f_{\sigma, z}(x)\et e^{-i \zeta H_{1}}
\varphi \right\| \leq \frac{\|f\|_\infty}{\pi^{1/4}} I_{ps}(\mu_{1},
\mu_{2}, \mu_{3}, \zeta), \ene

where, \beq
\begin{array}{l}
 I_{ps}(\mu_{1}, \mu_{2}, \mu_{3}, \zeta) :=
  \pi^{1/4}\, z_{\sqrt{2},\nu,\mu_3}(h(\mu_{\max}))
\max_{\mu_{i} \in \{ \mu_1 , \mu_2 \}}e^{-\frac{r_1^2}{2} \rho(\mu_{i},
  z_{\sqrt{2},\nu,\mu_3}(h(\mu_{\max})))^2} +
  \pi^{1/4}\, \max\{- \zeta , 0  \} \\\\ \max_{\mu_{i} \in \{ \mu_1 , \mu_2 \}}e^{-\frac{r_1^2}{2} \rho(\mu_{i},
   \zeta)^2} +
\sum_{\mu_{i} \in \{ \nu, \mu_{3}  \}}
\int_{z_{\sqrt{2},\nu,\mu_3}(h(\mu_{\max}))}^Z \,
\Upsilon(\mu_i,\tau,Z, h(\mu_{\max}))^{1/2} e^{-\frac{r_1^2}{2}
\rho(\mu_{i}, \tau)^2} d\tau.
\end{array}
\label{a.31}
\ene
\end{lemma}

\noindent {\it Proof:}
It follows from equation (\ref{a.28}) that $  z(\sigma) \leq
z_{\tilde{\omega}(\sigma),\sigma }(h(\mu_{\max}))   $. If $ \mu_{i} \geq \sigma_{0}  $
 then  $ \tilde{\omega}(\sigma)= \tilde{\omega}(\mu_{\max})  $  for $ \sigma \in   [\mu_{\min}, \mu_{\max}  ]  $. It follows
from Remark \ref{rem-a.7} that $ z(\sigma) \leq   \max_{i \in \{ 1 ,2
\}}\{z_{\tilde{\omega}(\mu_{\max}),
  \mu_{i}}(h(\mu_{\max})) \} = Z  $. If   $ \mu_{i} \leq \sigma_{0}  $ then from
formula (\ref{a.28}) and the definition of $ \tilde{\omega}(\sigma)  $  we have that
$ z_{\tilde{\omega}(\sigma),\sigma }(h(\mu_{\max})) \leq  z_{\tilde{\omega}(\mu_{\max}),\mu_{\max}
  }(h(\mu_{\max})) = z(\mu_{\max})  $. We conclude that
\beq
 \label{a.31biprima}
z(\sigma) \leq Z,
\ene
and then,
\beq\label{a.31triprima}
\int_0^{z(\sigma)-\zeta}\, dz\, \left\| f_{\sigma, z}(x)\et e^{-i \zeta H_{1}}
\varphi \right\|  \leq \int_0^{Z -\zeta} \, dz\, \left\| f_{\sigma, z}(x)\et e^{-i \zeta H_{1}}
\varphi \right\|.
\ene

\noindent As in (\ref{a.11}) we prove that

\beq \left\| f_{\sigma,z}(x)\et
e^{-i\zeta H_{1}} \varphi \right\|^2 \leq
\frac{\|f\|_\infty^2}{\pi^{1/2}}\, \Upsilon(\sigma,z + \zeta,Z,
h(\mu_{\max})) \, e^{-r^2_1 \rho(\sigma,z + \zeta)^{2}}. \label{a.32}
\ene Then, \beq
\begin{array}{l}
\int_0^{Z-\zeta} \left\| f_{\sigma,z}(x)\et e^{-i \zeta H_{1}}
\varphi \right\| dz  \leq
\frac{\|f\|_\infty}{\pi^{1/4}} \Big{[}
 \max(
\int_\zeta^0  \Upsilon(\sigma,z ,Z,
h(\mu_{\max}))^{1/2}e^{-\frac{r^2_1}{2} \rho(\sigma,z )^{2}}, 0) +
\\\\
\int_0^{z_{\sqrt{2},\nu,\mu_3}(h(\mu_{\max}))}
 \Upsilon(\sigma,z,Z,
h(\mu_{\max}))^{1/2} \, e^{-\frac{r^2_1}{2} \rho(\sigma,z)^{2}}
+\int_{ z_{\sqrt{2},\nu,\mu_3}(h(\mu_{\max}))}^{Z}
\Upsilon(\sigma,z ,Z,
h(\mu_{\max}))^{1/2} \, e^{-\frac{r^2_1}{2} \rho(\sigma,z)^{2}} \Big{]}  \leq
\\\\
  \frac{\|f \|_\infty} {\pi^{1/4}} \left[     \pi^{1/4}\, \max(-\zeta, 0)
e^{-\frac{r_1^2}{2} \rho(\sigma,\zeta)^{2}}     +
\pi^{1/4}\, z_{\sqrt{2},\nu,\mu_3}(h(\mu_{\max}))
e^{-\frac{r_1^2}{2} \rho(\sigma,z_{\sqrt{2},\nu,\mu_3}(h(\mu_{\max})))^{2}}
+\right.
\\ \\ \left. \int_{ z_{\sqrt{2},\nu,\mu_3}(h(\mu_{\max}))}^{Z}
\Upsilon(\sigma,z ,Z,
h(\mu_{\max}))^{1/2} \, e^{-\frac{r^2_1}{2} \rho(\sigma,z)^{2}}   \right]  ,
\end{array}
\label{a.33} \ene where we used that, $\Upsilon \leq \sqrt{\pi}$.
If $ z \geq   z_{\sqrt{2},\nu,\mu_3}(h(\mu_{\max})) $, it
follows from Remark \ref{rem-a.7} that $ z \geq z_{
\sqrt{2},\sigma}(h(\mu_{\max})) $  for every  $ \sigma $ belonging to the
interval limited by $\nu $ and $ \mu_3 $. We
complete the proof of the lemma using (\ref{a.24}) in the integral
in the right-hand side of (\ref{a.33}), and for the other two terms we argue as
in the proof of (\ref{a.24}).

\bull

Using the proof of the preceding lemma, we  prove the following,

\begin{lemma}\label{lemm-a.8.5}
Suppose that the hypothesis of the Lemma  \ref{lemm-a.8} are fulfilled and
furthermore, assume that the support  of  $f_{\sigma, z} $  is  contained in
$K$ for every $ \sigma \in \ere_{+} $ and every $ z \in \ere   $. Then, for every
$ \zeta \in \ere $ with $ |\zeta| \leq z(\sigma)$,
\beq \label{a.33prima}
\int_0^{z(\sigma)-\zeta} \left\| f_{\sigma, z}(x)\et e^{-i \zeta H_{1}}
\varphi \right\| \leq \frac{\|f\|_\infty}{\pi^{1/4}} 2 I_{pp}(\mu_{1},
\mu_{2}, \mu_{3}), \ene
where,

\beq \label{a.33prima.1.menos1}
 I_{pp}(\mu_{1},
\mu_{2}, \mu_{3}):= I_{ps}(\mu_{1},
\mu_{2}, \mu_{3}, 0).
\ene
and

\beq \label{a.33prima.1}
\int_0^{z(\sigma)} \left\| f_{\sigma, z}(x)\et
\varphi \right\| \leq \frac{\|f\|_\infty}{\pi^{1/4}} I_{pp}(\mu_{1},
\mu_{2}, \mu_{3}). \ene

\end{lemma}

\begin{lemma} \label{lemm-a.9}
Let $ \mu_{i}, \mu_{max}, \mu_{min}, \nu  $ and $ Z $ be as in Lemma
\ref{lemm-a.8}. We suppose furthermore that $   Z
\geqq z_{\sqrt{\frac{2}{3}},\nu, \mu_{3}  }(h(\mu_{max}))  $ and
 $ r_{1}
\rho(\mu_i,  z_{\sqrt{2},\nu, \mu_{3}  }(h(\mu_{max})))\geq 1 $ for $ i \in \{ 1, 2  \} $.
  Let $ g : \ere^{3} \times \ere_{+} \times
\ere \to \mathbb{C}^{3} $  be a complex vector valued function and
we take $g_{\sigma, z}(x) : = g(x, \sigma, z) $. Suppose that the
support of $g_{\sigma, z}  $  is contained in $K-[0, z(\sigma)- \zeta - z]\hv$ for some
$ \zeta \in \ere $,
all $ \sigma \in \ere_{+}  $ and for all $  z  $ with $ z + \zeta \leq z(\sigma)   $. Then, for every gaussian wave function $
\varphi  $ with variance $ \sigma \in [\mu_{min}, \mu_{max}] $,

\beq \int_0^{z(\sigma)-\zeta} \left\| g_{\sigma, z}(x)\cdot \mo \, \et
e^{-i\zeta H_{1}} \varphi \right\|\, dz \leq
\frac{\|g\|_\infty}{\pi^{1/4}\,\sigma} I_{ss}(\mu_{1},\mu_{2},
\mu_{3}, \zeta) \label{a.33biprima}
\ene
where,
\beq
\begin{array}{l}
 I_{ss}(\mu_{1}, \mu_{2}, \mu_{3}, \zeta) :=
  \frac{\pi^{1/4}}{\sqrt{2}}\, z_{\sqrt{\frac{2}{3}},\nu,\mu_3}(h(\mu_{\max}))
\max_{\mu_i \in \{ \mu_1 , \mu_2 \}}(e^{-\frac{r_1^2}{2} \rho(\mu_{i},
  z_{\sqrt{\frac{2}{3}},\nu,\mu_3}(h(\mu_{\max})))^2}) + \\\\
  \frac{\pi^{1/4}}{\sqrt{2}}\, \max\{- \zeta , 0  \}
  \max_{\mu_i \in \{ \mu_1 , \mu_2
\}}(e^{-\frac{r_1^2}{2} \rho(\mu_{i},
   \zeta)^2}) + \\\\
\pi^{1/4}\, z_{\sqrt{2},\nu,\mu_3}(h(\mu_{\max}))
   \max_{\mu_i \in
\{ \mu_1 , \mu_2 \}} ( r_1\rho(\mu_{i},
  z_{\sqrt{2},\nu,\mu_3}(h(\mu_{\max})))   e^{-\frac{r_1^2}{2}
\rho(\mu_{i},
  z_{\sqrt{2},\nu,\mu_3}(h(\mu_{\max})))^2})     + \\\\

\pi^{1/4}\, \max(-\zeta, 0) \left\{ \begin{array}{cc}
 \max_{\mu_i \in
\{ \mu_1 , \mu_2 \}} ( r_1\rho(\mu_{i},
  \zeta)   e^{-\frac{r_1^2}{2}
\rho(\mu_{i},
  \zeta)^2}), \, \,  if  \, \, |\zeta | \leq r_{\mu_1, \mu_2} \\\\
e^{-1/2}, \, \, if  \, \,  |\zeta | > r_{\mu_1, \mu_2}  \end{array} \right\} + \\\\
\pi^{1/4}\, z_{\sqrt{2},\nu,\mu_3}(h(\mu_{\max}))
 \max_{\mu_i \in
\{ \mu_1 , \mu_2 \}} (    e^{-\frac{r_1^2}{2} \rho(\mu_{i},
  z_{\sqrt{2},\nu,\mu_3}(h(\mu_{\max})))^2}) + \\\\

\pi^{1/4}\, \max(-\zeta, 0)
 \max_{\mu_i \in
\{ \mu_1 , \mu_2 \}} (   e^{-\frac{r_1^2}{2} \rho(\mu_{i},
  \zeta)^2}) + \\\\

 \sum_{\mu_i \in \{ \nu, \mu_3 \}}
\int_{z_{\sqrt{\frac{2}{3}},\nu,\mu_3}(h(\mu_{\max}))}^Z \,
\Theta(\mu_i,\tau,Z, h(\mu_{\max}))^{1/2} e^{-\frac{r_1^2}{2}
\rho(\mu_{i}, \tau)^2} d\tau + \\\\

 \sum_{\mu_i \in \{ \nu, \mu_3 \}} \max(
\int_{z_{\sqrt{2},\nu,\mu_3}(h(\mu_{\max}))}^{\min (r_{\nu,\mu_3},Z)} \,
\Upsilon(\mu_i,\tau,Z, h(\mu_{\max}))^{1/2}  r_1 \rho(\mu_{i}, \tau)
e^{-\frac{r_1^2}{2}
\rho(\mu_{i}, \tau)^2} d\tau, 0) +\\\\

 \sum_{\mu_i \in \{ \nu, \mu_3 \}}
\int_{\min (r_{\nu,\mu_3}, Z)}^{Z} \, \Upsilon(\mu_i,\tau,Z,
h(\mu_{\max}))^{1/2} e^{-1/2} d\tau +\\\\
\sum_{\mu_i \in \{ \nu, \mu_3 \}}
\int_{z_{\sqrt{2},\nu,\mu_3}(h(\mu_{\max}))}^Z \,
\Upsilon(\mu_i,\tau,Z, h(\mu_{\max}))^{1/2} e^{-\frac{r_1^2}{2}
\rho(\mu_{i}, \tau)^2}] d\tau.

\end{array}
\label{a.31.1.1.1}
\ene
\end{lemma}
\noindent{\it Proof:} By (\ref{a.31biprima}),
\beq
\int_0^{z(\sigma)-\zeta} \left\| g_{\sigma, z}(x) \cdot \mo \et e^{-i \zeta H_{1}}
\varphi \right\|  \leq \int_0^{Z -\zeta} \left\| g_{\sigma, z}(x) \cdot \mo \et e^{-i \zeta H_{1}}
\varphi \right\|.
\ene

Estimating as in the proof of (\ref{a.17}) we
prove that,
\beq
\begin{array}{l}
\left\| g_{\sigma,z}(x)\cdot \mo \et e^{-i\zeta
H_1} \varphi \right\|^2 \leq
 \frac{ \|g\|_\infty^2 }{\pi^{1/2}
\sigma^2}\left[ \Theta(\sigma, z +
 \zeta,Z, h(\mu_{max}))\,+  \right. \\\\ \left.
\Upsilon(\sigma, z + \zeta,Z, h(\mu_{\max}))(1+ r^2_1  \rho(\sigma, z
+ \zeta)^{2})\right] e^{- r_1^2 \rho(\sigma, z + \zeta)^{2}}.
\end{array}
\label{a.36}
 \ene

We have that,
\beq \int_0^{Z-\zeta} \left\| g_{\sigma,z}(x)\cdot \mo
\, \et e^{-i\zeta H_{1}} \varphi   \right\|\, d\tau \leq
\frac{\|g\|_\infty}{\pi^{1/4}\,\sigma} \sum_{j=1}^7 I_j,
\label{a.37}
\ene
where,

 \beq
\begin{array}{l}
I_1:=  max(\int_{\zeta}^0\, \left( \Theta(\sigma, \tau,Z, h(\mu_{\max}))\,e^{-
r_1^2 \rho(\sigma, \tau )^2}\right)^{1/2}\, d\tau, 0) +\\\\
\int_0^{z_{\sqrt{\frac{2}{3}},  \nu, \mu_3}(h(\mu_{\max}))}\, \left( \Theta(\sigma,
\tau,Z, h(\mu_{\max}))\,e^{- r_1^2 \rho(\sigma, \tau)^2}\right)^{1/2}\,
d\tau,
\end{array}
\label{a.38}
\ene

 \beq
I_2:=
\int_{z_{\sqrt{\frac{2}{3}}, \nu, \mu_3}(h(\mu_{\max}))}^Z \, \left( \Theta(\sigma,
\tau,Z, h(\mu_{\max}))\,e^{- r_1^2 \rho(\sigma,
\tau)^{2}}\right)^{1/2}\, d\tau,
\label{a.39}
\ene

\beq
\begin{array}{l} I_3:= max(\int_{\zeta}^0\, \left( \Upsilon(\sigma, \tau,Z,
h(\mu_{\max}))\,  r_1^2 \rho(\sigma, \tau)^2 e^{- r_1^2 \rho(\sigma,
\tau)^2}\right)^{1/2}\, d\tau, 0)   \\\\  + \int_0^{z_{\sqrt{2},
\nu, \mu_3}(h(\mu_{\max}))}\, \left( \Upsilon(\sigma, \tau,Z, h(\mu_{\max}))\, r_1^2
\rho(\sigma, \tau)^2  e^{- r_1^2 \rho(\sigma, \tau)^2}\right)^{1/2}\,
d\tau,
\end{array}
\label{a.40}
\ene

\beq
I_4:=\int_{z_{\sqrt{2}, \nu, \mu_3}(h(\mu_{\max}))}^{\max(z_{\sqrt{2}, \nu,
\mu_3}(h(\mu_{\max})), \min(r_{\nu,\mu_3},Z))}\, \left( \Upsilon(\sigma, \tau,Z,
h(\mu_{\max}))\, r_1^2 \rho(\sigma, \tau)^2  e^{- r_1^2 \rho(\sigma,
\tau)^2}\right)^{1/2}\, d\tau,
\label{a.41}
\ene

\beq
I_5:=\int^{Z}_{\max(z_{\sqrt{2}, \nu, \mu_3}(h(\mu_{\max})),
\min(r_{\nu,\mu_3},Z))}\, \left( \Upsilon(\sigma, \tau,Z,
h(\mu_{\max}))\, r_1^2 \rho(\sigma, \tau)^2  e^{- r_1^2 \rho(\sigma,
\tau)^2}\right)^{1/2}\, d\tau,
\label{a.42}
\ene

\beq
\begin{array}{l}
I_6:=   \max( \int_{\zeta}^{0}\, \left( \Upsilon(\sigma,
\tau,Z, h(\mu_{\max}))\,e^{- r_1^2 \rho(\sigma,
\tau)^{2}}\right)^{1/2} \, d\tau , 0)   + \\\\ \int_0^{z_{\sqrt{2},
\nu,\mu_{3}}(h(\mu_{\max}))}\, \left( \Upsilon(\sigma, \tau,Z, h(\mu_{\max}))\,e^{-
r_1^2 \rho(\sigma, \tau)^{2}}\right)^{1/2} \, d\tau,
\end{array}
\label{a.43}
\ene

\beq
I_7:=\int_{z_{\sqrt{2}, \nu, \mu_{3}}(h(\mu_{\max}))}^{Z}\, \left(
\Upsilon(\sigma, \tau,Z, h(\mu_{\max}))\,e^{- r_1^2 \rho(\sigma,
\tau)^{2}}\right)^{1/2} \, d\tau.
\label{a.44}
\ene

Since $\Upsilon \leq \sqrt{\pi}$ and $\Theta \leq \sqrt{\pi}/2$ we
have that,

\beq
\begin{array}{l}
 I_1+ I_6 \leq \pi^{1/4}\left( \frac{1}{\sqrt{2}}\,
\max(-\zeta,0)\,e^{-\frac{r_1^2}{2}
 \rho(\sigma, \zeta )^{2}}
+
 \, \frac{1}{\sqrt{2}}\,
z_{\sqrt{\frac{2}{3}}, \nu, \mu_{3}}(h(\mu_{\max}))\,e^{-\frac{r_1^2}{2}
 \rho(\sigma, z_{\sqrt{\frac{2}{3}}, \nu, \mu_{3}}(h(\mu_{\max})) )^{2}} \right) \\ +

\pi^{1/4}\left( \max(-\zeta, 0)\, e^{- \frac{r_1^2}{2}
\rho(\sigma,\zeta )^{2}} + z_{\sqrt{2}, \nu, \mu_3}(h(\mu_{\max}))\, e^{-
\frac{r_1^2}{2} \rho(\sigma,z_{\sqrt{2}, \nu, \mu_3}(h(\mu_{\max})) )^{2}}  \,
\right).

\end{array}\label{a.45}
\ene

By Remark \ref{rem-a.6}
\beq I_2\leq \sum_{\mu_i \in \{ \nu, \mu_3  \}}
\int_{z_{\sqrt{\frac{2}{3}}, \nu, \mu_3}(h(\mu_{\max}))}^Z \, \left( \Theta(\mu_i,
\tau,Z, h(\mu_{\max}))\,e^{- r_1^2 \rho(\mu_i,
\tau)^2}\right)^{1/2}\, d\tau, \label{a.46} \ene

\beq I_7\leq \sum_{\mu_i \in \{ \nu, \mu_3  \}} \int_{z_{\sqrt{2}, \nu, \mu_{3}}(h(\mu_{\max}))}^{Z}\,
\left( \Upsilon(\mu_i, \tau,Z, h(\mu_{\max}))\,e^{- r_1^2 \rho(\mu_i,
\tau)^{2}}\right)^{1/2} \, d\tau. \label{a.47} \ene

Moreover, since $xe^{-x^2/2}$ is increasing for $ 0 \leq x < 1$ and
decreasing for $ x \geq 1$,

\beq \begin{array}{l} I_3 \leq \pi^{1/4}\,\max(-\zeta, 0) \left\{ \begin{array}{cc}
  r_1\rho(\sigma,
  \zeta)   e^{-\frac{r_1^2}{2}
\rho(\sigma,
  \zeta)^2}, \, \,  if  \, \, |\zeta | \leq r_{\mu_1, \mu_2} \\
e^{-1/2}, \, \, if  \, \,  |\zeta | > r_{\mu_1, \mu_2}  \end{array} \right\} \\ +

\pi^{1/4}\,z_{\sqrt{2}, \nu, \mu_3}(h(\mu_{\max}))\,r_1\, \rho(\sigma,z_{\sqrt{2},
\nu, \mu_3})\, e^{-\frac{r_1^2}{2} \rho(\sigma, z_{\sqrt{2}, \nu,\mu_3}(h(\mu_{\max})))^2}.

\end{array}
\label{a.48} \ene

By Remark\,  \ref{rem-a.6}, if $\tau \geq z_{\sqrt{2}, \nu, \mu_3}(h(\mu_{\max}))$,

\beq
\Upsilon(\sigma,\tau,Z, h(\mu_{\max}) )\leq \max_{\mu_{i} \in \{
\nu, \mu_3 \}}\Upsilon(\mu_i,\tau,Z, h(\mu_{\max}))  . \label{a.49}
\ene

By (\ref{a.49}) and as $x\,e^{-x}$ takes its maximum at $x=1$,

\beq I_5 \leq \sum_{\mu_{i} \in \{ \nu, \mu_3
\}}\int_{\max(z_{\sqrt{2}, \nu, \mu_3 }(h(\mu_{\max})), \min(r_{\nu,\mu_3},Z))}^Z\,
\left(\Upsilon(\mu_i, \tau,Z, h(\mu_{\max}))\,e^{-1}\right)^{1/2}
d\tau. \label{a.50} \ene

Note that if $ r_1^2 \rho(\mu_i, \tau)^2\geq 1, \mu_{i} \in \{ \nu,
\mu_{3} \}$ then, $ r_1^2 \rho(\sigma, \tau)^2\geq 1, \forall \sigma
$ between $ \nu $  and $ \mu_3  $ . Hence, as in the proof of
Remark \ref{rem-a.6} we prove that,

\beq
\begin{array}{l}r_1^2 \rho(\sigma, \tau)^2 \Upsilon(\sigma,
\tau,Z, h(\mu_{\max}))\,e^{- r_1^2 \rho(\sigma, \tau)^2} \chi_{
\cap_{\mu_i \in \{ \nu, \mu_3  \}} \{r_1^2 \rho(\mu_i, \tau)^2 \geq
1\}}(\tau) \\\\
 \leq \max_{\mu_i \in \{  \nu, \mu_3  \}}\,r_1^2 \rho(\mu_i, \tau)^2
\Upsilon(\mu_i, \tau, Z, h(\mu_{\max}))\,e^{- r_1^2 \rho(\mu_i,
\tau)^2} \chi_{\cap_{\mu_{i} \in \{ \nu, \mu_3 \}}\, \{r_1^2
\rho^{2}(\mu_i, \tau)\geq 1\}}(\tau),
\end{array} \label{a.51} \ene

and then,

\beq I_4 \leq \sum_{\mu_{i}\in  \{ \nu, \mu_3  \}}
\int_{z_{\sqrt{2}, \nu, \mu_3}(h(\mu_{\max}))}^{\max( z_{\sqrt{2}, \nu, \mu_3}(h(\mu_{\max})) ,
\min(r_{\nu,\mu_3},Z))}\, \left(r_1^2 \rho(\mu_i, \tau)^2\,
\Upsilon(\mu_{i}, \tau,Z, h(\mu_{\max}))\,e^{- r_1^2 \rho(\mu_i,
\tau)^2}\right)^{1/2}\, d\tau. \label{a.52} \ene

To obtain equation (\ref{a.33biprima}), we use (\ref{a.37},
\ref{a.45}--\ref{a.48}, \ref{a.50}, \ref{a.52}) and we argue as in
the proofs of   Remark \ref{rem-a.6} to estimate equations (\ref{a.45}) and (\ref{a.48}).

\bull

Using the proof  of the preceding lemma we prove the following,

\begin{lemma}\label{lemm-a.9.5}
Suppose that the hypothesis of the Lemma \ref{lemm-a.9} are
fulfilled, assume furthermore, that the support of $g_{\sigma, z} $
is contained in $K$,  for all $ \sigma \in \ere_{+}  $ and for all $
z   $. Then, for every $ \zeta \ \in \ere $ with $ |\zeta | \leq z(\sigma)  $ and every gaussian wave
function $ \varphi  $ with variance $ \sigma \in [\mu_{min},
\mu_{max}] $,
\beq
\int_0^{z(\sigma)-\zeta} \left\| g_{\sigma, z}(x)\cdot \mo \, \et
e^{-i\zeta H_{1}} \varphi \right\|\, dz \leq
\frac{\|g\|_\infty}{\pi^{1/4}\,\sigma}2 I_{sp}(\mu_{1},\mu_{2},
\mu_{3}), \label{a.52prima} \ene

where,

\beq
\label{a.31.1.1.1.1}
 I_{sp}(\mu_{1},\mu_{2}, \mu_{3}):=
I_{ss}(\mu_{1},\mu_{2}, \mu_{3}, 0).
\ene

And
\beq \int_0^{z(\sigma)} \left\| g_{\sigma, z}(x)\cdot \mo \, \et
 \varphi \right\|\, dz \leq
\frac{\|g\|_\infty}{\pi^{1/4}\,\sigma} I_{sp}(\mu_{1},\mu_{2},
\mu_{3}).
\label{a.52prima.1}
\ene
\end{lemma}

\begin{lemma}\label{lemm-a.10}
Let $f: \ere^3 \rightarrow \CE$ be bounded and with support
contained in $D$. Then, for  $Z \geq h$, and $ \zeta  $ such that $
  \zeta  \leqq Z  $,

\beq \int_0^{Z- \zeta} \, \left\| f(x+(Z-(z + \zeta))\hv )\et e^{-i
\zeta H_1} \varphi\right\|\, dz\leq (Z
-\zeta)\frac{\|f\|_\infty}{\sqrt{2}} \,
e^{-\frac{1}{2}\theta_{inv}(\sigma, Z)^2}. \label{a.53} \ene
\end{lemma}

\noindent {\it Proof:} Estimating as in the proof of(\ref{a.11}) we prove that,
\beq
\begin{array}{l}
 \left\| f(x+(Z-(z+ \zeta))\hv )\et  \varphi\right\|^2 \leq \frac{\|f\|_\infty^2}{\pi^{3/2}} \int_{(-Z-h)\rho(\sigma, z + \zeta) }^{(-Z+h)
 \rho(\sigma, z + \zeta)} \, e^{-z^2}\, dz\,  \pi\, \int_0^{r_2 \rho(\sigma, z + \zeta)}\, e^{- r^2} 2r\, dr\\\\
 \leq
\frac{\|f\|_\infty^2}{2}\, e^{- \theta_{inv}(\sigma, Z)^2},
\end{array}
\label{a.54}
\ene
where we used (\ref{a.5}).

\begin{lemma}\label{lemm-a.11}
Let $g: \ere^3 \rightarrow C^3$ be bounded and with support
contained in $D$, suppose that $ \theta_{inv}(\sigma, Z)^2 \geq
\frac{1}{2} $. Then, for $ Z \geq h$, and $ \zeta $ such that $
\zeta  \leq Z $, we have that,

\beq \int_0^{Z - \zeta} \left\|
g(x+(Z-(z + \zeta))\hv)\cdot \mo \,\et e^{-i \zeta H_{1}}\varphi  \right\| \, dz \leq
(Z- \zeta )\frac{ \|g\|_\infty }{\pi^{1/4} \sigma}\, e^{ -
\frac{1}{2}\theta_{inv}(\sigma, Z)^2}\, \left[
\frac{-\theta_{inv}(\sigma,Z)}{2}+ \frac{3\sqrt{\pi}}{4}
\right]^{1/2}. \label{a.55} \ene

\end{lemma}

\noindent {\it Proof:} The lemma is proven estimating as in the
proof of (\ref{a.15}) using Remark \ref{rem-a.1}.

\section{Appendix B. Upper Bounds for the Integrals}
\sss
In this appendix we prove upper bounds for the integrals appearing in the terms
$I_{ps}, I_{pp},  I_{ss}  $ and $ I_{sp} $  (see
(\ref{a.31}), (\ref{a.33prima.1.menos1}), (\ref{a.31.1.1.1}), (\ref{a.31.1.1.1.1})).

Suppose that $ Z \geq s \geq \zeta, \delta_0 >0$. Designate $ \sqrt{\mathbb N}:=\{
0,1,\sqrt{2}, \sqrt{3}, \cdots\}$.
We denote,
\beq
\{Z_1,Z_2,\cdots, Z_K\}:= \sqrt{\delta_0} \sqrt{\mathbb N} \cap \left[
-\theta_{inv}(\sigma, s, s, \zeta), -\theta_{inv}(\sigma, Z, Z, \zeta)\right],
\label{b.1}
\ene
where $ Z_1 < Z_2 < \cdots < Z_K$.
 As $ -\theta_{inv}(\sigma, \tau, \tau, \zeta)  $  is increasing as a
 fuction of $ \tau  $ we have that,
\beq
s \leq z_{Z_1^{-1},\sigma}(\zeta) <  z_{Z_2^{-1},\sigma}(\zeta) <
z_{Z_3^{-1},\sigma}(\zeta)  < \cdots <  z_{Z_K^{-1},\sigma}(\zeta) \leq Z,
\label{b.2}
\ene

\begin{lemma}\label{lemm-b.1}

Suppose that $ Z \geq s \geq \zeta, r >0$, and let $ f: \ere \to \ere  $
satisfy $ f(\tau)\geq  \tau - 2\zeta  $. Then,
\beq
\begin{array}{l}
\int_s^Z \, d\tau \Upsilon (\sigma, \tau, f(\tau), \zeta)^{1/2} \leq \\\\
\frac{\pi^{1/4}}{\sqrt{2}}\, \big[
e^{- \frac{1}{2}\theta_{inv}(\sigma, s, s, \zeta)^2} (z_{Z_1^{-1},\sigma}(\zeta)-s)
+\sum_{j=1}^{K-1}
e^{-\frac{1}{2}Z_j^2}\, (z_{Z_{j+1}^{-1},\sigma}(\zeta)-z_{Z_j^{-1},\sigma}(\zeta))+
e^{-\frac{1}{2}Z_K^2}\, (Z- z_{Z_k^{-1},\sigma}(\zeta))\big],
\end{array}
\label{b.3}
\ene
\beq
\begin{array}{l}
\int_s^Z \, d\tau \Upsilon (\sigma, \tau, f(\tau), \zeta)^{1/2} \, e^{- \frac{r_1^2}{2}
  \rho(\tau)^2 } \leq
 \frac{\pi^{1/4}}{\sqrt{2}}[
 e^{- \frac{r_1^2}{2}  \rho( z_{Z_1^{-1},\sigma}(\zeta))^2
 }\,e^{-\frac{1}{2}\theta_{inv}(\sigma, s, s, \zeta)^2}  (z_{Z_1^{-1},\sigma}(\zeta)-s)
 \, +\\\\
\sum_{j=1}^{K-1}
  e^{- \frac{r_1^2}{2}\rho(z_{Z_{j+1}^{-1}, \sigma}(\zeta))^2 }
\, e^{-\frac{1}{2}Z_j^2}\,
 (z_{Z_{j+1}^{-1},\sigma}(\zeta)-z_{Z_j^{-1},\sigma}(\zeta)) \, +
 e^{- \frac{r_1^2}{2}{ \rho(Z)^2 } }
\, e^{-\frac{1}{2}Z_k^2}\,
 (Z-z_{Z_k^{-1},\sigma}(\zeta)) \,
],
\end{array}
\label{b.4}
\ene
\beq
\begin{array}{l}
\int_s^Z \, d\tau \Theta (\sigma, \tau, f(\tau), \zeta)^{1/2} \, e^{- \frac{r_1^2}{2}
  \rho(\tau)^2 } \leq
 \frac{1}{\sqrt{2}}[
 e^{- \frac{r_1^2}{2}  \rho( z_{Z_1^{-1},\sigma}(\zeta))^2
 }\,e^{-\frac{1}{2}\theta_{inv}(\sigma, s, s, \zeta)^2} ( Z_1+ \sqrt{\pi}/2)^{1/2} (z_{Z_1^{-1},\sigma}(\zeta)-s)
 \, +
\\\\
\sum_{j=1}^{K-1}
  e^{- \frac{r_1^2}{2}\rho(z_{Z_{j+1}^{-1}, \sigma}(\zeta))^{2} }
\, e^{-\frac{1}{2}Z_j^2}\,   (Z_{j+1}+ \sqrt{\pi}/2)^{1/2}\,
 (z_{Z_{j+1}^{-1},\sigma}(\zeta)-z_{Z_j^{-1},\sigma}(\zeta)) \, +  \\\\
 e^{- \frac{r_1^2}{2}{ \rho(Z)^2 }}
\, e^{-\frac{1}{2}Z_k^2}\, ( \theta_{inv}(\sigma,Z,Z,\zeta) + \sqrt{\pi}/2)^{1/2}\,
 (Z-z_{Z_k^{-1},\sigma}(\zeta)) \,
].
\end{array}
\label{b.5}
\ene
If moreover, $ r_1 \rho(Z)\geq 1$,
\beq
\begin{array}{l}
\int_s^Z \, d\tau \Upsilon (\sigma, \tau, f(\tau), \zeta)^{1/2} \,r_1 \rho(\tau) e^{- \frac{r_1^2}{2}
  \rho(\tau)^2 } \leq
 \frac{\pi^{1/4}}{\sqrt{2}}[
 r_1 \rho( z_{Z_1^{-1},\sigma}(\zeta))\\\\ e^{- \frac{r_1^2}{2}  \rho( z_{Z_1^{-1},\sigma}(\zeta))^2
 }\,e^{-\frac{1}{2}\theta_{inv}(\sigma, s, s, \zeta)^2}  (z_{Z_1^{-1},\sigma}(\zeta)-s)
 \, +
\sum_{j=1}^{K-1}
r_1 \rho(z_{Z_{j+1}^{-1}, \sigma}(\zeta)) e^{- \frac{r_1^2}{2}\rho(z_{Z_{j+1}^{-1}, \sigma}(\zeta))^2 }
\, e^{-\frac{1}{2}Z_j^2}\,
 (z_{Z_{j+1}^{-1},\sigma}(\zeta)-z_{Z_j^{-1},\sigma}(\zeta)) \, +  \\\\
r_1 \rho(Z) e^{- \frac{r_1^2}{2}{ \rho(Z)^2 }}
\, e^{-\frac{1}{2}Z_k^2}\,
 (Z-z_{Z_k^{-1},\sigma}(\zeta)) \,
]. \\\\
\end{array}
\label{b.6}
\ene
\end{lemma}

\noindent {\it Proof:} We split the integral in the left-hand side of (\ref{b.3}) as follows,
\beq
\begin{array}{l}
\int_s^Z \, d\tau \Upsilon (\sigma, \tau, f(\tau), \zeta)^{1/2}=
\int_s^{z_{Z_1^{-1},\sigma}(\zeta)} \, d\tau \Upsilon (\sigma, \tau, f(\tau), \zeta)^{1/2}+\sum_{j=1}^{K-1}
\int_{z_{Z_j^{-1},\sigma}(\zeta)}^{z_{Z_{j+1}^{-1},\sigma} (\zeta)} \, d\tau \Upsilon
(\sigma, \tau, f(\tau), \zeta)^{1/2}+
\\\\
\int_{z_{Z_K^{-1},\sigma}(\zeta)}^Z \, d\tau \Upsilon (\sigma, \tau, f(\tau), \zeta)^{1/2},
\end{array}
\label{b.7}
\ene
and we apply (\ref{a.5}). This proves (\ref{b.3}). (\ref{b.4}) is proved in a
similar way.  Equation (\ref{b.5}) is
proven in the same way,
 but using (\ref{a.6}).
Finally, we prove (\ref{b.6}) as above, using (\ref{a.5}) and observing  that
the function $x \,e^{-x^2/2}$ is decreasing for $ x \geq 1$.

\noindent {\bf Acknowledgement}

\noindent This work was partially done while we were visiting the project POems at Institut National de Recherche en Informatique et en
Automatique (INRIA) Paris-Rocquencourt. We thank Patrick Joly for his kind hopitality. We thank Luis Carlos Vel\'azquez for his help in writing the
Matlab code.

 \newpage

\begin{figure}
\begin{center}
\includegraphics[width=17cm]{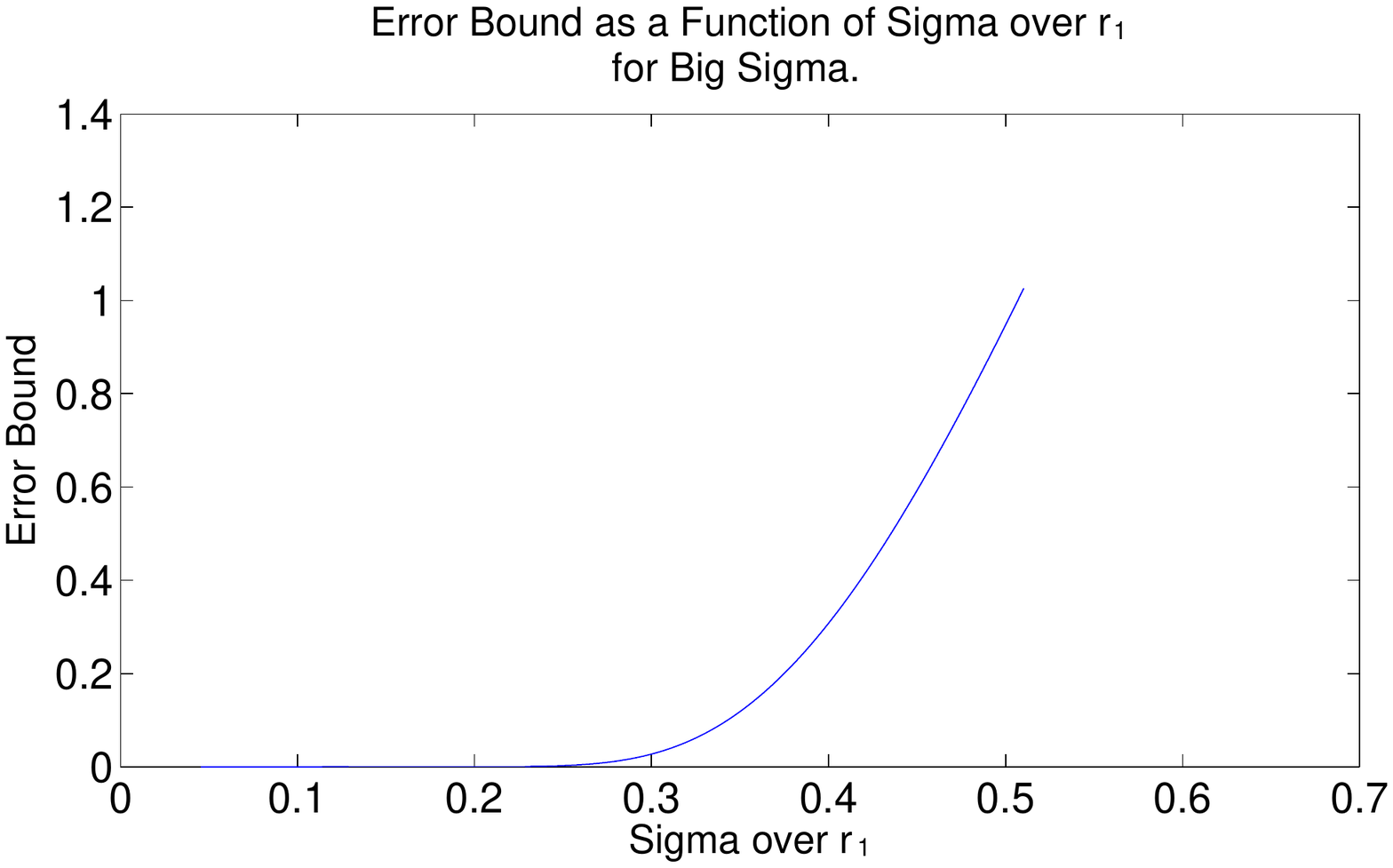}
\caption{Error bound as a function of $\sigma$ over $r_1$ for big sigma}\label{big-sigma-figure}
\end{center}
\end{figure}

\newpage

\begin{figure}
\begin{center}
\includegraphics[width=17cm]{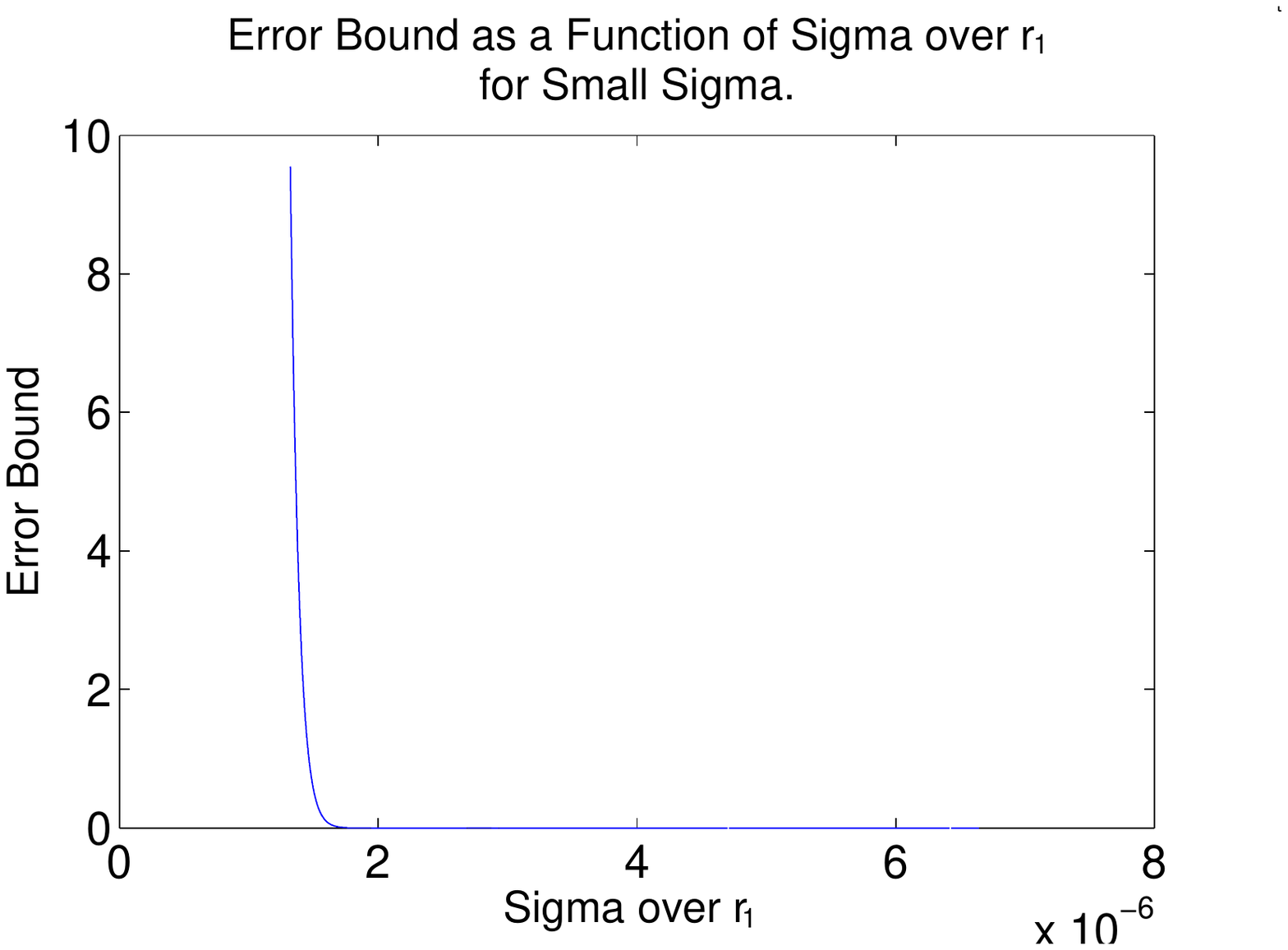}
\caption{Error bound as a function of $ \frac{\sigma}{r_1} \times 10^{-6} $ for small sigma} \label{small-sigma-figure-prima}
\end{center}
\end{figure}

\newpage

\begin{figure}
\begin{center}
\includegraphics[width=17cm]{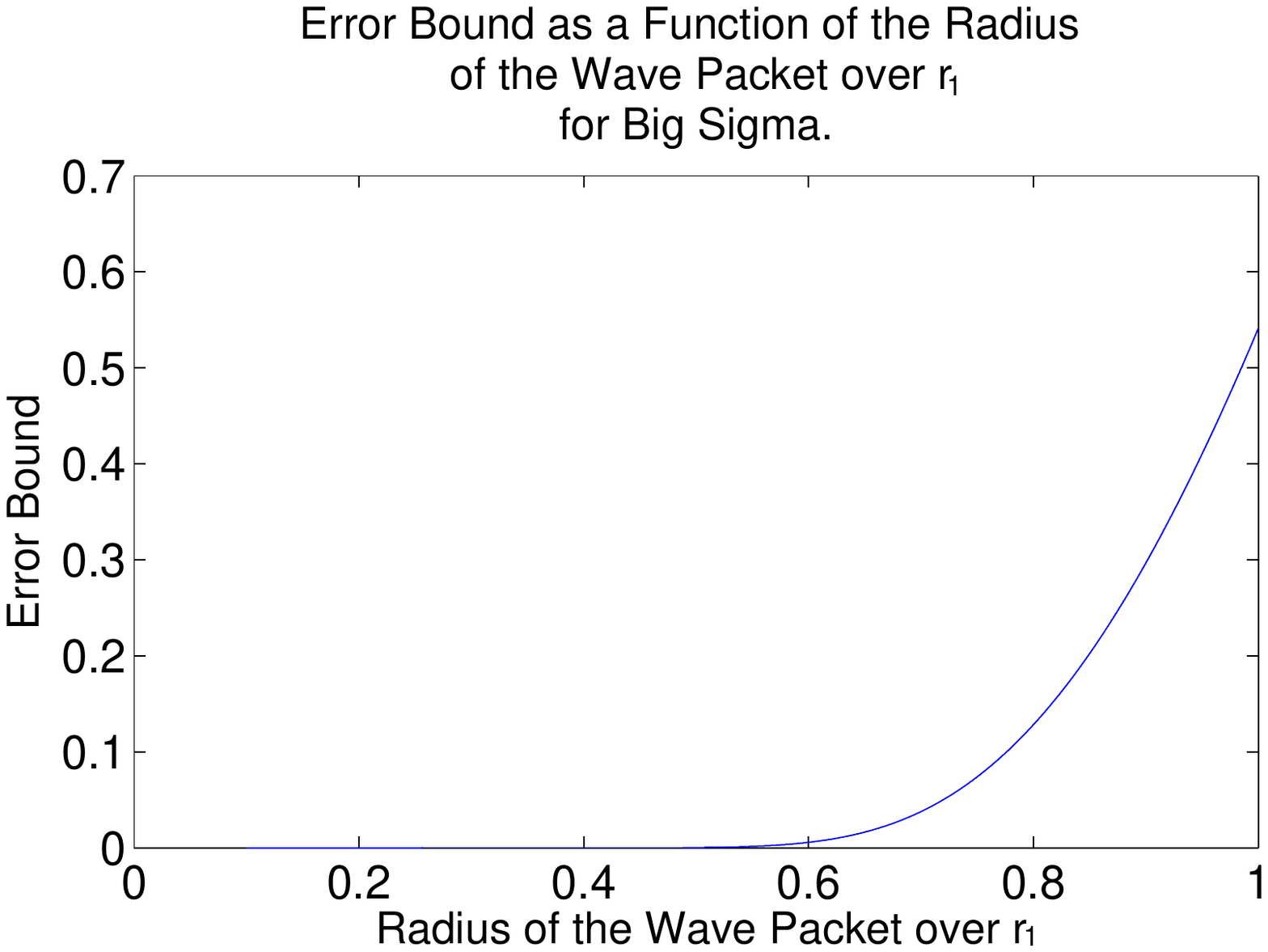}
\caption{Error bound as a function of the radius of the wave packet over $r_1 $
  for big sigma} \label{radius-figure}
\end{center}
\end{figure}


\newpage
\begin{figure}
\begin{center}
\includegraphics[height=20cm]{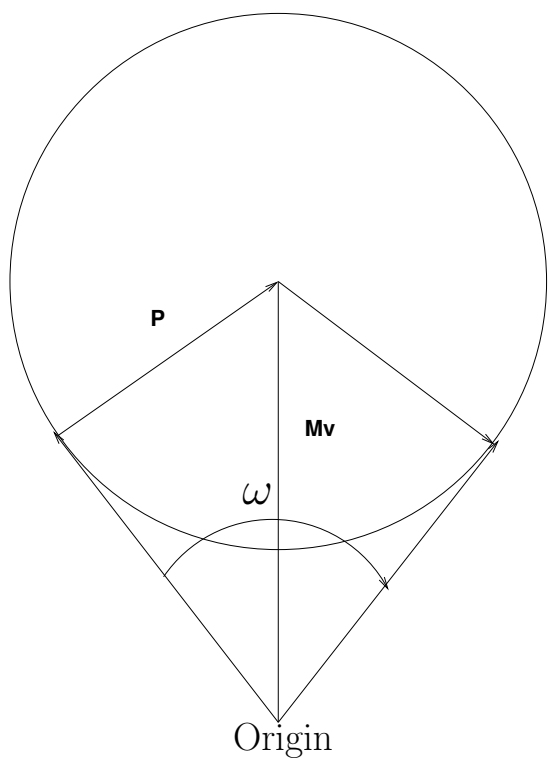}
\caption{Opening angle}\label{spreading-angle-draw}
\end{center}
\end{figure}


\newpage

\begin{figure}
\begin{center}
\includegraphics[width=17cm]{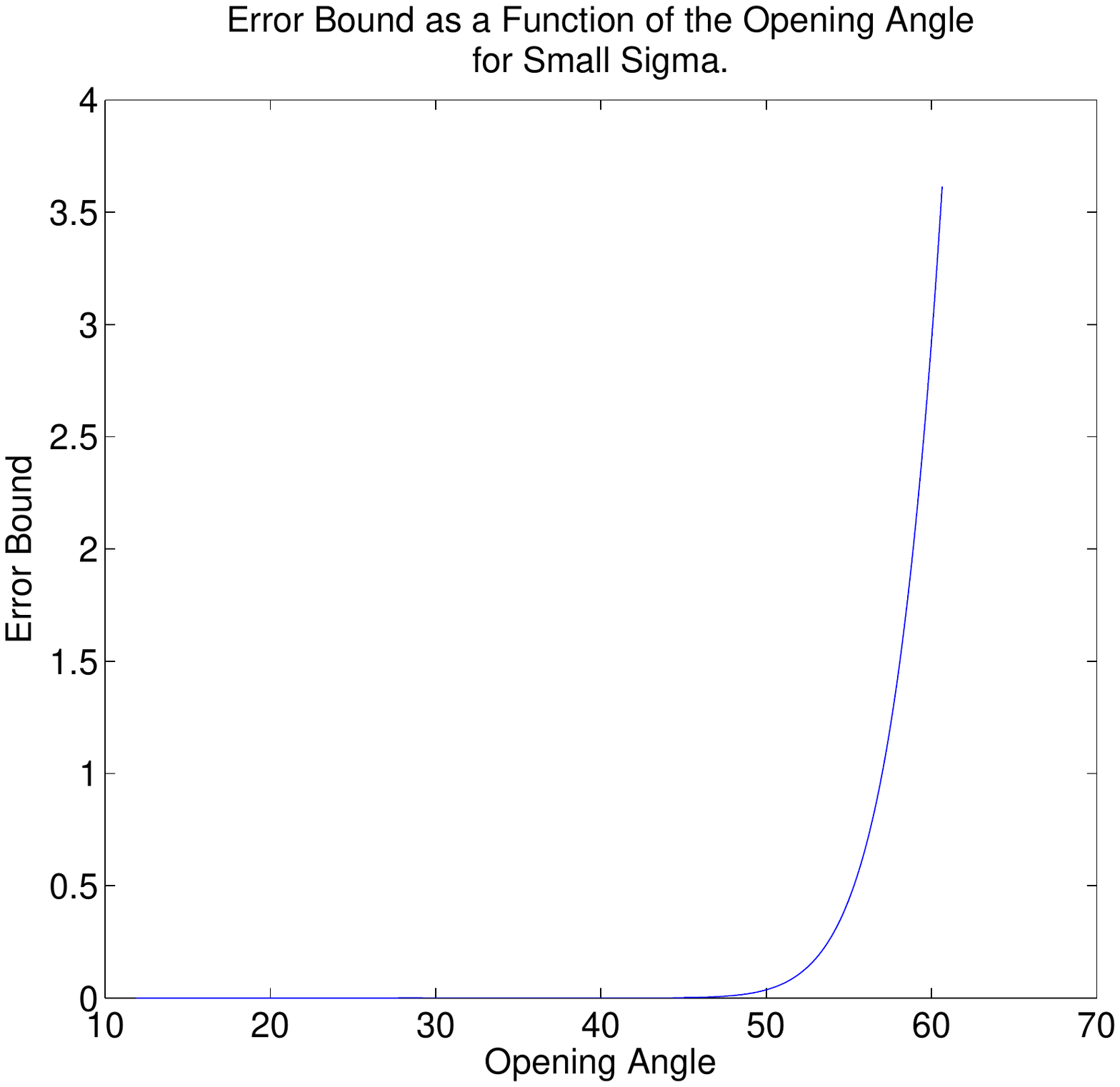}
\caption{Error bound as a function of the opening angle for small sigma}\label{spreading-angle-figure}
\end{center}
\end{figure}

\end{document}